\documentclass[a4paper,fleqn,usenatbib]{mnras}
\usepackage{mathptmx}
\usepackage[T1]{fontenc}
\usepackage{ae,aecompl}
\usepackage{graphicx}	
\usepackage{amsmath}	
\usepackage{amssymb}	
\usepackage{natbib}
\usepackage{bm}			

\title[Tidal oscillations: predicted observables]{Tidally induced stellar oscillations: converting modelled oscillations excited by hot Jupiters into observables}

\author[A. Bunting, C. Terquem]{
Andrew Bunting$^{1}$\thanks{andrew.bunting@physics.ox.ac.uk} and
Caroline Terquem$^{1,2}$\thanks{caroline.terquem@physics.ox.ac.uk}
\\
$^{1}$Department of Physics, Oxford University, Keble Road, Oxford OX1 3RH, UK\\
$^{2}$ Institut d'Astrophysique de Paris, Sorbonne Universit\'e, CNRS,
  UMR 7095, 98 bis boulevard Arago, F-75014, Paris, France \\
}

\pubyear{2020}

\begin{document}
\label{firstpage}
\pagerange{\pageref{firstpage}--\pageref{lastpage}}
\maketitle

\begin{abstract}
We calculate the conversion from non--adiabatic, non--radial oscillations tidally induced by a hot Jupiter on a  star to observable spectroscopic and photometric signals.  Models with both frozen convection and an approximation for a perturbation to the convective flux are discussed. Observables are calculated for some real planetary systems to give specific predictions. Time--dependent line broadening and the radial velocity signal during transit are both investigated as methods to provide further insight into the nature of the stellar oscillations. The photometric signal is predicted to be proportional to  the inverse square of the orbital period, $P^{-2}$, as in the equilibrium tide approximation.  However, the radial velocity signal is predicted to be proportional to $ P^{-1}$, and is therefore much larger at long orbital periods than the signal corresponding to the equilibrium tide approximation, which is proportional to $P^{-3}$. The prospects for detecting these oscillations and the implications for the detection and characterisation of planets are discussed.
\end{abstract}


\section{Introduction}

Stars have been known to vary for millenia, with both binary systems \citep{Jetsu2015} and variable stars \citep{Hoffleit1997} being observed to periodically change in brightness. More recently, the Sun was found to exhibit periodic variation in its surface velocity, with velocity fields being detected across the solar surface \citep{Leighton1962}. Many other stars have since been found to exhibit similar oscillations, detected both through radial velocity (RV) measurements (such as \citet{Brown1991, Kjeldsen2003}) and photometrically (see \citet{Chaplin2013} and \citet{DiMauro2017} for reviews), and such oscillations have even been observed in Jupiter \citep{Markham2018}.

Such oscillations are excited by internal processes, whether convection in the case of solar--type stars \citep{Kjeldsen1995} or `rock storms' on Jupiter \citep{Markham2018}. In these cases, the information content of the oscillations is primarily held within their frequencies, which has been used to great effect in investigating the structure of the Sun \citep{Deubner1984}. The presence of an external perturber can similarly drive oscillations, though the information is contained within the amplitude and phase of the response, as the frequency is determined by the orbit of the perturber \citep{Burkart2012}.

Tidally excited oscillations have been studied in the context of orbital evolution (\citet{Savonije1983}; \citet{Goldreich1989}; \citet{Smeyers1998}), which occurs because  energy is dissipated in the stellar interior as the oscillations are damped, and angular momentum is transferred from the perturber's orbit to the star's rotation, or vice versa.

Tidal oscillations have also been investigated more directly, as they lead to  the surface of the star varying periodically. This can result in both photometric and spectroscopic variations, and has been investigated in the context of stellar binaries \citep{Quataert1995} and planetary companions, for eccentricities both small (\citet{Terquem1998}; \citet{Arras2012}) and large (\citet{Burkart2012}; \citet{Fuller2017}; \citet{Penoyre2018}).

Previous work has set out the framework in which to convert the behaviour of the stellar surface into observable signals for both RV (\citet{Dziembowski1977}; \citet{Arras2012}) and photometric (\citet{Dziembowski1977}; \citet{Pfahl2008}) signals. This work will build upon and extend these formalisms, particularly in asserting the importance of considering the non--radial components of the displacement of the stellar surface, and investigating the full spectroscopic RV signal in terms of time--dependent line--broadening.

These variations have been observed photometrically (\citet{Welsh2010}; \citet{Mazeh2010}; \citet{Mislis2012}) and spectroscopically (\citet{Maciejewski2020}), though it is possible that other tidal RV signals may have been mistakenly attributed to non--zero orbital eccentricities \citep{Arras2012}.

Observations of tidal oscillations could be used to derive  the parameters of the system under investigation. Transiting planets can have their masses directly inferred, whilst planets discovered using the RV method could have the degeneracy between the inclination and mass broken; if both transit and RV data already exist, an independent measurement of the planetary mass can be made.   Providing tighter constraints on the system parameters can then be used to test our models of both planetary atmospheres, formation and orbital evolution.

In order for this to be done accurately, the oscillations must be well modelled. The equilibrium tide approximation, whilst found to be reasonable throughout the bulk of the star \citep{Pfahl2008}, breaks down at the stellar surface, where non--adiabatic effects become prominent (\citet{Henyey1965}; \citet{Savonije1983}; \citet{Arras2012}; \citet{Houdek2017}; \citet{Fuller2017}). The fully non--adiabatic stellar oscillation equations are solved here for the case of a periodic, tidal perturbation, as set out in \citet{Bunting2019}, with particular focus given to modelling the response at the surface.

Section~\ref{sec:Methods} addresses the procedure for converting the modelled oscillations into an observable signal, for both the observed flux variation (section~\ref{sec:Methods:lum_var}) and radial velocity variation (section~\ref{sec:Methods:RV_var}). Some alternative methods for observing the tidal oscillations are discussed in section~\ref{sec:Methods:alternative}. The predicted observable signals are explored for a test case in section~\ref{sec:test_cases}, beginning with a summary of relevant results from \citet{Bunting2019} in section~\ref{sec:recap}, then addressing behaviour for long periods (\ref{sec:test_cases:long_P}), short periods (\ref{sec:test_cases:short_P}), resonances (\ref{sec:test_cases:resonances}), general trends in behaviour (\ref{sec:test_cases:general}) and variation with stellar mass (\ref{sec:test_cases:mass}). The observable signals for observed systems are presented in section~\ref{sec:observed_systems}, and discussed in section~\ref{sec:Discussion}, whilst section~\ref{sec:Conclusion} concludes this work.

\section{Methods}
\label{sec:Methods}

In this section, the conversion from the calculated behaviour of the stellar surface into an observable signal is detailed. Initially, the set-up of the problem and the conventions used in the calculations are described in section~\ref{sec:Methods:set-up}. The  observed flux variation calculation is addressed in section~\ref{sec:Methods:lum_var}, followed by the signals due to spectroscopic variation in section~\ref{sec:Methods:RV_var}.

\subsection{Set-up}
\label{sec:Methods:set-up}

We follow \cite{Bunting2019} and assume a non--rotating star with polar coordinates $(r, \theta_{*}, \phi_{*})$ centred on the star, with the planetary companion existing in a circular orbit with $\theta_{*} = \pi / 2$. In this frame, the observer is taken to be in the direction given by $(\theta_{0}, \phi_{0})$.

In the observer's frame, described by $(r, \theta_{\text{ob}}, \phi_{\text{ob}})$, the observer is at $\theta_{\text{ob}} = 0$, with $\theta_{\text{ob}} < \pi / 2$ visible to the observer.  The unit vectors associated with these coordinate systems are denoted with a {\em hat}.  The epoch of inferior conjunction is used to define the origin of the time coordinate and for the orbital phase.

In the frame of the star, the leading order non--constant term of the tidal perturbation has the form of a spherical harmonic with $l = m = 2$. Therefore the response of the star will have the form:
\begin{equation}
\label{eq:spherical_form}
q'(r, \theta_{*}, \phi_{*}, t) = \Re \left( q'(r) 3 \sin^{2} \theta_{*} e^{2 \text{i} (\omega t - \phi_{*})} \right),
\end{equation}
where $q'(r)$ varies only with the radial coordinate, $r$, and $\Re$ denotes the taking of the real part.  Here, $\omega$ is the orbital frequency,  and it can be seen from the above expression that the frequency of the perturbation is twice the orbital frequency.
 This form applies to $\xi_{r}$, the radial displacement, $F'_{r}$, the perturbation to the radial flux, and $V$, which gives the horizontal displacement, $\bm{\xi}_{h} = \bm{\xi} - \xi_{r} \bm{\hat{r}}$, through $\bm{\xi}_{h} = r \bm{\nabla}_{h} V$, with $\bm{\nabla}_{h}$ being the non-radial component of the gradient operator.

Here, we are interested in close binary systems for which the orbital period, and hence the period of the tidal perturbation, is in the range of a fraction of a day to a few days.   Forcing at such frequencies excites gravity modes in the star, which propagate in the radiative zone and are evanescent in the convective envelope.   Additional inertial modes would be excited in the convective envelope of the star if it rotated  with an angular velocity (assumed uniform) $\Omega_{\rm rot}$ such that $\omega \le 2  \Omega_{\rm rot}$  (see, e.g., \citet{Ogilvie2004}).  These modes are not be taken into account in this paper, and our results therefore only apply to systems in which the star is slowly rotating, with a period larger than twice the orbital period of the planet.  This assumption is expected to hold for most solar--type stars hosting planets.  However, stellar rotation may have to be taken into account for more massive stars, as discussed in \citet{Arras2012}.  In that case, inertial modes may contribute to signals of the type discussed here, as shown by \citet{Lanza2019}, who considered   purely toroidal inertial modes.  

To convert between the frame of the observer and that of the star, we use the properties of spherical harmonics and Euler angles, guided by \citet{Morrison1987} and detailed in Appendix~\ref{App:angles}. For integrations over the visible disc we need only keep track of one spherical harmonic (as the integral over $\phi_{\text{ob}}$ will eliminate terms with $e^{i \mu \phi_{\text{ob}}}$ where $\mu \ne 0$). This allows us to convert the expression of the star's tidal response into the coordinates of the observer's frame, as (eq.~[\ref{eq:App:spherical_useful}]): \begin{align} \label{eq:spherical_conversion}
  \int_{0}^{2 \upi} 3 \sin^{2} \theta_{*}  e^{2 \text{i} (\omega t - \phi_{*})} {\rm d}\phi_{\rm ob}  & \nonumber \\
 =  3 \upi \sin^{2} \theta_{0} &  (3 \cos^{2}  \theta_{\text{ob}} - 1 ) e^{2 \text{i} (\omega t - \phi_{0})}.
\end{align}

To account for $\bm{\xi}_{h}$ we use the fact that the relation $\bm{\xi}_{h} = r \bm{\nabla}_{h} V$ is true independent of the orientation of the coordinate system used.
This gives the expression for the horizontal displacement as:
\begin{align}
  \int_{0}^{2 \upi} {\xi}_{\theta_{\rm ob}} (r, & \theta_{\text{ob}},   \phi_{\text{ob}}, t) {\rm d}\phi_{\rm ob}    \nonumber \\
 = & \Re \left[ \frac{\partial}{\partial \theta_{\rm ob} } \int_{0}^{2 \upi} V(r) 3 \sin^{2} \theta_{*}  e^{2 \text{i} (\omega t - \phi_{*})} {\rm d}\phi_{\rm ob} \right] ,
   \nonumber \\
= & \Re \left[ - 18 \upi V(r) \sin^{2} \theta_{0} 
    \cos  \theta_{\text{ob}} \sin  \theta_{\text{ob}} e^{2 \text{i} (\omega t - \phi_{0})} \right],
\label{eq:xi_h_disc_integral}
\end{align}
where we have retained only the component which will be observed, as the displacement perpendicular to the direction towards the observer does not contribute since $\hat{\mathbf{\phi}}_{\rm ob} \mathbf{\cdot} \hat{\mathbf{n}}_{\rm ob} = 0$, where $\hat{\bm{n}}_{\text{ob}}$ is the unit vector towards the observer.

\subsection{Variation of the observed stellar flux}
\label{sec:Methods:lum_var} 

The  observed stellar flux is given by
\begin{equation}
\label{eq:Methods:full_lum}
L = \int \int h \bar{F} \hat{\mathbfit{n}}_{\text{ob}} \mathbf{\cdot} \hat{\mathbfit{n}} \text{d} S
\end{equation}
where $h$ is the limb-darkening (note that this is wavelength dependent), $\bar{F}$ is the emergent flux, equal to $\bm{F} \bm{\cdot} \hat{\bm{n}}$, $\hat{\bm{n}}$ is the unit vector normal to the surface, and $\text{d} S$ is the surface area element. 

In order to evaluate the first--order change in observed flux, each of these terms must be evaluated as a function of perturbations to the equilibrium state, which will result in first order changes in  observed flux due to limb--darkening, flux, surface normal, and surface area, corresponding respectively to $\Delta L_{\text{h}}$, $\Delta L_{F}$, $\Delta L_{n}$, and $\Delta L_{S}$. Full details of the derivation can be found in Appendix~\ref{App:Lum}.

The equilibrium observed flux is given by
\begin{equation}
\label{eq:Methods:lum_var:equilibrium}
L_{0} = \int_{0}^{2 \pi} \int_{0}^{\frac{\pi}{2}} h_{0} \bar{F}_{0} \hat{\mathbfit{r}} \mathbf{\cdot} \hat{\mathbfit{n}}_{\text{ob}} \text{d} S_{0} ,
\end{equation}
where the subscript $0$ indicates that it is the equilibrium value; $\bar{F}_{0} = \mathbfit{F}_{0} \mathbf{\cdot} \hat{\mathbfit{r}}$ where $\mathbfit{F}_{0}$ is the vector equilibrium radiative flux at the surface (equal to $F_{0} \hat{\mathbfit{r}}$, where $F_{0}$ is its magnitude); $\hat{\mathbfit{r}}$ is the radial unit vector, which is the surface normal for the equilibrium case; $\text{d}S_{0} = R^{2} \sin \theta_{\rm ob} \text{d}\theta_{\rm ob} \text{d}\phi_{\rm ob}$, where $R$ is the equilibrium radius of the star. The integral is calculated in the observer's frame, so that $\hat{\mathbfit{r}} \mathbf{\cdot} \hat{\mathbfit{n}}_{\text{ob}} = \cos \theta_{\rm ob}$.   Using a quadratic limb darkening law, we have: $h_{0} = c[1 - a(1 -  \hat{\mathbfit{r}} \mathbf{\cdot} \hat{\mathbfit{n}}_{\text{ob}} ) - b ( 1 - \hat{\mathbfit{r}} \mathbf{\cdot} \hat{\mathbfit{n}}_{\text{ob}} )^{2} ]$, where $a$ and $b$ parametrise the limb darkening and $c$ normalises it, such that $\int_{0}^{1} \mu h \text{d}\mu = 1$, with $\mu = \cos \theta_{\rm ob}$. As found by \citet{Arras2012}, the disc-integrated values vary weakly with the choice of limb darkening law, so for simplicity Eddington limb-darkening is used here, with $c=5/2$, $a=3/5$ and $b = 0$ \citep{Dziembowski1977}, although the full quadratic limb darkening coefficients are retained in the general form of the equations.

The explicit expression for the equilibrium observed flux is then found to be
\begin{equation}
\label{eq:Methods:lum_var:equilibrium:explicit}
L_{0} = 2 \pi R^{2} F_{0} ,
\end{equation}
where the factor of 2 comes from the definition of the normalisation of the limb darkening.

The first order perturbations are calculated by expanding the four terms in the integrand into their equilibrium and first order terms, as detailed in Appendix~\ref{App:Lum}, resulting in the following expressions:

\begin{align}
\label{eq:Methods:lum_var:DeltaL_h:explicit}
\Delta L_{h} & =  \Re \left[ \frac{12 \pi}{5} R c  F_{0} (a + \frac{3 b}{4}) \left[ V(R) - \xi_{r}(R) \right] \sin^{2} \theta_{0} \text{e}^{2 \text{i}(\omega t - \phi_{0})} \right], \\
  \Delta L_{F} & = \Re \left[ \frac{3 \pi}{4} R^{2} c \left( 1 + \frac{a + 2b}{15} \right) \left( F_{r}'(R)+ \xi_{r}(R) \frac{{\rm d} F_{0}}{{\rm d} r} \right) \right.  \nonumber \\
             & \hspace*{4.5cm} \left.  \times \sin^{2} \theta_{0} \text{e}^{2 \text{i}(\omega t - \phi_{0})} \right],  \label{eq:Methods:lum_var:DeltaL_F:explicit} \\
\label{eq:Methods:lum_var:DeltaL_n:explicit}
\Delta L_{n} & =  \Re \left[ \frac{9 \pi}{2}  R c \left( 1 - \frac{7a + 4b}{15} \right) F_{0} \left[ V(R) - \xi_{r}(R) \right] \right.  \nonumber \\
             & \hspace*{4.5cm} \left.  \times
             \sin^{2} \theta_{0} \text{e}^{2 \text{i}(\omega t - \phi_{0})} \right] , \\
\label{eq:Methods:lum_var:DeltaL_S:explicit}
\Delta L_{S} & = \Re \left[ \frac{3 \pi}{2} R c \left( 1 + \frac{a + 2b}{15} \right) F_{0} \left[ \xi_{r}(R) - 3 V(R) \right] \right.  \nonumber \\
             & \hspace*{4.5cm} \left.  \times
             \sin^{2} \theta_{0} \text{e}^{2 \text{i}(\omega t - \phi_{0})} \right],
\end{align}
where $F_{r}'$ is the Eulerian perturbation to the flux, as defined in appendix~\ref{App:Lum:F}. These can be combined to give the fractional total change in  observed flux as:
\begin{equation}
\label{eq:Methods:lum_var:Delta_L_TOT}
\frac{\Delta L}{L_{0}} = \Re \left[ \frac{3}{8} c \left( 1 + \frac{a + 2b}{15} \right) \left( \frac{\Delta F_r}{F_0} - 4 \frac{\xi_{r}(R)}{R} \right) \sin^{2} \theta_{0} \text{e}^{2 \text{i}(\omega t - \phi_{0})} \right] .
\end{equation}
where:  
\begin{equation}
\Delta F_r = F_{r}'(R)+ \xi_{r}(R)  \frac{{\rm d} F_{0}}{{\rm d} r}(R).
\label{eq:lagFr}
\end{equation}
Whilst the individual effects may involve the horizontal components of displacement, as noted by \citet{Heynderickx1994} they cancel out exactly to first order.

\subsection{Radial velocity variation}
\label{sec:Methods:RV_var}

The periodic change in shape of the star results in a periodic change in the velocity of any given surface element. Projecting this along the observer's line of sight gives the radial velocity (RV) which is proportional to the shift in wavelength caused by the motion (for the very non-relativistic motions considered here). Expressing this formally gives
\begin{equation}
\label{eq:Methods:RV:RV_original}
{\rm v}_{\text{RV}} = - \dot{\mathbfit{r}} \mathbf{\cdot} \hat{\mathbfit{n}}_{\text{ob}},
\end{equation}
where $\mathbfit{r}$ is the vector from the centre of the star to the surface element in question.

To first order in perturbed quantities, this becomes

\begin{equation}
\label{eq:Methods:RV:RV_first_order}
{\rm v}_{\text{RV}} = - \dot{\mathbf{\xi}} \mathbf{\cdot} \hat{\mathbfit{n}}_{\text{ob}} = \Re \left( - 2 \text{i} \omega \mathbf{\xi} \mathbf{\cdot} \hat{\mathbfit{n}}_{\text{ob}} \right).
\end{equation}

This can be encapsulated by a single curve by integrating over the disc, weighted by the observed flux, as done by \citet{Dziembowski1977}:
\begin{equation}
\label{eq:Methods:RV:disk_integrated:general}
\rm{v}_{\text{disc}}  = \frac{\iint \mathnormal{h} \hat{\mathbfit{r}} \mathbf{\cdot} \hat{\mathbfit{n}}_{\text{ob}}\bar{F}_{0} v_{RV} {\rm d}S }{\iint \mathnormal{h} \hat{\mathbfit{r}} \mathbf{\cdot} \hat{\mathbfit{n}}_{\text{ob}} \bar{F}_{0} {\rm d}S} = \frac{1}{2 \pi R^2} \int \int \mathnormal{h} \hat{\mathbfit{r}} \mathbf{\cdot} \hat{\mathbfit{n}}_{\text{ob}} v_{RV} {\rm d}S,
\end{equation} 
which can be analytically solved.

\noindent Evaluating this integral (see appendix~\ref{App:RV_var} for details) gives the final expression for the disc--integrated radial velocity as:
\begin{multline}
  \rm{v}_{\text{disc}} = \Re \left\{ - \frac{4}{5} \rm{i} \omega c \bigg[
  \left( 1 - \frac{\mathnormal{a}}{16} + \frac{b}{56} \right) \xi_{r}(R)+  \right.  \\ 3    \left.  \left( 1 - \frac{3 a}{8} - \frac{5 b}{28} \right) V(R) \bigg]   \sin^{2} \theta_{0} \rm{e}^{2 \rm{i} (\omega t - \phi_{0})}  \right\} .
  \label{eq:Methods:RV:disk_integrated:final}
\end{multline} 

It is worth noting the dependencies on the orientation of the system: $\sin^{2} \theta_{0}$ differs from the $\sin \theta_{0}$ dependence in standard RV detections, and $\phi_{0}$ introduces a phase difference, as expected, but non--adiabaticity at the surface (exhibited by non--zero imaginary components of variables) can also lead to a phase difference.

Alternatively, the full spectrum of the line--broadening can be computed by numerically integrating over the visible disc to calculate the  observed flux as a function of radial velocity. This enables the time--dependent line--broadening to be evaluated, and a non--trivial signal is present at all orientations of the system (although, for a system where $\theta_{0} = 0$, the signal is time--independent).
This is further discussed in  appendix~\ref{App:non-disc:inhomogeneous}.

\section{Alternative method of observation: signal during transit}
\label{sec:Methods:alternative}

Above, we have described the standard methods used for detecting the tidal signal at the stellar surface.  In this section, we set out  an alternative method which would provide opportunities to deduce further information about the oscillations.  Further details can be found in Appendix~\ref{App:non-disc}.


As the planet occludes part of the stellar disc, comparing the signal at a point during transit to the signal during the secondary eclipse (when the planet is blocked from view by the star) provides the opportunity to isolate the signal originating from a given location on the stellar surface. This enables the horizontal and radial displacements to be disentangled, as the relative contributions to the RV signal from $\xi_{r}$ and $V$ vary differently across the face of the stellar disc, with $V$ and $\xi_r$ being more prominent at the edges and towards the centre, respectively. 

As the planet crosses the disc, the RV signal at a particular radial velocity will be reduced, with the reduction in brightness depending on the limb--darkening and how much of the rest of the disc is also producing a signal at the same radial velocity.

During a transit,  we can approximate $\theta_{0} \approx \upi / 2$ and, by taking the midpoint of the transit to occur at $t=0$, $\phi_{0} \approx 0$, enabling an analytical approximation for the value of $\rm{v_{RV}}$ as a function of position on the stellar disc. Approximating the  velocity of the planet as being constant over the course of the transit then gives an analytical expression for the value of $\rm{v_{RV}}$ blocked as a function of time. More details can be found in Appendix\,\ref{App:non-disc:transit:RV}.

This method could also be applied to the brightness variation in order to calculate the extra change in brightness due to blocking that part of the perturbed flux signal, although this contribution would be a factor of $F' / F_{0}$ smaller than the transit signal, so would be very difficult to detect.


In principle, another method of observation could use the fact that the differential motion of the stellar surface produces an inhomogeneous broadening effect \citep{Cunha2007}. Throughout the oscillation, the shape of the broadening will change, depending on  $\xi_{r}$, $V$ and the inclination. 

As a constant line--broadening would be indistinguishable from other sources of constant line--broadening (such as Doppler broadening due to the non--zero temperature at the stellar surface), we would only expect to be able to discern a periodically changing line--width, due to the change in line--shape from the surface motion.

However, a realistic calculation has to take into account all sources of line broadening, and the observable signal is a convolution of the broadening kernel due to the tidal motion of the surface with any other broadening kernels.  This is discussed in more details in Appendix~\ref{App:non-disc:inhomogeneous}.  
It is not clear whether the contribution from the tidal motion to line broadening could be detected, as  it is likely to be dominated by  thermal broadening.

\section{Results for test cases}
\label{sec:test_cases}

In order to understand the implications and limitations of these tidal oscillations more generally, the test cases used in \citet{Bunting2019} are explored as examples here. Extreme cases are used to show where this approach may break down, and why, and general relationships are also demonstrated.

\begin{figure*}
       \centering
        \includegraphics[width=8cm]{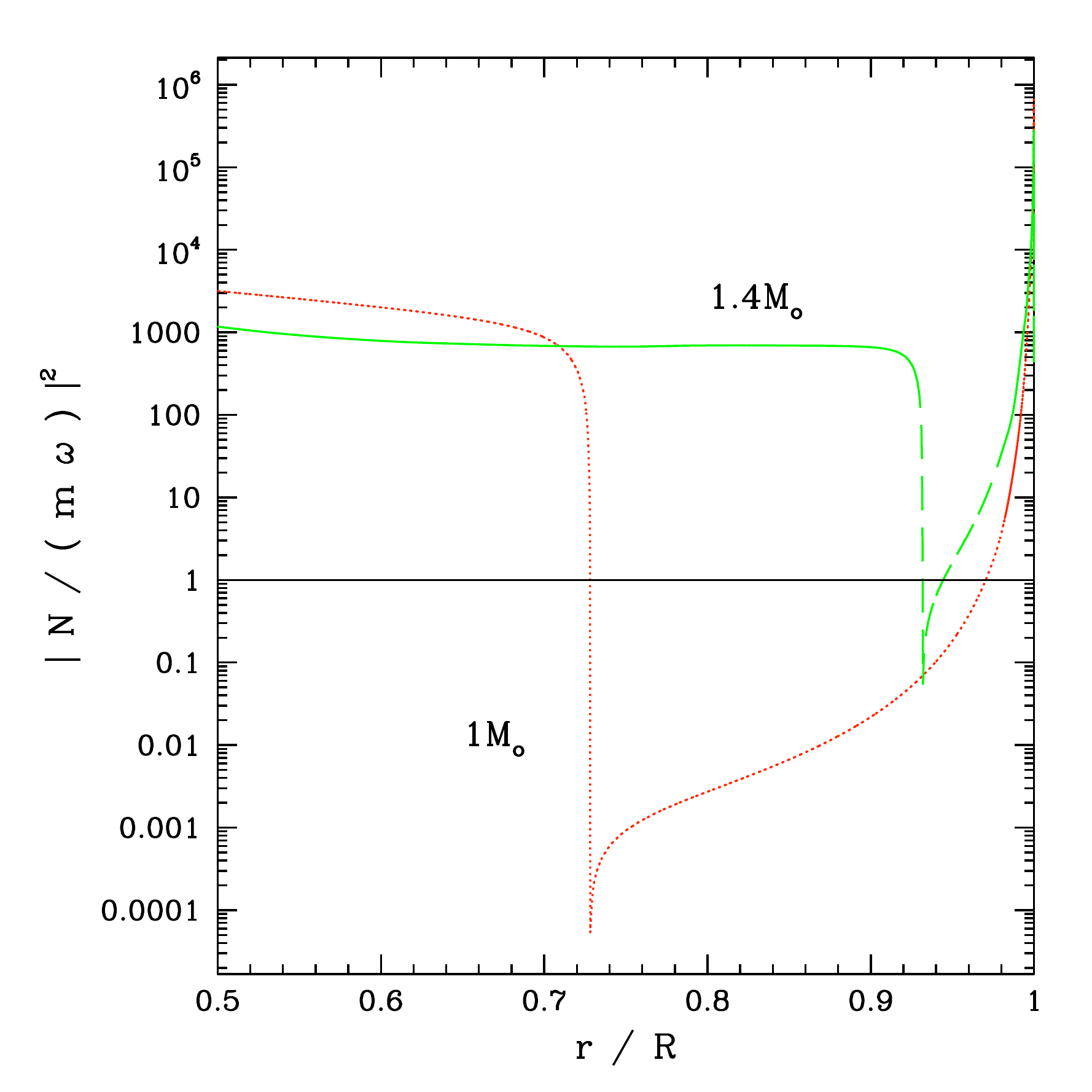}
        \caption{This figure shows the absolute value of $N^{2}/(m^{2} \omega^{2})$ as a function of $r/R$, where $R$ is the stellar radius, $\omega$ is the orbital frequency and $m=2$.  Here the orbital period is $4.23$ days.  Only  the outer 50\% of the stellar radius is shown. The red dotted line corresponds to a $1$~$\text{M}_{\odot}$ star, and the green dashed line corresponds to a $1.4$~$\text{M}_{\odot}$ star.  The quantity $N^2 /(m^{2} \omega^{2})$ indicates the structure of the star: a negative value corresponds to an imaginary frequency and implies convection; a positive value indicates a real frequency and therefore a stratified, radiative region.  This quantity passes through zero near $ r/R = 0.73$  and $0.94$  for the lower and higher mass, respectively, which correspond to the inner boundaries of their  convective envelopes. This behaviour is not fully resolved in the plots.}
    \label{fig:N2}
\end{figure*}

The basic model used in \citet{Bunting2019} was roughly modelled on  51 Pegasi b, with a solar mass star, and a Jupiter mass planet orbiting with a period of $4.23$~days, modelled using both frozen and perturbed convection. This will be the model to which we are comparing the results discussed in this section. In the reference model, the Brunt--V\"ais\"al\"a frequency is much greater than the oscillation frequency within the radiative core, as seen in  Figure~\ref{fig:N2}, with a sharp transition at the bottom of the convection zone to being much less than the oscillation frequency. By changing the orbital frequency of the planet, we are able to change the locations in the star at which the Brunt--V\"ais\"al\"a frequency and the forcing frequency will be comparable.  This is important because, as shown in \citet{Bunting2019}, the equilibrium tide approximation fails when the forcing frequency is small but larger than the Brunt--V\"ais\"al\"a frequency.

\subsection{Background and brief summary of relevant results from \citet{Bunting2019}}
\label{sec:recap}

This work is primarily an exploration and application of the modelling work as set out in \citet{Bunting2019}, where the non--adiabatic stellar oscillation equations are solved for the case of a tidal perturbation. The equilibrium tide solution was found to be approximately valid within the body of the star, but broke down towards the surface where the equilibrium tide is inconsistent with non-adiabaticity. This leads to behaviour in a thin region at the surface which can deviate significantly from the equilibrium tide.

The exact response at the surface depends upon the model used for convection, though some general trends emerge whether convection is treated as frozen (that is, the convective flux is assumed to be unchanged by the perturbation) or allowed to be perturbed. The radial displacement and flux perturbation were found to scale with the equilibrium tide, proportional to $P^{-2}$ (where $P$ is the orbital period), whilst the horizontal displacement was approximately constant, independent of $P$. At long periods, the horizontal displacement is therefore likely to dominate the stellar response.

Compared to the equilibrium tide at the surface, $\xi_{r}$  is decreased by an order of magnitude in the frozen convection case, whereas allowing a perturbation to the convective flux gives a similar value of $\xi_{r}$.  The flux at the surface is found to be different from the equilibrium value in both cases, with the perturbed convection case giving a greatly amplified value.

The different models of convection present different obstacles, but together show the dependence of the response on the choice of model for convection, as well as highlighting which features persist independent of that choice. When convection is frozen, non--adiabatic behaviour within the convection zone is artificially suppressed, and it is found that the stellar response exhibits large changes over a short scale in and around the radiative skin of the star. Modelling the perturbation to the convective flux is motivated by the desire to accurately model the non--adiabatic effects towards the surface of the star, and to avoid these problems.

The model used to perturb the convective flux sets:
\begin{equation}
F'_{\rm conv} = A \frac{\partial s'}{\partial r} ,
\end{equation}
where $F'_{\rm conv}$ is the perturbation to the convective flux, $s'$ is the perturbation to the entropy, and $A$ is the coefficient of proportionality. This assumes that perturbation to the convective flux is dominated by the perturbation  to the entropy gradient, so that $A$ is kept constant, and that the gradient is dominated by the radial component.  Note that results for the case where perturbation of $A$ is taken into account have also been presented in \citet{Bunting2019}.
Deep within the convection zone, $A$ is large and $\partial s_{0} / \partial r $ (the background entropy gradient) is small, so a very small perturbation to the entropy gradient can produce a large perturbation to the flux. This approximation may then give rise to anomalously large perturbations to the convective flux in this region. Towards the surface, and particularly within the superadiabatic zone, $A$ becomes small, as convection becomes inefficient, and $\partial s_{0} / \partial r$ is large.  Therefore,  in this region, the approximation will hold better as errors are less likely to have a significant impact, and the non--adiabatic behaviour of the displacement can be calculated more reliably.  Overall, this model may produce anomalous values in the surface flux due to inaccuracies in modelling convection deep within the convection zone, whilst being able to model the displacement within the superadiabatic zone and towards the surface reasonably well.

Further details can be found throughout \citet{Bunting2019}, particularly in their section~3.

\begin{figure*}
	\centering
	\includegraphics[width=8cm]{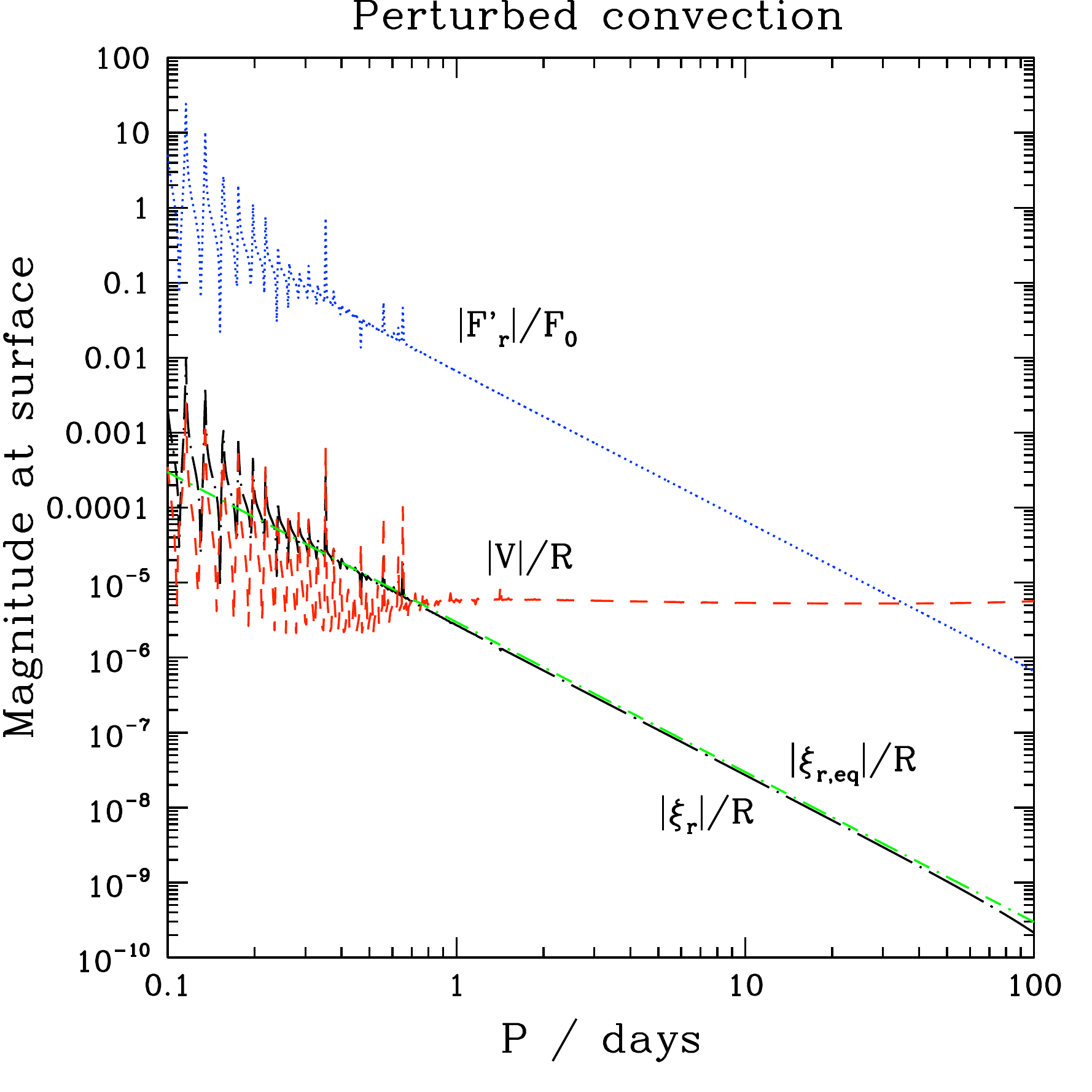}
	\includegraphics[width=8cm]{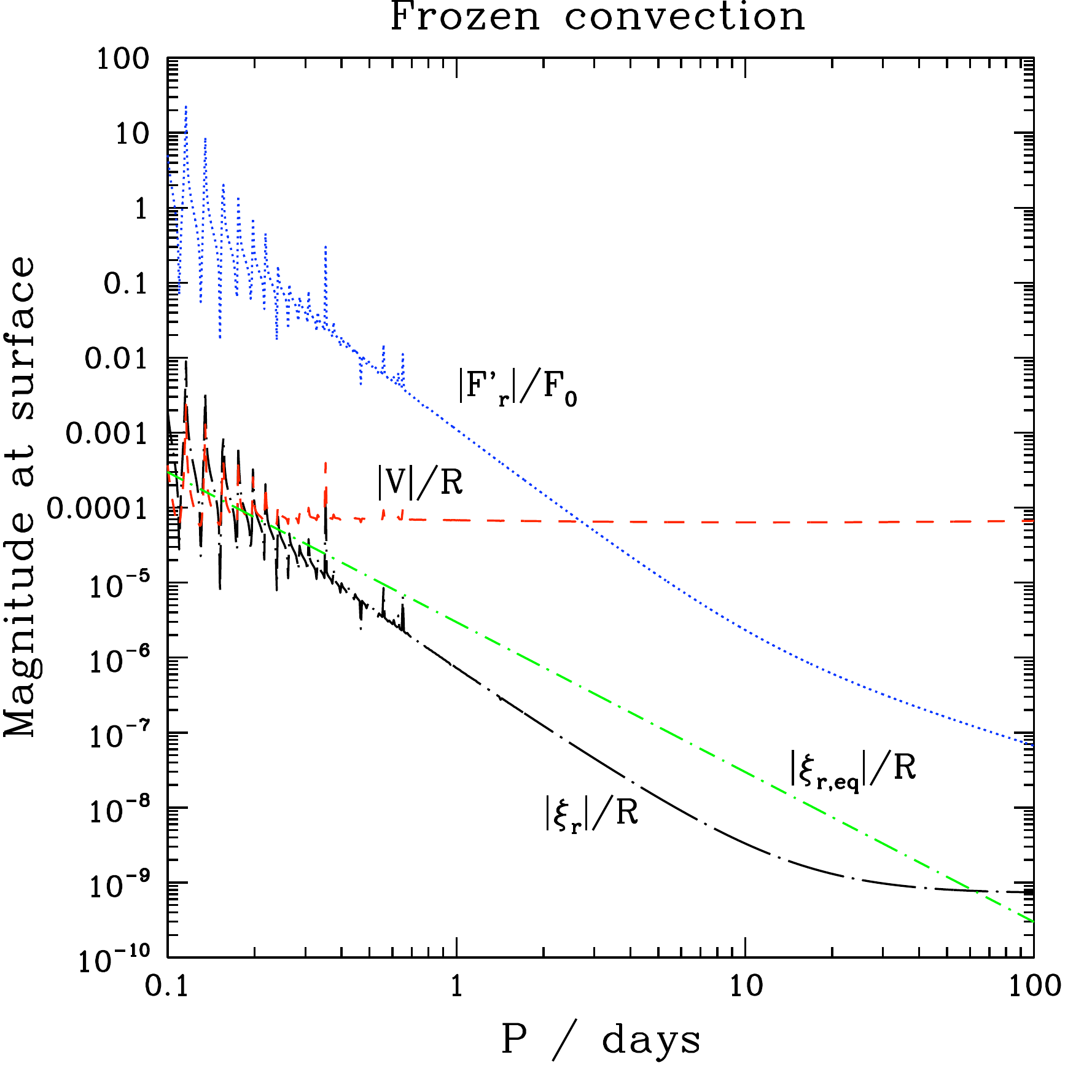} \\
	\caption{The magnitudes of the surface values of the radial displacement $\left| \xi_r \right| / R$ (long dashed--dotted black curve),
          horizontal displacement $|V|/R$ (dashed red curve), equilibrium tide radial displacement
          $\left| \xi_{r ,{\rm eq}} \right| / R$ (short dashed--dotted green curve), and flux perturbation $ \left| F'_r
            \right| / F_0$ (dotted blue curve), as a function of the orbital periods $P$ in days, in logarithmic scale.  The left panel shows the results for a star with perturbed convection, and the right panel shows the results for the case of assuming frozen convection. Both cases show a constant $|V|$  outside of the region where resonances are prevalent. In the perturbed convection case,   $|F'_{r}|$ and $|\xi_{r}|$, like $\xi_{r,\text{eq}}$, are proportional to  $P^{-2}$.  In the frozen convection case and for periods less than 10 days, $\xi_{r}$ and $F'_{r}$ are  roughly proportional to $P^{-3}$. For periods of less than a day, resonances become prominent, and the background scaling is somewhat obscured. There are some differences between the two cases -- the perturbed convection value of $\xi_{r}$ closely matches the equilibrium tide, whereas the frozen convection $\xi_{r}$ is suppressed by about an order of magnitude; the frozen convection $F'_{r}$ is approximately an order of magnitude smaller than for the perturbed convection case; and $V$ attains a constant value which is an order of magnitude larger in the frozen convection case than in the perturbed convection case.}
	\label{fig:log_fractional}
\end{figure*}

\subsection{Long period behaviour}
\label{sec:test_cases:long_P}

\subsubsection{Perturbed convection}

As the period of the orbit increases, the proportion of the convection zone in which $|N|^{2} \ll (m \omega)^{2}$ decreases, and the transition into the super--adiabatic region occurs deeper in the star (see Figure \ref{fig:N2}). This gives rise to increasingly large deviations in $\xi_{r}$ from the equilibrium tide displacement within the convection zone, but does not produce a significant change in the response of the displacement at the very surface -- that is, $\xi_{r} \approx \xi_{r,eq} \propto P^{-2}$ and $V$ remains approximately constant there. This holds up to very long periods of $\sim 100$ days.

The perturbation to the flux depends on the forcing frequency, with the magnitude of the response, both within the convection zone and at the very surface, proportional to $P^{-2}$. This holds from periods of a couple of days up to periods of hundreds of days.

At long periods, the wavelength of the spatial oscillations in the stellar core becomes very small. As a result of this, resolving the oscillations computationally becomes difficult. The presence of a thick convection zone may reduce the impact that these unresolved oscillations have on the surface response, which is suggested by the fact that the surface response remains consistent with the expected behaviour even once the core is very poorly resolved. Resolution issues become apparent in the centre of the star in this model for orbital periods above $\sim 20$ days, and become apparent at $r = 0.5 R$ for periods above $\sim 100$ days.

\subsubsection{Frozen convection}

In this case, the behaviour within the body of the star is fairly similar to that obtained when convection is perturbed, but the surface response is very different. The frozen convection case displays much larger changes over a thin surface region of the star, resulting in $|\xi_{r}|$ being suppressed by around a factor of 10 compared to $\xi_{r, eq}$.  On the other hand, $|V|$ is still constant, but at a much larger value than in the perturbed convection case.  This value agrees well  with the prediction for the low--frequency limit in the non--adiabatic case which is $|V/R| = m_{p} / (4 m^{2} M)$, where $M$ is the stellar mass, $m_p$ is the planet's mass and $m=2$ \citep{Bunting2019}.

The surface value of the  perturbation of the flux is reduced by an order of magnitude by freezing convection, compared to the perturbed convection case. Within the convection zone, the assumption of frozen convection causes the perturbation of the flux to be greatly suppressed before growing over a very small scale when approaching the thin radiative skin, as opposed to growing over a large scale throughout the convection zone as in the perturbed convection case.

The results for both perturbed and frozen convection are illustrated in Figure~\ref{fig:log_fractional}.

\subsection{Short period behaviour}
\label{sec:test_cases:short_P}

At very short periods, the Brunt--V\"ais\"al\"a frequency in the radiative core can become comparable to the oscillation frequency. When this occurs, the behaviour of $\xi_{r}$ in the core can deviate significantly from the equilibrium tide as opposed to oscillating around it as in the reference case with $P=4.23$~days. This deviation then persists throughout the convection zone and produces surface behaviour which doesn't resemble the equilibrium tide prediction, as seen in Figure~\ref{fig:log_fractional}. This would have an impact on the behaviour of this system at periods of $\sim 0.3$ day or less.

At such short periods, the assumption of small perturbations can break down. In the very centre of the star, $F'$ can become large such that it is no longer negligible compared to the background flux when considering the stellar structure. This can occur in this model for periods up to $\sim 0.8$~day. Whilst this is primarily a consideration at the centre of the star, large deviations can occur in the radiative skin at the surface for orbital periods up to $\sim 0.3$~day.

\subsection{Resonances}
\label{sec:test_cases:resonances}

Whilst resonances do have a significant effect on the response of the stellar interior (such that the assumption of small perturbations may no longer be valid when very close to resonance), the response at the surface is much less pronounced.  When going through a resonance, the RV signals for orbital periods of a couple of days change by $\sim 10\%$, whilst the photometric signals change by around $\sim 1\%$. Resonances for this system are therefore not likely to greatly impact the observed signals, in the unlikely event that an on--resonance system is found.

For periods in the range $1$--$10$~days, the resonances are very narrow, with a quality factor $\sim 10^{5}$ around $P \approx 4$ days. As the period decreases, the resonances become less sharp, and have a more prominent impact for periods less than a day, and particularly once $\xi_{r} \sim V$, as seen in Figure~\ref{fig:log_fractional}.

The choice of model for the convective flux does not affect the location or quality factor of resonances, as it is the radiative zone which acts as the resonant cavity. The effect at the surface is changed, however, as the extent to which this resonant cavity is isolated from the surface depends on how the response within the convective zone is modelled. By freezing convection, any deviation from equilibrium present at the base of the convection zone is maintained throughout the convection zone and is able to produce greater deviations from equilibrium in the thin radiative skin at the surface. By contrast, including a perturbation to the convective flux allows the non--equilibrium behaviour of the radiative zone to be attenuated before reaching the surface.

\subsection{General trends}
\label{sec:test_cases:general}

The calculation of the response of the star is expected to be valid within the limits stated above. For orbital periods between $1$ and $20$ days, the system is well modelled, with the results likely to hold for longer periods up to $100$ days.  The change in the amplitude and phase of the response when the orbital period is varied yields a change in the  observable signals.  Here we describe some general behaviours.

\begin{figure*}
       \centering
        \includegraphics[width=8cm]{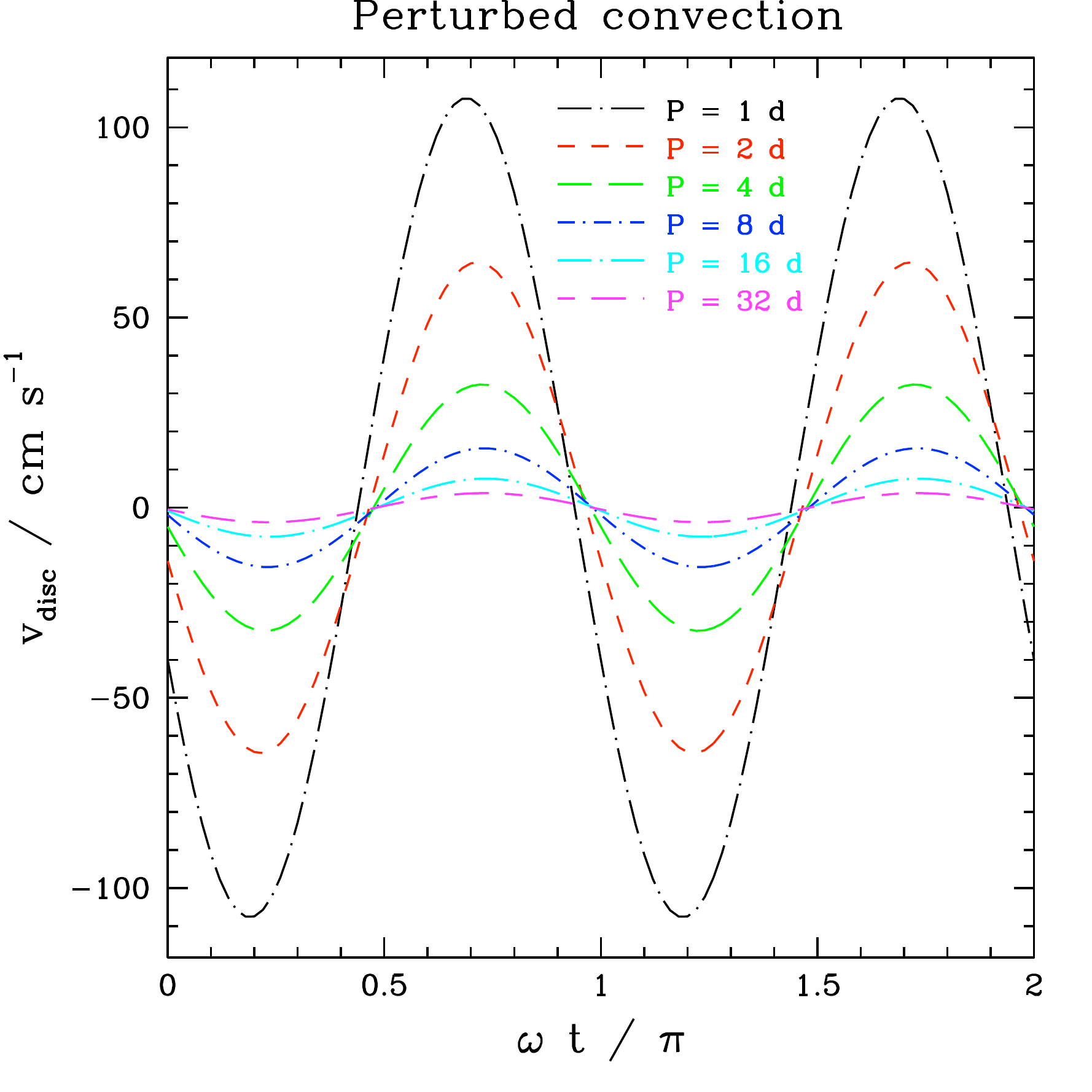}
        \includegraphics[width=8cm]{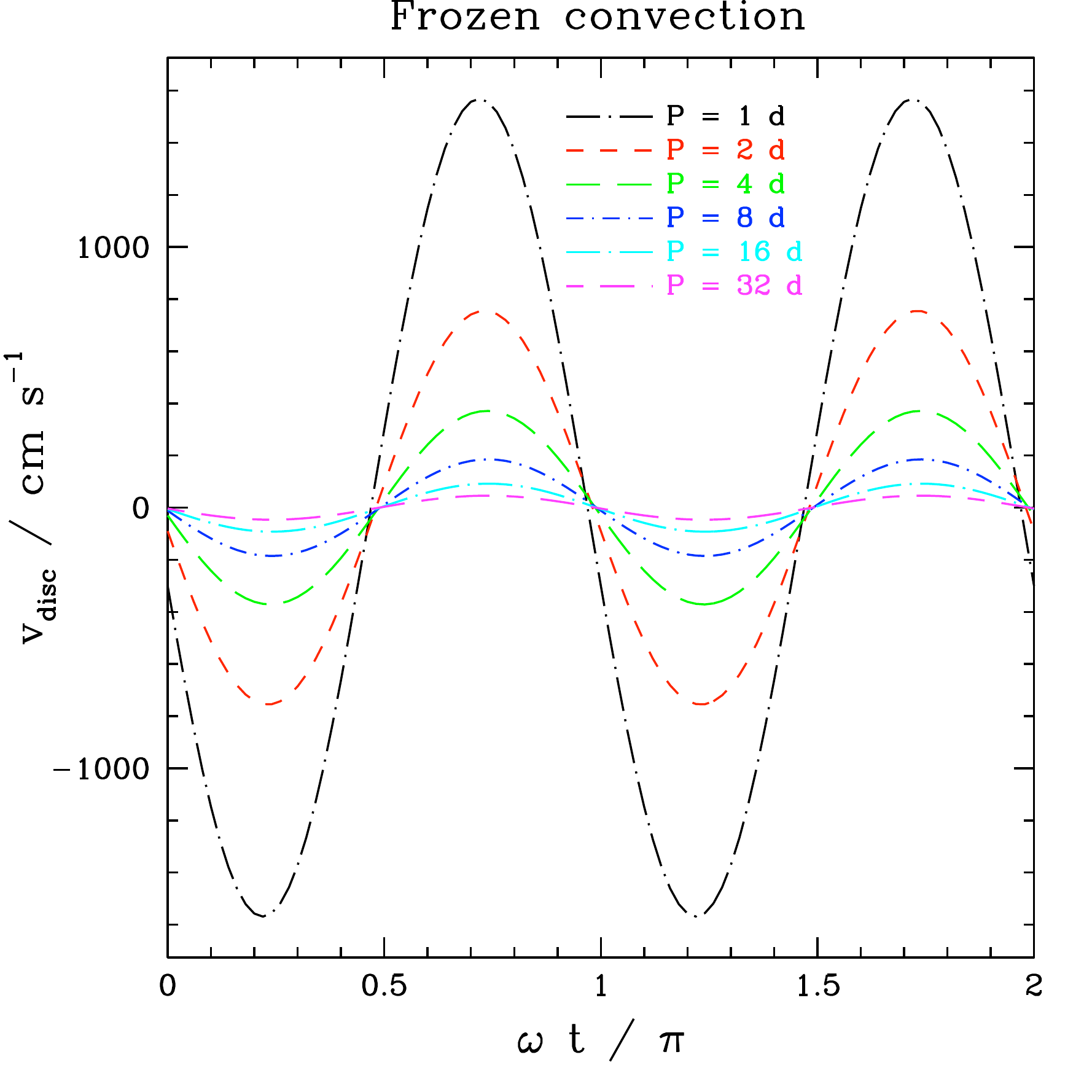}
        \caption{
       	The disc--integrated radial velocity $ \rm{v}_{\text{disc}}$ (in units of cm s$^{-1}$) is shown against orbital phase $\omega t / \pi$,  the origin of which is given by the epoch of inferior conjunction, for a range of orbital periods for the case of perturbed convection (on the left) and frozen convection (on the right).  From the largest to the smallest amplitude, the different curves correspond to $P$=1, 2, 4, 8, 16 and 32~days. In both cases, the amplitude scales as $P^{-1}$, which is different than the scaling corresponding to the equilibrium tide prediction.  This is due to $V$ dominating the signal, as it is much larger than $\xi_{r}$, and being constant. This will lead to RV signals which are larger than predicted by the equilibrium tide, particularly at longer orbital periods. As the orbital period changes, the phase of the RV signal remains fairly constant.  At short periods however,  $\xi_{r}$ becomes comparable to $V$ and the phase is no longer constant. Comparing the two convection models shows that the frozen model produces a signal which is an order of magnitude larger than predicted by the perturbed convection model. This follows directly from the constant value of $V$ being an order of magnitude larger in the frozen convection case.}
    \label{fig:RV_signals}
\end{figure*}

\begin{figure*}
       \centering
        \includegraphics[width=8cm]{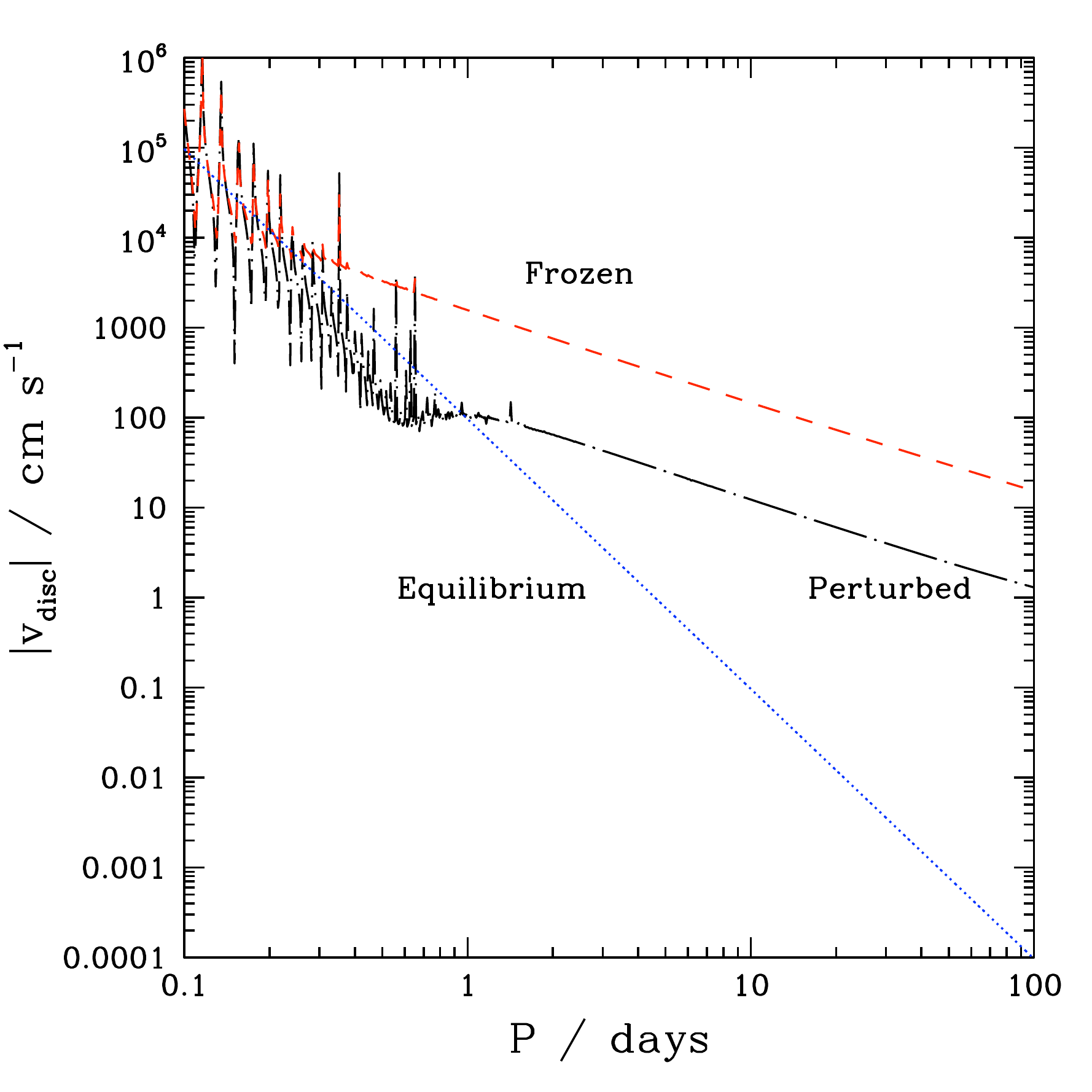}
        \includegraphics[width=8cm]{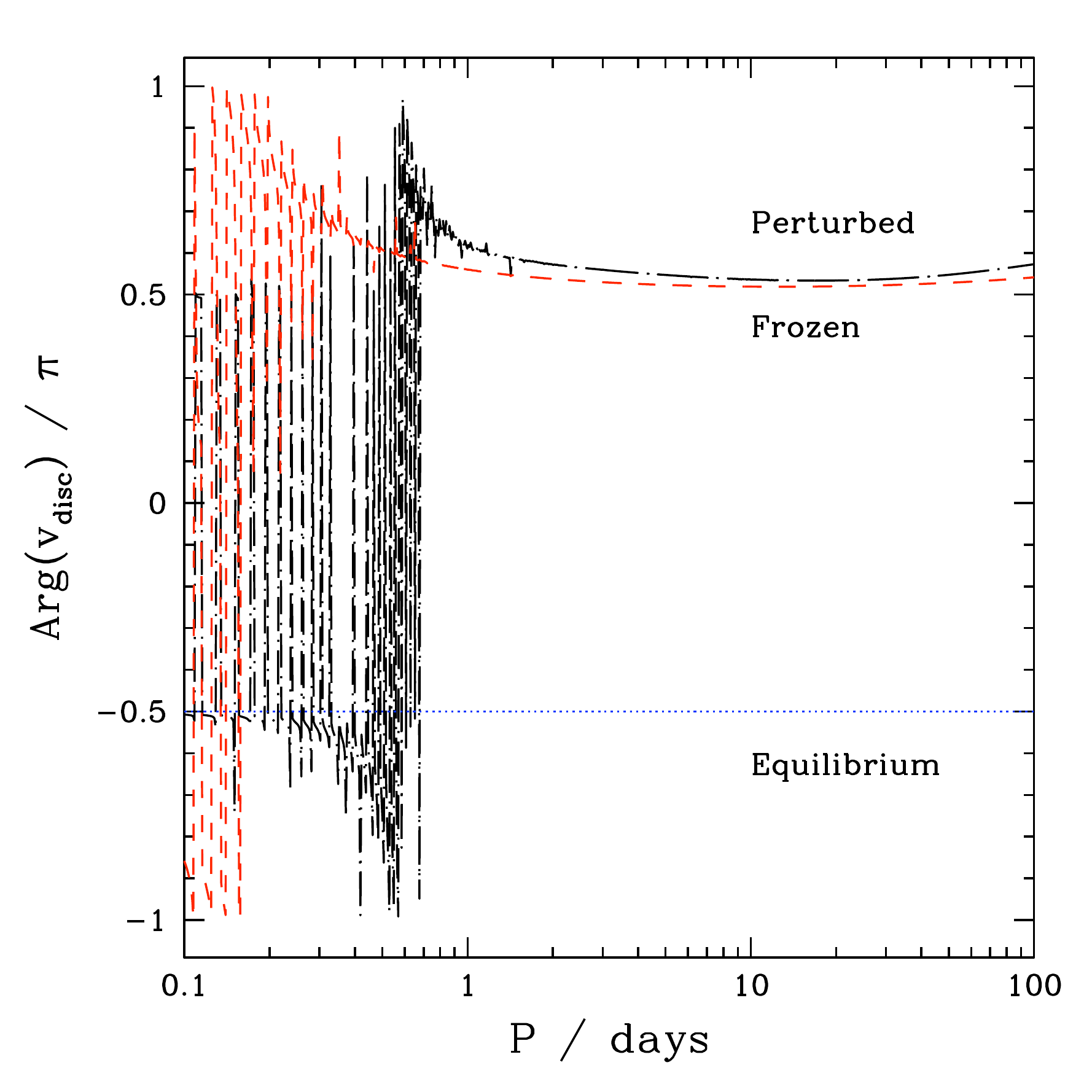}
      \caption{The magnitude (left panel) and phase (right panel) of  the disc--integrated radial velocity $ \rm{v}_{\text{disc}}$  is shown against orbital period for a Jupiter--mass planet orbiting a solar--mass star, comparing three models: frozen convection (dashed red line), perturbed convection (dashed dotted black line), and the equilibrium tide (dotted blue line). For periods greater than $\sim 1$ day, the frozen and perturbed convection models both scale as $P^{-1}$, whereas the equilibrium tide scales as $P^{-3}$. This leads to large deviation in the predicted signal amplitude for long period orbits: at $P = 10$\,days the equilibrium tide prediction is $| \rm{v}_{\text{disc}}| = 0.1$\,cm\,s$^{-1}$, whereas the perturbed convection and frozen convection models are larger than this by a factor $10^{2}$ and $10^{3}$, respectively. At short periods, once $| \xi_{r} | \sim |V|$, resonances become significant, and the scaling of $ \rm{v}_{\text{disc}}$ changes to follow the equilibrium tide scaling (as $\xi_{r}$ scales as $\xi_{r, \text{eq}}$). The phase of the signals remains fairly consistent for orbital periods greater than 1 day, with resonances clearly present below that value. The perturbed and frozen convection models are $\upi$ out of phase with the equilibrium tide prediction.}
    \label{fig:RV_comparisons}
\end{figure*}

Figures~\ref{fig:RV_signals} and~\ref{fig:RV_comparisons} show  the disc--integrated radial velocity, 
$  \rm{v}_{\text{disc}}$, calculated using equation~(\ref{eq:Methods:RV:disk_integrated:final}).  As can be seen on these figures,
$ \left| \rm{v}_{\text{disc}} \right| \propto P^{-1}$, which is a result of the signal being dominated by $V$, which remains fairly constant. The phase shift of the response is similarly constant, and does not change as the period increases further.

At short periods $|\xi_{r}| \sim |V|$ is approached, and  $V$ is no longer dominant. At this point, $V$ starts to deviate from the expected constant value, and scales roughly as $ P^{-2}$, just as the radial displacement does. This change in scaling is reflected in the scaling of the RV signal, which goes as $|v_{\rm disc}| \propto P^{-3}$. However,  at these very short periods, the presence of wider resonances somewhat obscures the background scaling relation.

At $P = 1$ day both the equilibrium tide and the perturbed convection model predict an amplitude of $\sim 100$ cm s$^{-1}$, whilst the frozen convection model predicts an amplitude of $\sim 1400$ cm s$^{-1}$. The phase for the equilibrium tide and perturbed convection models are however significantly different, with the signals being almost in anti--phase with each other.  At $P = 2$ days, the equilibrium tide prediction is reduced to $\sim 15$ cm s$^{-1}$, whilst the responses including convection are both reduced by a factor of two, at $\sim 60$ cm s$^{-1}$ for perturbed convection and $\sim 760$ cm s$^{-1}$ for frozen convection.

As the orbital period increases further, the disparity between the equilibrium tide prediction and those including convection widens further, with the amplitude of the equilibrium tide prediction at $P = 8$ days being less than $1$ cm s$^{-1}$, compared to $\sim 16$ cm s$^{-1}$ for the perturbed convection case, and $190$ cm s$^{-1}$ for the frozen convection case.

As changing the model used for convection does not affect the scaling of the response with orbital period, the frozen convection case consistently predicts an amplitude which is larger by an order of magnitude than the prediction from the perturbed convection case.

\begin{figure*}
        \centering
        \includegraphics[width=8cm]{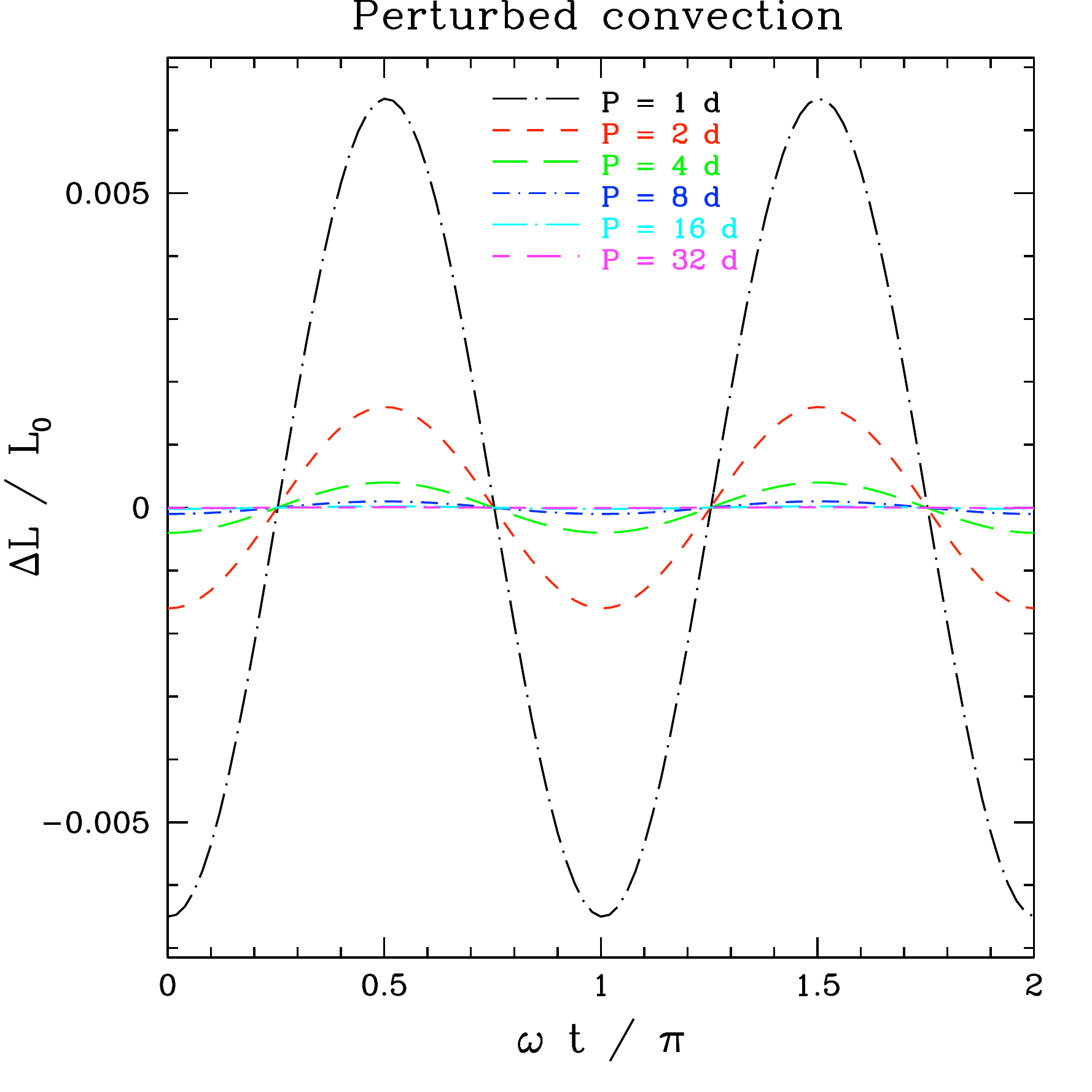}
        \includegraphics[width=8cm]{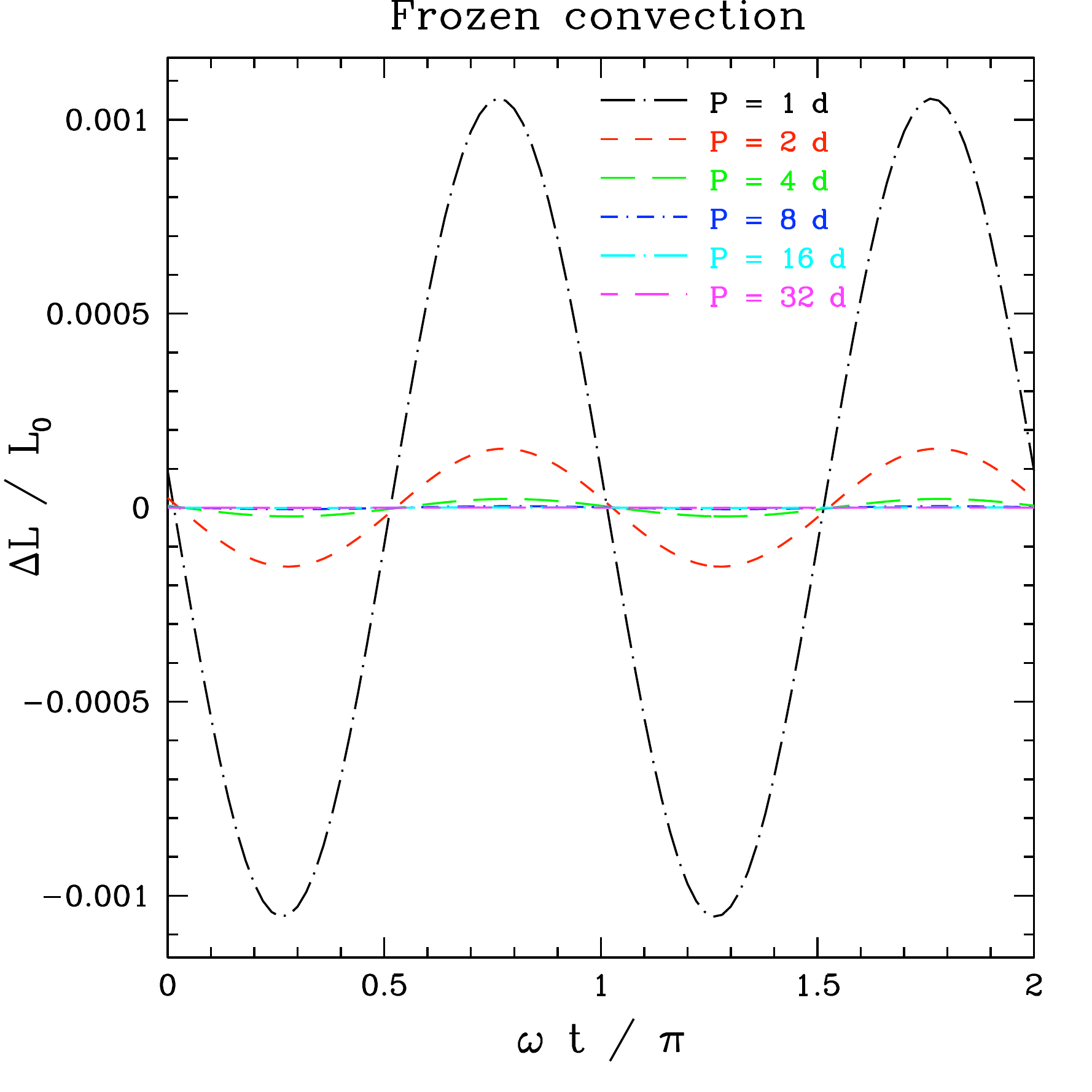}
        \caption{The photometric signal  as a fraction of total  observed flux, $\Delta L / L_0$, is shown against orbital phase,  the origin of which is given by the epoch of inferior conjunction, for a range of orbital periods for both the perturbed convection model (left panel) and the frozen convection model (right panel),  for a Jupiter--mass planet orbiting a solar--mass star.  The Eddington limb-darkening is used here, with $c=5/2$, $a=3/5$ and $b = 0$. From the largest to the smallest amplitude, the different curves correspond to $P$=1, 2, 4, 8, 16 and 32~days. In both models, the phase remains fairly constant over this range of periods.  In the perturbed convection case, the amplitude scales as $P^{-2}$, whereas in the frozen convection case the amplitude scaling is closer to $P^{-3}$. The perturbed convection model produces a prediction which is a factor of $\sim 5$ greater than the frozen convection model and, at very short periods, has a large peak--to--peak amplitude, being  $\sim 1$~per~cent at $P = 1$ day, which is comparable to the transit depth of a planet of similar size to Jupiter.  Because of the scaling with $P$, this quickly drops to a very small signal for long period orbits, being  $\sim 100$ ppm at $P = 8$ days.}
    \label{fig:photometric_signals}
\end{figure*}

Figure~\ref{fig:photometric_signals}  shows the fractional change in   observed flux, $\Delta L / L_0$, calculated using equation~(\ref{eq:Methods:lum_var:Delta_L_TOT}).
In the perturbed convection case, the photometric signal scales with the perturbation, and so is proportional to $P^{-2}$, as can be seen on the figures. For the frozen convection case, and for periods under 10 days, the scaling with $P$ is steeper than this, being closer to $P^{-3}$. 

\begin{figure*}
        \centering
        \includegraphics[width=8cm]{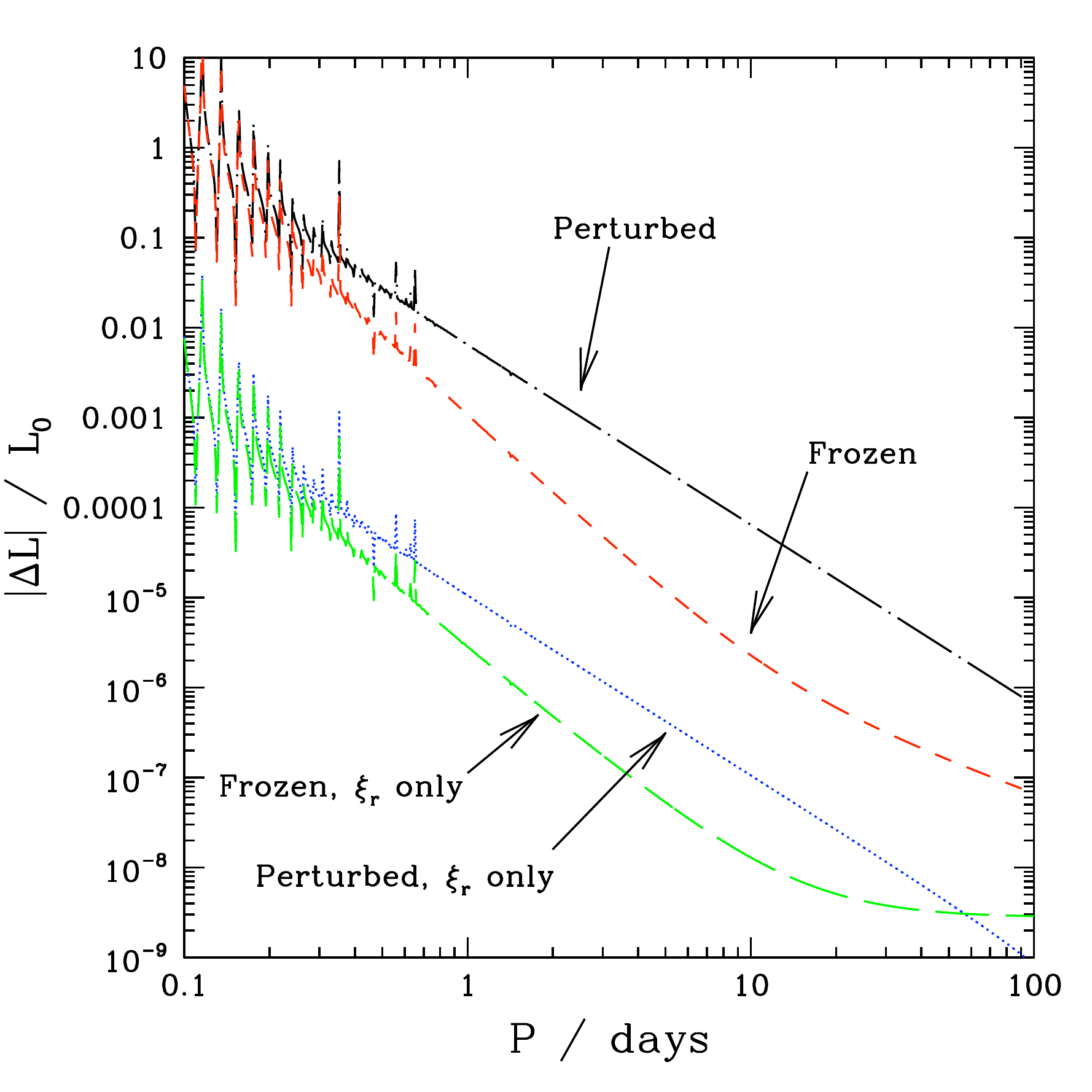}
        \includegraphics[width=8cm]{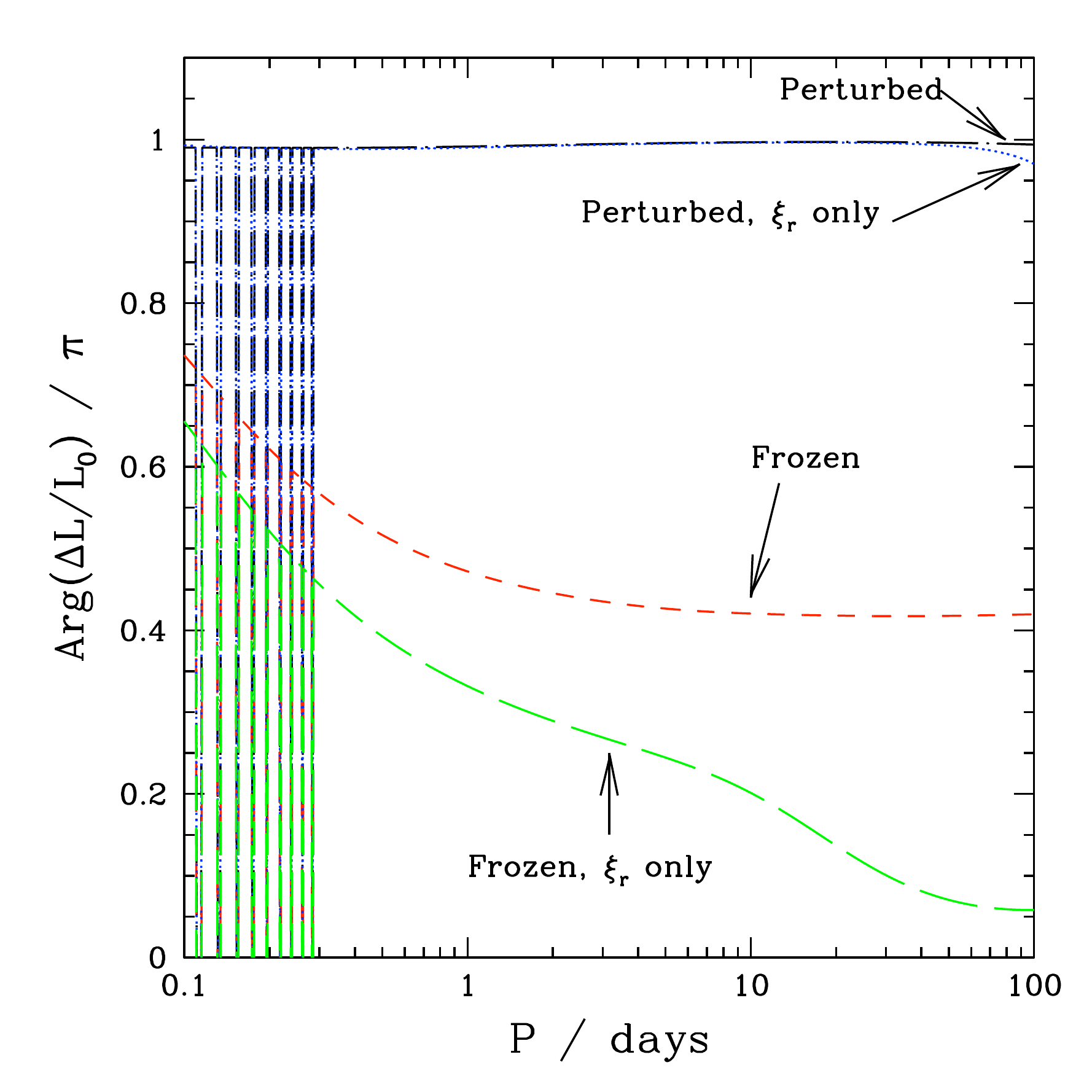}
       \caption{The magnitude (left panel) and phase (right panel) of the photometric signal $\Delta L / L_0$  is shown against orbital period for a Jupiter--mass planet orbiting a solar--mass star.  The Eddington limb-darkening is used here, with $c=5/2$, $a=3/5$ and $b = 0$.
The four different curves correspond to the full photometric signal and  perturbed convection (dashed dotted black  lines) or frozen convection (small dashed red lines) and to the signal arising from $\xi_r$ only and  perturbed convection (dotted blue lines) or frozen convection (long dashed green lines). 
In the perturbed convection case, the magnitude is proportional to $P^{-2}$, whereas for the frozen convection case, and for periods under 10 days, the scaling with $P$ is steeper than this, being closer to $P^{-3}$. For periods under 1\,day, resonances are seen to have a significant effect.
Outside of this region, the full signal from the frozen model is an order of magnitude smaller than that from the perturbed convection model.  At very short periods, both models predict unreasonably large values, and the assumption of small perturbations is clearly no longer valid there. The signals arising only as a result of $\xi_{r}$ show a similar relationship, but are smaller than the full signal by a factor of $10^{2} - 10^{3}$ (with the magnitude of the full signal  being $\sim 10^{-2}$ and that of the signal due only to $\xi_{r}$ being $\sim 10^{-5}$ at $P = 1$\,day). The phase of the signals arising from the perturbed convection model remain constant outside of the resonances.  By contrast, the frozen convection model does exhibit some change in phase as both the real and imaginary components are of roughly similar size, with some variation as the orbital period changes.
        }
    \label{fig:photometric_comparisons}
\end{figure*}

  At very short periods, both models with perturbed or frozen convection predict unreasonably large values of $\Delta L / L_0$.  The assumption of small perturbations is clearly no longer valid there, and the calculation of $F'$  in this regime may not be reliable, as mentioned in section~\ref{sec:test_cases:short_P}.  To test the effect of the perturbed flux on the photometric signal, we compare, in Figure~\ref{fig:photometric_comparisons}, $\Delta L / L_0$ obtained from equation~(\ref{eq:Methods:lum_var:Delta_L_TOT}) with this quantity calculated by setting $F'_r=0$ and keeping only the contribution from $\xi_r$.  If the perturbed flux were overestimated in our model, the realistic value of the change in  observed flux would probably be bracketed by these two estimates.  As can be seen from the figure, the perturbed flux  $F'_r$ totally dominates the change in  observed flux, which is larger  by a factor of a few hundred when $F'_r$ is taken into account compared to the case where only $\xi_{r}$ contributes. 

In the perturbed convection case, the amplitude of the brightness variation is proportional to $ P^{-2}$ in both the calculations with and without $F'_r$. For the frozen convection case and for periods under 10 days, the scaling of the amplitude is closer to $\propto P^{-3}$, although the calculation only including $\xi_{r}$ is slightly shallower than the full calculation.

The transit depth of a planet of similar radius to Jupiter would be $\sim 1$~per~cent, which is a similar order of magnitude to the amplitude of the signal for $P = 1$ day for the perturbed convection model (including $F'_r$ in the brightness calculation). At $P = 4$ days, the amplitude of the tidal signal is a factor of $\sim 25$ smaller than the transit depth, at $\sim 400$ ppm. The frozen convection model predicts an amplitude of $\sim 0.1$~per~cent at $P = 1$ day, and $\sim 20$ ppm at $P = 4$ days. This order of magnitude difference between the two models of convection is maintained as the orbital period changes.
This is due to the fact that, within the convection zone, the assumption of frozen convection causes the flux to be greatly suppressed before growing over a very small scale when approaching the thin radiative skin.  By contrast, in the perturbed convection case,  the flux grows over a large scale throughout the convection zone.

This difference in the change in  observed flux between the frozen and perturbed convection models is still present when only $\xi_r$ contributes to $\Delta L / L_0$, as can be seen in Figure~\ref{fig:photometric_comparisons},  although it is smaller than when $F'_r$ contributes.  
When only $\xi_r$ contributes, the signal  at $P = 1$ day has an amplitude of $\sim 10$ ppm in the perturbed convection model, and $\sim 3$ ppm in the frozen convection model. At $P = 4$ days, these are reduced to $0.6$ and $0.1$ ppm respectively, where the frozen convection model has slightly deviated from the $P^{-2}$ scaling.

The phase of the brightness variations does not vary hugely with orbital period, with the perturbed convection model predicting that the real component of the observed flux variation will dominate at all orbital periods. The frozen convection case differs from this, with the imaginary component dominating at long period orbits for the signal from $F'_r$, with the signal from $\xi_{r}$ gradually changing in phase as the orbital period changes.

\subsection{The effect of stellar mass}
\label{sec:test_cases:mass}

A more massive star of 1.4~M$_{\odot}$ was also investigated, as the stellar structure changes non--linearly with mass, leading to changes in the behaviour of the oscillations both throughout the star and at the surface. Structurally, the primary difference is the distribution of convective regions -- there is a small convective region at the centre of the star, and the outer convective region is significantly thinner than in the 1~M$_{\odot}$ case (its base is at $r = 0.94 R$ compared to $r = 0.73 R$ in the lower mass star), as shown in Figure~\ref{fig:N2}.

Computationally,  the presence of the small convective core gives rise to resolution issues at the inner boundary of the radiative zone, which leads to an unphysically large response at that point in the star for short period orbits. However, since the very centre of the star is convective, short--wavelength spatial oscillations are eliminated  in that region, contrary to the lower stellar mass case.

The star's response to a resonance is also affected. Changing the resonant cavity affects the location and spacing of the resonances, though they still have a similarly high quality factor. The response at the surface is much more sensitive to the presence of a resonance in the 1.4~M$_{\odot}$ case, as the thinner outer convective region means that the surface is much less insulated from the resonating radiative zone. In this case, a resonance can give an order of magnitude increase in both the RV and photometric signals.

In general, for a given period, the photometric response is a factor of $3$ smaller than in the 1~M$_{\odot}$ case, whilst the RV signal is approximately an order of magnitude greater. The change in  observed flux $\Delta L$ still scales with the perturbation, and is proportional to $P^{-2}$ in the perturbed convection case. Other than for very short periods, $V$ is constant, and dominates the RV signal, so that ${\rm v}_{\rm disc} \propto P^{-1}$, as in the 1~M$_{\odot}$ case.

\subsection{The effect of planetary mass}

In order to use observations to constrain the planetary mass, we must address how the observable signals are impacted by a change in the mass of the planet, $m_{p}$. The stellar response is found to scale linearly with the planetary mass, just as the tidal perturbation itself does. This linear scaling is found to hold well at the stellar surface even if the planetary mass is changed by up to an order of magnitude.

For a change in $m_{p}$ with a constant orbital separation, the orbital frequency will also be changed. The fractional change in frequency will be smaller than the fractional change in planetary mass by a factor of $m_{p}/M$. The change in frequency will therefore be very small, though this can have an impact on the response of the stellar interior if the oscillation frequency is near a resonance. However, as addressed in section~\ref{sec:test_cases:resonances}, the surface response is likely to be much less susceptible to these changes.

\section{Application to observed systems}
\label{sec:observed_systems}

We now model some specific systems in order to produce specific predictions. Table~\ref{tab:system_parameters} lists the key parameters of the systems, which have been approximately recreated in the model. These cases were chosen in order to show the behaviour of observable signals over a range of system parameters.  Since the eccentricity of these systems is negligible, the semi--major axis $a$ is equal to what has been called the separation $D$ in this paper.  The systems modelled here are WASP-19 \citep{Hebb2010}, WASP-18 \citep{Hellier2009}, WASP-12 \citep{Hebb2009}, Qatar 5 \citep{Alsubai2017}, and CoRoT-17 \citep{Csizmadia2011}.

\begin{table*}
	\centering
	\caption{The parameters of the modelled systems, as derived from observations. Listed are the name of the system, the stellar mass M in solar mass, the planet's mass $m_p$ in Jupiter masses, the stellar radius $R$ in solar radius, the age of the system in Gyr, the orbital semi--major axis $a$ in au, the orbital period $P$  and the rotational period $P_{\rm rot}$ of the star in days.  Other than the rotational periods, the data were taken from \citet{Arras2012}.  Rotational periods came from \citet{Brown2011} (WASP-19 and WASP-18), \citet{Alsubai2017} (Qatar 5), \citet{Csizmadia2011} (CoRoT-17, although this number is uncertain it was found to be ``typical of a main-sequence slow--rotating star'').   For WASP-12, we use the projected spectroscopic rotational velocity $v_{\rm rot} \sin i=3.4$~km~s$^{-1}$ \citep{Torres2012} and stellar radius $R_{\star}=1.14 \times 10^6$~km \citep{Maciejewski2011}, and assume $\sin i=1$ to calculate $P_{\rm rot}=2 \pi R_{\star} / v_{\rm rot}$.
 }
	\label{tab:system_parameters}
	\begin{tabular}{lccccccr} 
		\hline
		System & $M$ & $m_{p}$ & R & age & a & P & P$_{\text{rot}}$ \\
		 & (M$_{\odot}$) & (M$_{J}$) & (R$_{\odot}$) & (Gyr) & (au) & (d) & (d) \\
		\hline
		WASP-19 & 0.97 & 1.17  & 0.99 & 11.5 & 0.016 & 0.79 & 10.5 \\
		WASP-18 & 1.24 & 10.11  & 1.36 & 0.63 & 0.020 & 0.94 & 5.6 \\
		WASP-12 & 1.4 & 1.47  & 1.66 & 1.7 & 0.023 & 1.09 & 24.3 \\
		Qatar 5 & 1.13 & 4.32 & 1.08 & 0.53 & 0.041 & 2.88 & 12.1 \\
		CoRoT-17 & 1.04 & 2.43 & 1.59 & 10.7 & 0.046 & 3.77 & 20 \\
		\hline
	\end{tabular}
\end{table*}

Each of the systems exhibits a rotational period which is at least a factor of 4 larger than the orbital period, such that the approximation of a non--rotating star is not unreasonable,  and inertial modes are not excited. 

The details of the response are given in Table~\ref{tab:system_response}, which gives the surface behaviour, and how it converts into disc--integrated observables assuming a perfectly edge--on orbit (which is approximately true, as all five systems were discovered by transit).  The results displayed in this section have been obtained assuming an Eddington limb-darkening with $c=5/2$, $a=3/5$ and $b = 0$.

\begin{table*}
	\centering
	\caption{The results of the model applied to the real systems, for both the perturbed convection and frozen convection cases. The columns, from left to right, give the system in question, the radial equilibrium  displacement $\xi_{r, {\rm eq}}$ at the surface, the radial  displacement $\xi_r$ at the surface, the tangential  displacement $V$ at the surface, the orbital radial--velocity (RV) semi--amplitude $K_{\rm orb}$, the disc--integrated RV velocity ${\rm v}_{\rm disc, eq}$ in the equilibrium tide case, the disc--integrated RV velocity ${\rm v}_{\rm disc}$ for the full solution, and the fractional change in disc--integrated  observed flux $\Delta L / L_0$, respectively.
        }
	\label{tab:system_response}
	\begin{tabular}{lccccccc} 
		\hline
		System & $10^{-3} \times \xi_{r, {\rm eq}}$ & $10^{-3} \times \xi_{r}$ & $10^{-3} \times V$ & $10^{-3} \times K_{\rm orb}$ & ${\rm v}_{\rm disc, eq}$ & ${\rm v}_{\rm disc}$ & $10^5 \times \Delta L / L_{0}$ 
	 \\
		 & (cm) & (cm) & (cm) & (cm s$^{-1}$) & (cm s$^{-1}$) & (cm s$^{-1}$) &  \\
		\hline
		WASP-19, perturbed & $440$ & $(400-9{\rm i})$ & $-(460+103{\rm i})$  & $26$ & $-270 {\rm i}$ & $-46 + 125 {\rm i}$ & $(-1400 + 25 {\rm i})$ \\
		WASP-19, frozen & $440$ & $-(70+190{\rm i})$ & $-(5770+2170{\rm i})$  & $26$ & $-270 {\rm i}$ & $-970 + 2490 {\rm i}$ & $(-40 + 500 {\rm i})$ \\
		\hline
		WASP-18, perturbed & $5800$ & $(5460-190{\rm i})$ & $-(1700+1580{\rm i})$  & $180$ & $-2900 {\rm i}$ & $-590 + 10 {\rm i}$ & $(-13000 + 400 {\rm i})$ \\
		WASP-18, frozen & $5800$ & $-(820+1950{\rm i})$ & $-(52400+15600{\rm i})$  & $180$ & $-2900 {\rm i}$ & $-5900 + 19000 {\rm i}$ & $(-50 + 3530 {\rm i})$ \\
		\hline
		WASP-12, perturbed & $910$ & $(1200-360{\rm i})$ & $(2100-2700{\rm i})$  & $23$ & $-400 {\rm i}$ & $-890 - 800 {\rm i}$ & $(-2050 + 600 {\rm i})$ \\
		WASP-12, frozen & $910$ & $-(62+93{\rm i})$ & $-(7400+700{\rm i})$  & $23$ & $-400 {\rm i}$ & $-230 + 2300 {\rm i}$ & $(12 + 110 {\rm i})$ \\
		\hline
		Qatar 5, perturbed & $120$ & $(109-3{\rm i})$ & $-(1170+440{\rm i})$  & $57$ & $-20 {\rm i}$ & $-52 + 132 {\rm i}$ & $(-350 + 9 {\rm i})$ \\
		Qatar 5, frozen & $120$ & $-(9+10{\rm i})$ & $-(18100+1400{\rm i})$  & $57$ & $-20 {\rm i}$ & $-160 + 2130 {\rm i}$ & $(4 + 18 {\rm i})$ \\
		\hline
		CoRoT-17, perturbed & $230$ & $(220+9{\rm i})$ & $-(840+607{\rm i})$  & $32$ & $-32 {\rm i}$ & $-58 + 71 {\rm i}$ & $(-330 + 13 {\rm i})$ \\
		CoRoT-17, frozen & $230$ & $-(20+26{\rm i})$ & $-(16300+1700{\rm i})$  & $32$ & $-32 {\rm i}$ & $-160 + 1560 {\rm i}$ & $(4 + 25 {\rm i})$ \\
		\hline
	\end{tabular}
\end{table*}

\begin{figure*}
    \centering
    \includegraphics[width=\columnwidth]{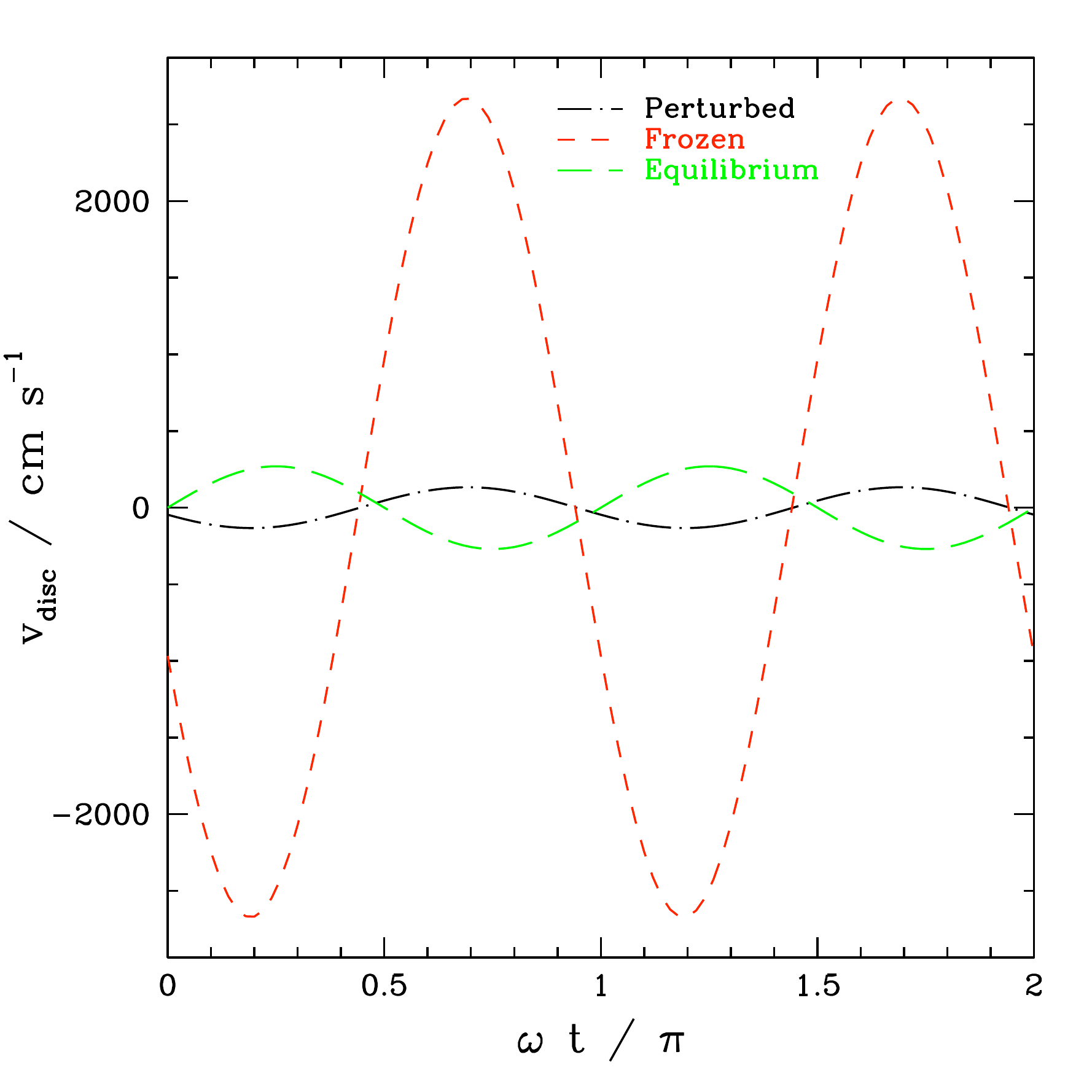}
    \includegraphics[width=\columnwidth]{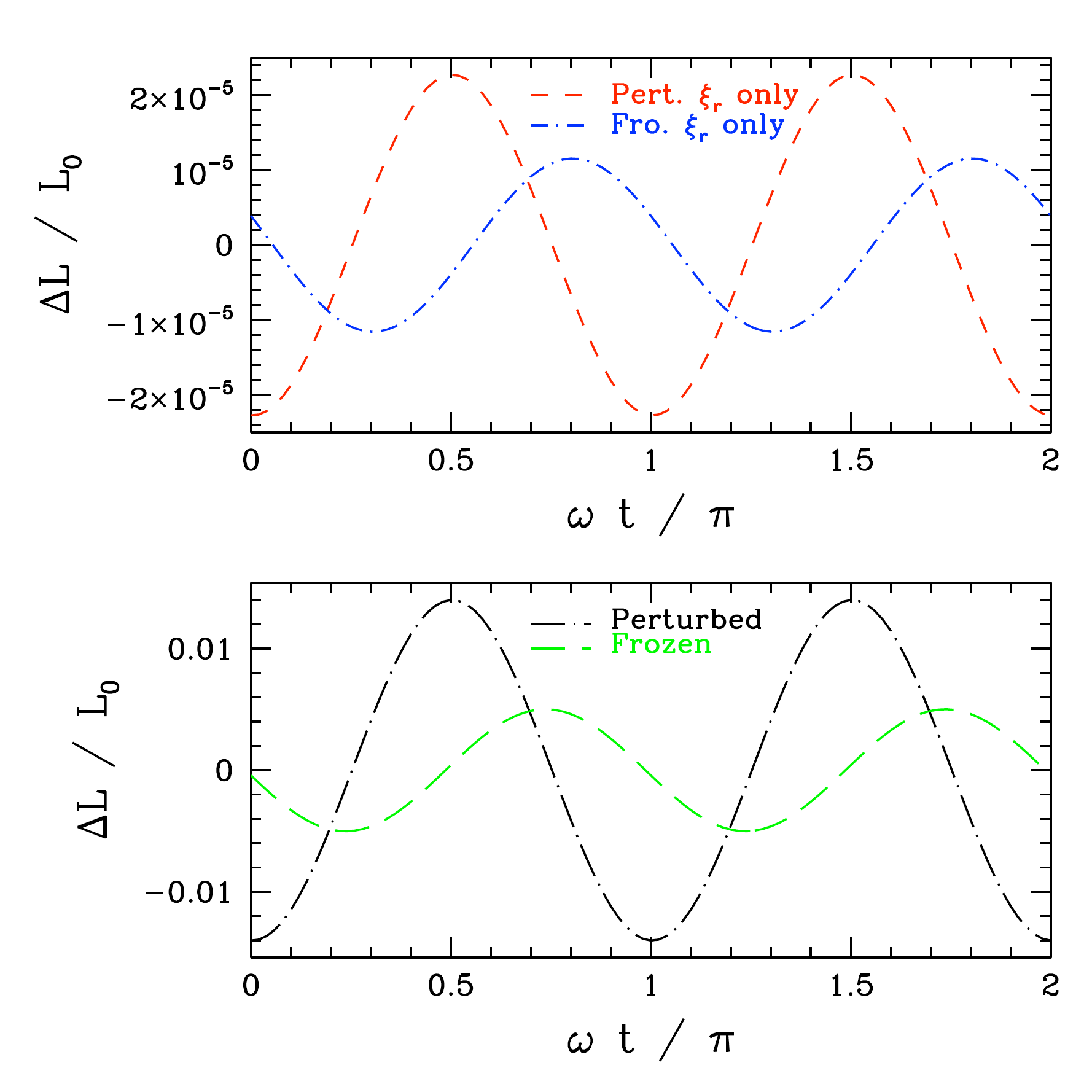}
    \caption{WASP-19: the left panel shows the RV signal in units of cm s$^{-1}$ for the cases of perturbed convection (black dot-dash line), frozen convection (red dash line), and the equilibrium tide (green long dash line); the right panels show the  observed flux variation. These quantities are plotted against the orbital phase,  the origin of which is given by the epoch of inferior conjunction.  The top right panel shows the  observed flux variation resulting only from the change in $\xi_{r}$ for the perturbed convection case (red dash line) and the frozen convection case (blue dot-dash line). The bottom right panel shows the complete  observed flux variation, including the flux perturbation, for the perturbed convection case (black dot-dash line) and the frozen convection case (green dash line).}
    \label{fig:WASP_19}
\end{figure*}

\subsection{WASP--19}
\label{sec:observed_systems:WASP-19}

Results for this system are displayed in Figure~\ref{fig:WASP_19}.
This system has a very short period orbit, such that $\xi_{r, {\rm eq}} \sim |V|$ for the perturbed convection model. The predicted RV signal for this case is a factor of two smaller than the equilibrium tide prediction, and is almost in anti--phase with it as a result of the fact that $V \approx -V_{\rm eq}$. Therefore, the radial and tangential components of the displacement counteract each other, reducing the disc--integrated signal.

The frozen convection case is dominated by a large value of $V$, and is approximately an order of magnitude greater than the equilibrium tide prediction. The phase matches the perturbed convection prediction, and is therefore also in anti--phase with the equilibrium tide prediction.

The variation in  observed flux predicted is very large, with an amplitude of $\sim 1$~per~cent for both the perturbed and frozen convection cases, which is the order of magnitude of the transit depth for a Jupiter analogue. Calculating the brightness variation arising only from $\xi_r$, that is to say neglecting the contribution from $F'_r$, predicts amplitudes which are smaller by a factor of $\sim 10^{3}$. In both cases, the perturbed convection case is dominated by the real component, whereas the frozen convection case is dominated by the imaginary component. The arguments therefore differ by a factor of $\upi / 2$, giving a clear phase difference in the observable signal.

\begin{figure*}
    \centering
    \includegraphics[width=\columnwidth]{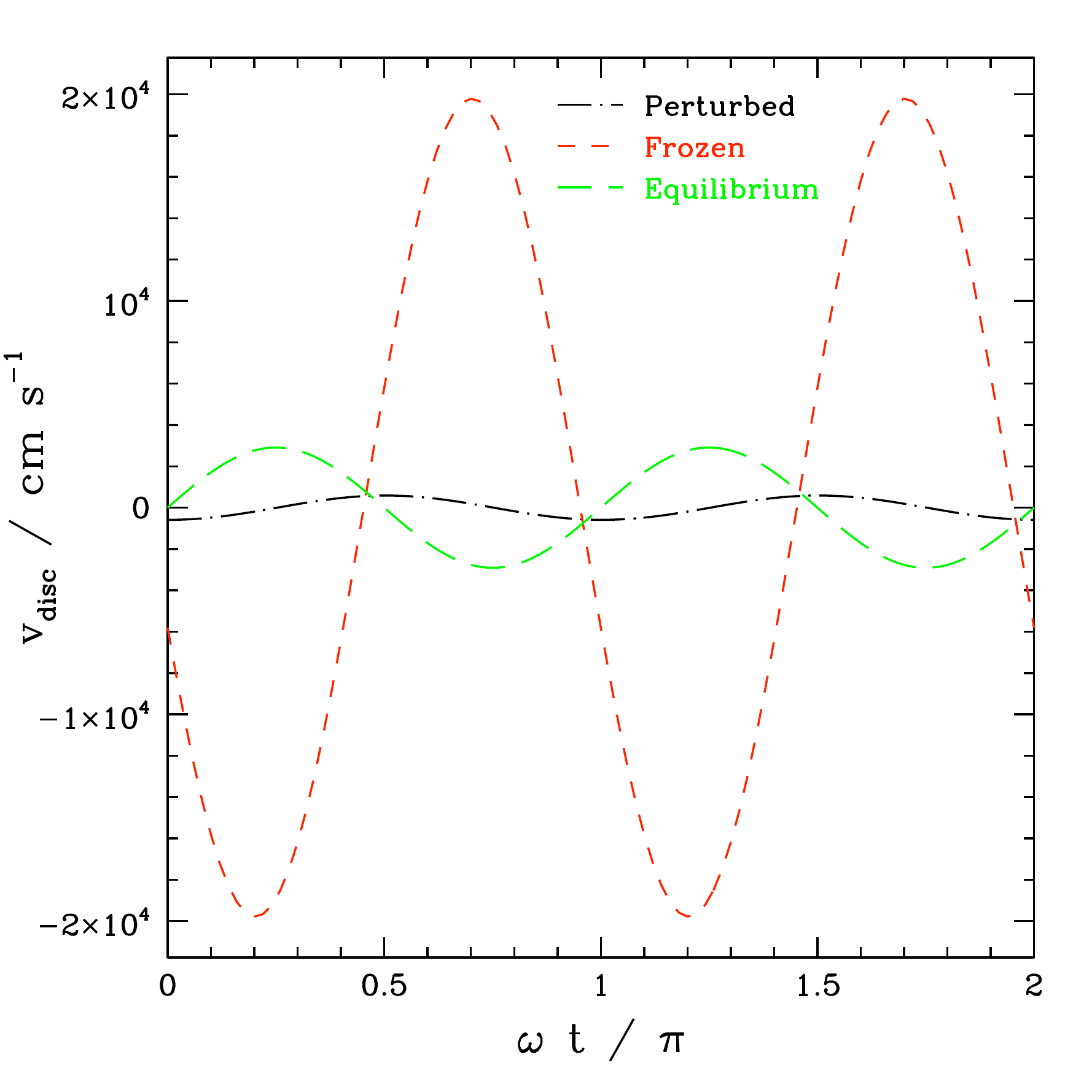}
    \includegraphics[width=\columnwidth]{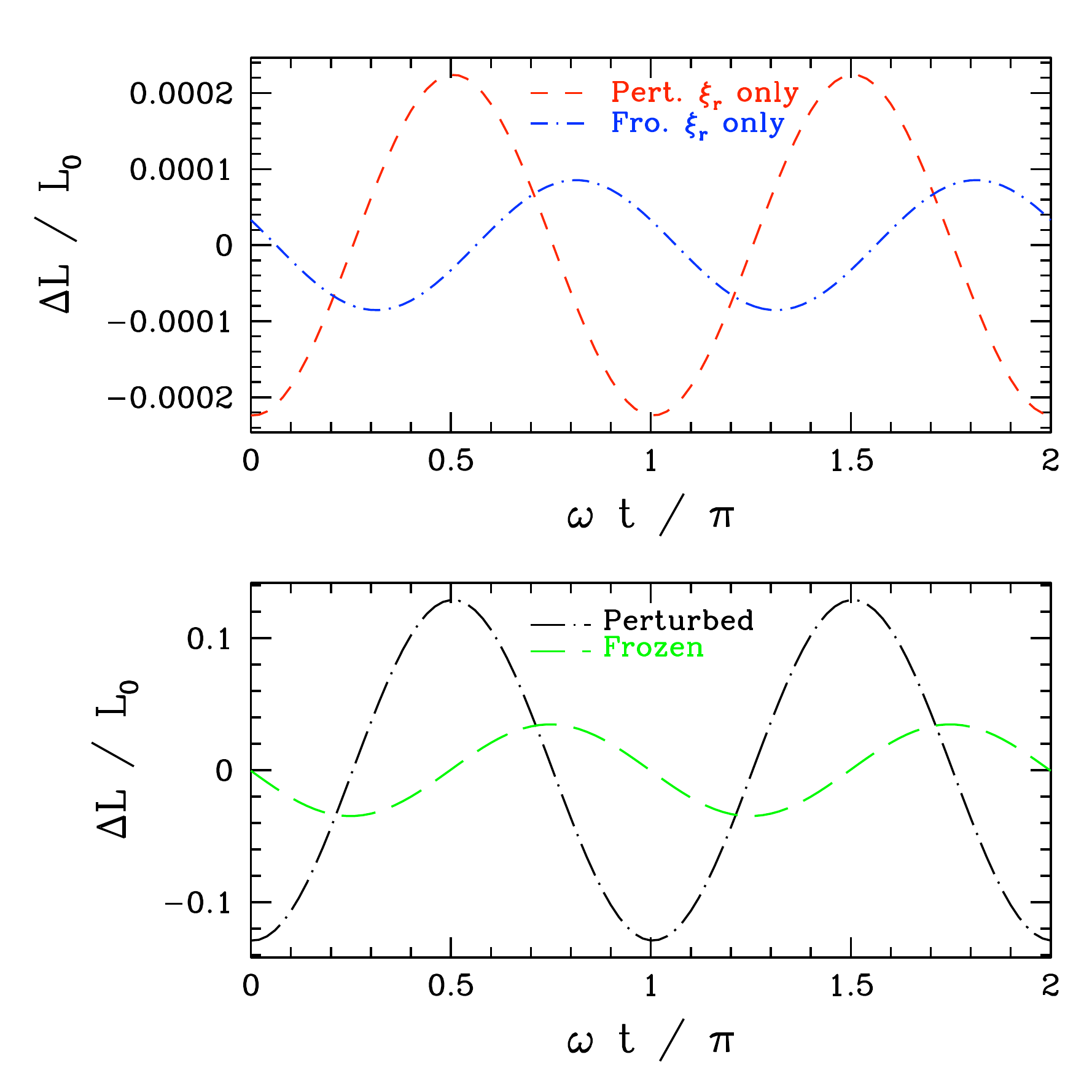}
    \caption{Same as Figure~\ref{fig:WASP_19} but for WASP--18.
    }
    \label{fig:WASP_18}
\end{figure*}

\subsection{WASP-18}
\label{sec:observed_systems:WASP-18}

Results for this system are displayed in Figure~\ref{fig:WASP_18}.
The very massive, short period planet gives rise to a large equilibrium tide RV response of $\sim 30$~m~s$^{-1}$. This is a factor of $\sim 5$ greater than the perturbed convection model's response ($\sim 6$ m s$^{-1}$), though it is still around an order of magnitude smaller than the frozen convection model's prediction ($\sim 200$ m s$^{-1}$). As the stellar model used here has a resonance close to orbital period, the values given in Table~\ref{tab:system_response} for this system could vary significantly for a small change in the orbital period modelled, and if on resonance could give very large values.

The perturbation to the flux within the star is very large in both the perturbed and frozen convection models, being $\sim 0.13$ and $\sim 0.04$, respectively. This would likely break the assumption of small perturbations, and therefore this model -- particularly the prediction for the variation in  observed flux -- may not be reliable.

If the contribution from only $\xi_{r}$ is taken into account, the  observed flux variation is still fairly large ($\sim 10^{-4}$). The perturbed convection model prediction is similar to the expectation from the equilibrium tide, as $\xi_{r} \approx \xi_{r, \text{eq}}$. The frozen convection prediction is smaller by a factor of 2, and is phase shifted compared to both the equilibrium tide and the perturbed convection prediction.

\begin{figure*}
	\centering
	\includegraphics[width=\columnwidth]{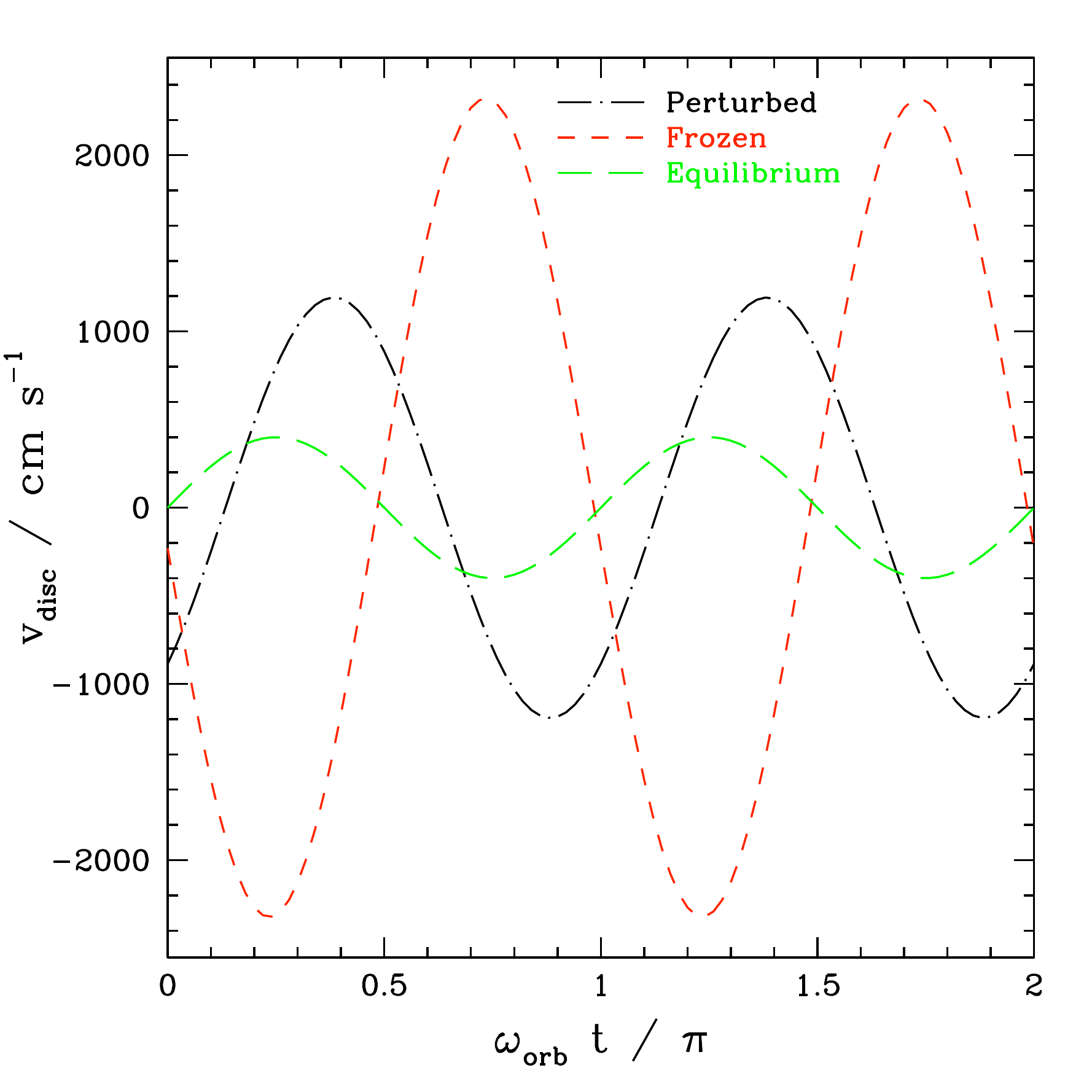}
	\includegraphics[width=\columnwidth]{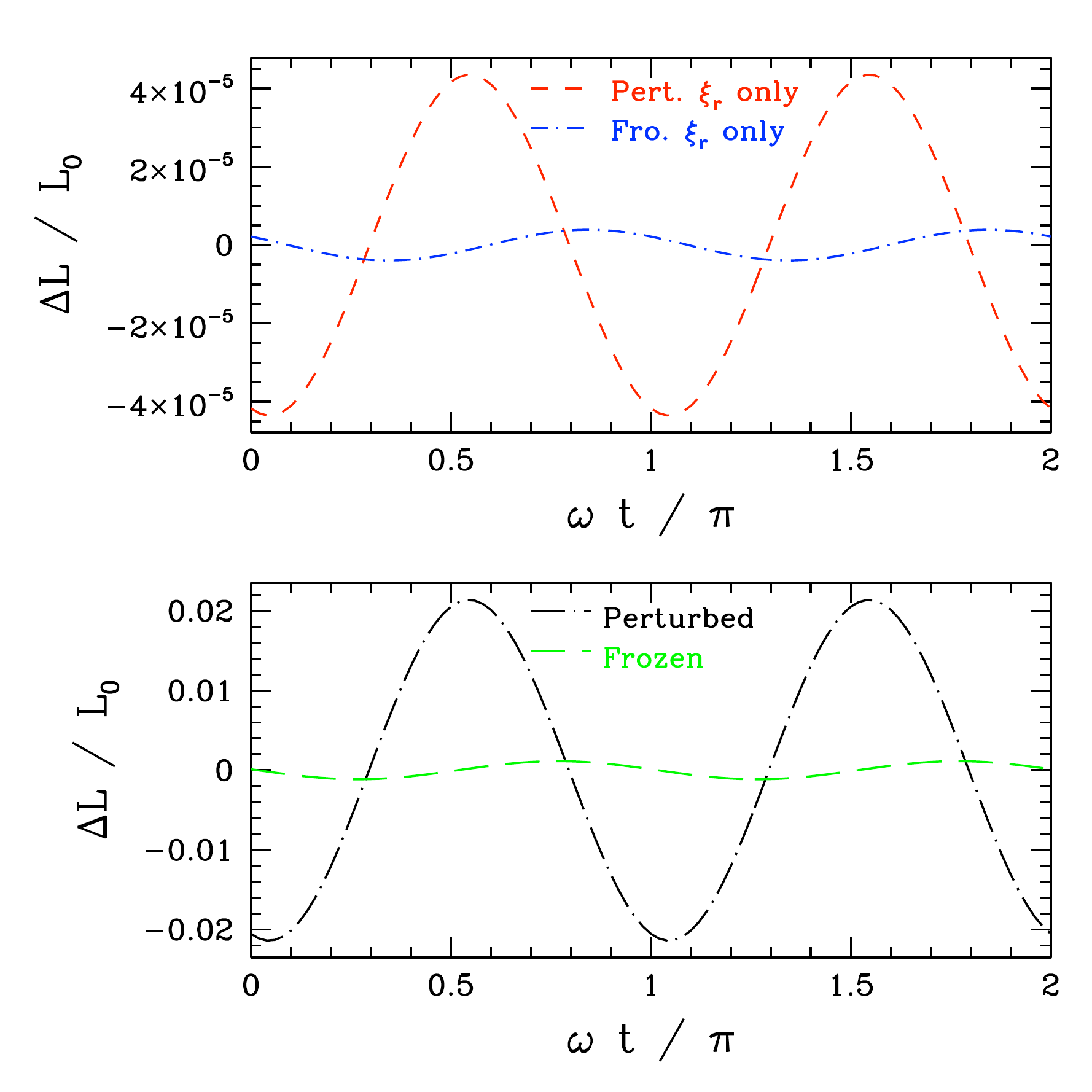}
	\caption{ Same as Figure~\ref{fig:WASP_19} but for WASP--12.
	}
	\label{fig:WASP_12}
\end{figure*}

\subsection{WASP-12}
\label{sec:observed_systems:WASP-12}

Results for this system are displayed in Figure~\ref{fig:WASP_12}. The star of this system is fairly massive, with $M \sim 1.4$~M$_{\odot}$, giving rise to a thin convective zone, starting at around $r \approx 0.93 R$. The behaviour of the surface is therefore much less insulated from the behaviour in the radiative zone than in the test case discussed in section~\ref{sec:test_cases}. This, coupled with the short orbital period of the planet, means that resonances may have a non--negligible effect on the surface behaviour.  Even if the system is not exactly in resonance, the magnitude, and particularly the phase, of the surface response could be affected. The accuracy of the modelled oscillations for this system is therefore likely to be more limited by the accuracy of the background stellar model than in cases with longer orbital periods of lower mass stars.

The radial velocity signals for this system are all fairly similar, with a difference of, at most, a factor of $6$ between them. The equilibrium tide predicts a magnitude of $4$~m~s$^{-1}$, and the perturbed convection model predicts a magnitude of $\sim12$~m~s$^{-1}$, lagging slightly behind the equilibrium prediction by approximately $20^{\circ}$ of orbital phase. The frozen convection amplitude is larger than both, at $\sim23$~m~s$^{-1}$, and is nearly in anti--phase with the equilibrium signal.

Whether considering the full observed flux variation, or only that arising due to $\xi_{r}$, the perturbed convection case predicts a semi--amplitude an order of magnitude larger than that predicted in the frozen convection case: $2$~per~cent compared to $0.1$~per~cent in the full observed flux variation, and $44$~ppm compared to $4$~ppm when considering only the effect arising due to $\xi_{r}$.





\begin{figure*}
    \centering
    \includegraphics[width=\columnwidth]{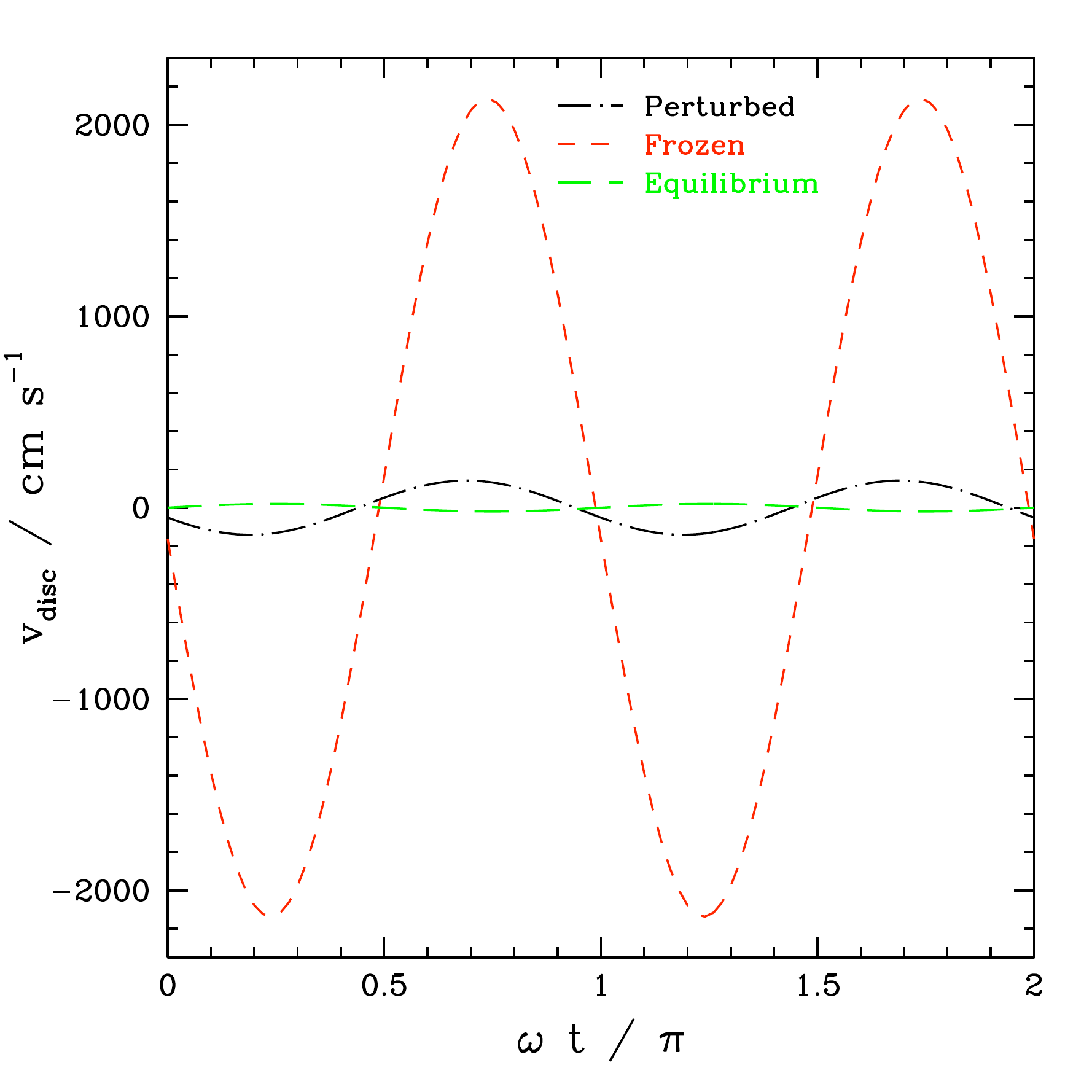}
    \includegraphics[width=\columnwidth]{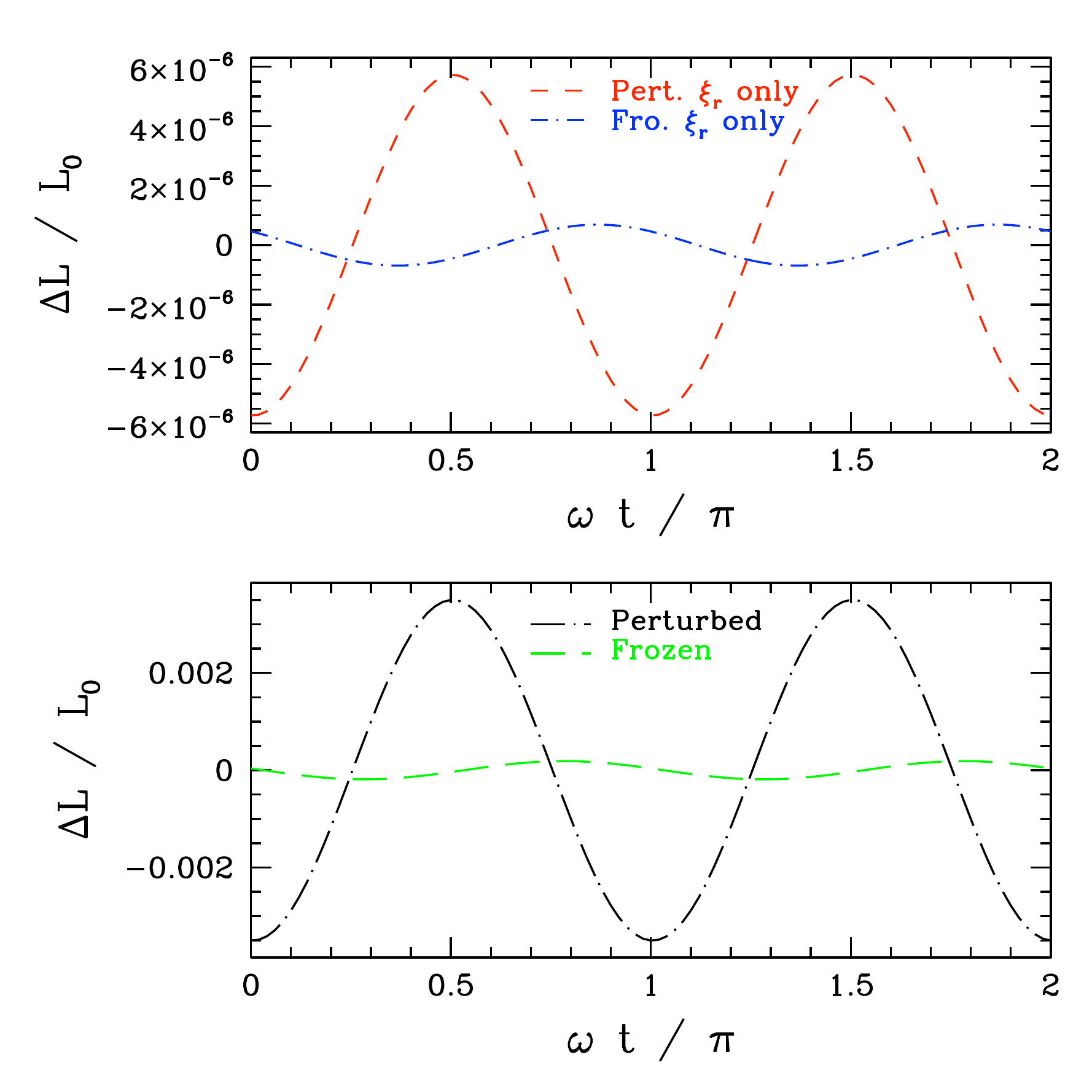}
    \caption{Same as Figure~\ref{fig:WASP_19} but for Qatar 5.
    }
    \label{fig:Qatar_5}
\end{figure*}

\subsection{Qatar 5}
\label{sec:observed_systems:Qatar_5}

Results for this system are displayed in Figure~\ref{fig:Qatar_5}.  The Qatar 5 system has a star which is very similar to the Sun in terms of structure, with a fairly massive planet on a short period orbit. As the orbital period is varied, the surface displacement behaves in a way similar to that described in section~\ref{sec:test_cases}. In the perturbed convection case, the radial displacement tracks the equilibrium tide well, whilst the frozen convection prediction is an order of magnitude smaller. In both cases, the tangential displacement is fairly insensitive to the forcing frequency. At the orbital period of the real system, we expect a greater RV signal than predicted by the equilibrium tide as a result of the tangential displacement being increased by a factor of $\sim 10$ for the perturbed convection case, and $\sim 100$ for the frozen convection case. The frozen convection amplitude of $20$~m~s$^{-1}$ would be much more easily detectable than the equilibrium tide prediction, and even the perturbed convection prediction of 2~m~s$^{-1}$ is potentially detectable.

The  observed flux variation $\Delta L$ for the perturbed convection model  is  an order of magnitude greater than, and out of phase with, the same quantity for the frozen convection model.   This applies whether or not $F'_r$ is taken into account in the calculation of $\Delta L$.  The phase difference could be
used to distinguish between the perturbed and frozen convection models. 

\begin{figure*}
    \centering
    \includegraphics[width=\columnwidth]{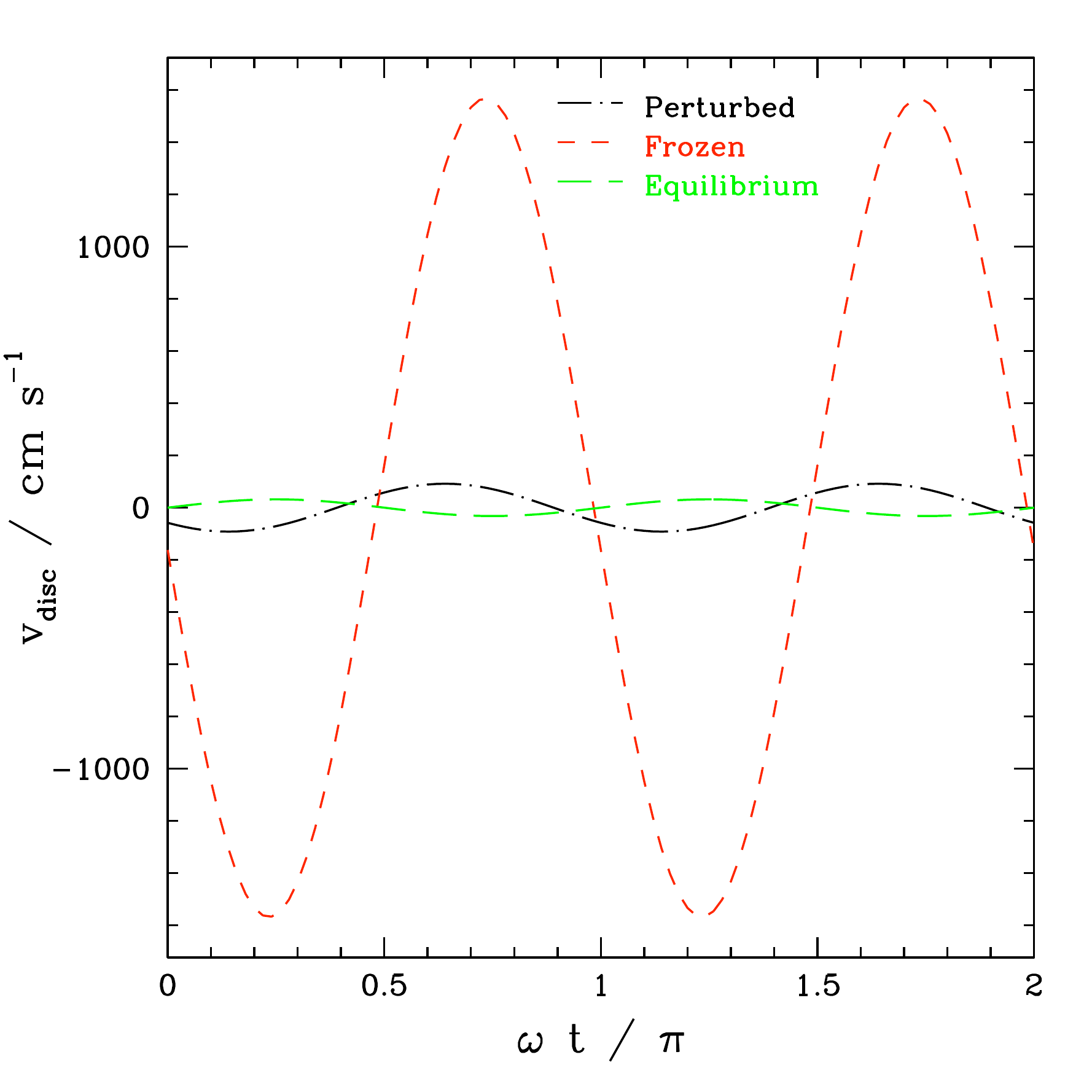}
    \includegraphics[width=\columnwidth]{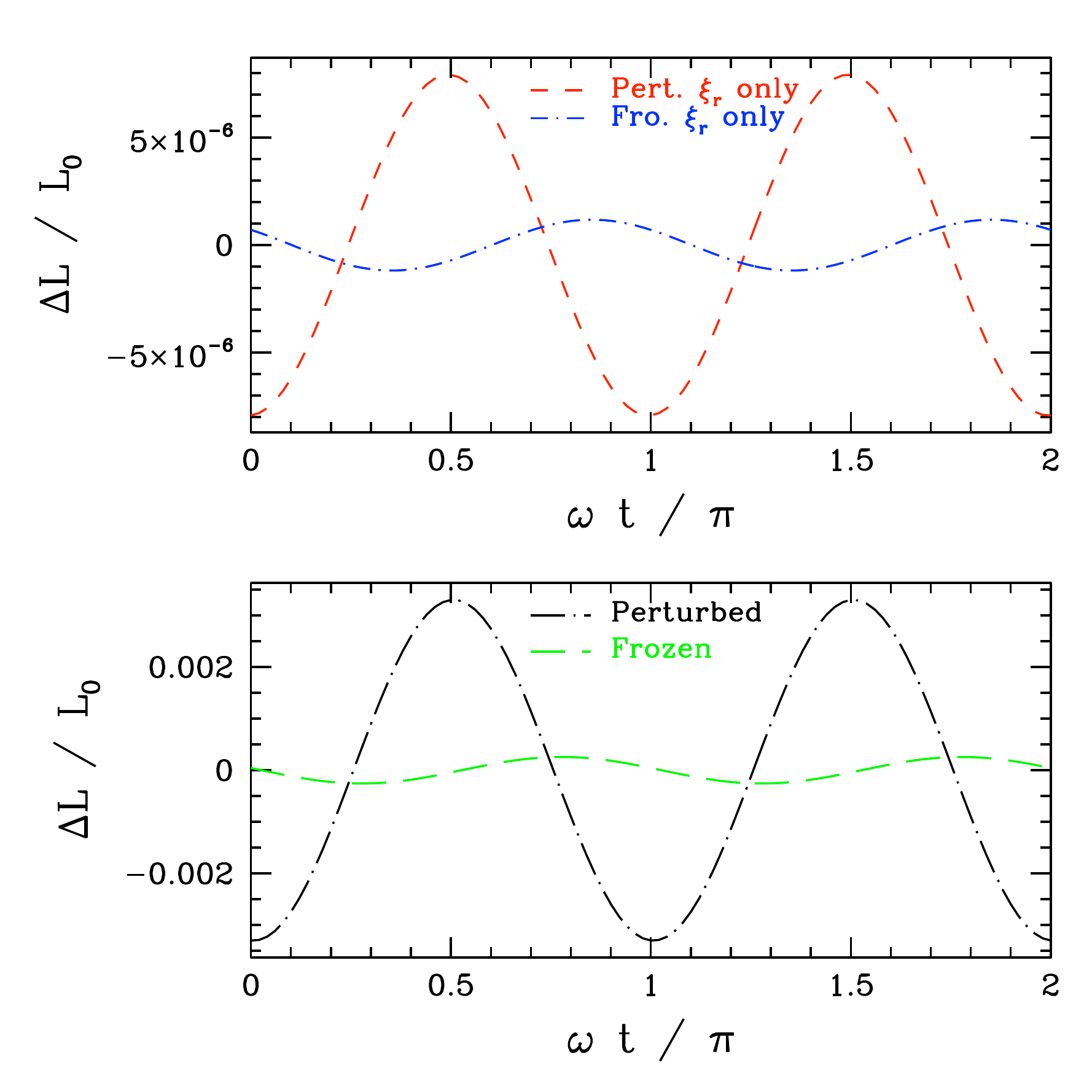}
    \caption{Same as Figure~\ref{fig:WASP_19} but for CoRoT--17.
    }
    \label{fig:CoRoT_17}
\end{figure*}

\subsection{CoRoT--17}
\label{sec:observed_systems:CoRoT-17}

The star in the CoRoT--17 system is similar to an aged Sun, which has expanded. The increased stellar radius would be expected to lead to a greater response to the tidal potential. The star maintains a radiative core surrounded by a convective envelope, with a thin radiative skin at the very surface. The radial displacement once again closely tracks the equilibrium tide prediction, whilst the tangential displacement remains fairly insensitive to the orbital period.

The large radius of the star gives a large prediction for the equilibrium radial displacement, given the comparatively long period orbit. The perturbed convection case matches this prediction fairly well, giving rise to a fairly small discrepancy between $\xi_{r}$ and $V$: only a factor of $\sim 5$. The frozen convection model predicts a much greater discrepancy, as the radial displacement is an order of magnitude smaller than the equilibrium tide prediction, and the horizontal displacement is a factor of $\sim 16$ greater than the perturbed convection model's prediction.

Overall, this results in a RV prediction from the perturbed convection model which is a factor of $\sim 3$ greater than the equilibrium tide prediction ($\sim 90$ cm s$^{-1}$ compared to $\sim 30$ cm s$^{-1}$). On the other hand, the frozen convection model predicts a signal a factor of $\sim 50$ greater than the equilibrium tide, at $\sim 16$ m s$^{-1}$. Each of the three cases has a different phase, which could help to distinguish between the different behaviours.

The photometric variation is very similar to that of Qatar 5 b, with the perturbed convection case predicting  an observed flux variation  which is an order of magnitude greater than the frozen convection prediction, whether including $F'_r$ or just $\xi_{r}$. The prediction taking $F'_r$ into account  should be detectable, at $\sim 10^{-3}$, and is three orders of magnitude greater than the prediction which includes  $\xi_{r}$ only, which would be very difficult to detect.

\subsection{Non--disc--integrated signal}
\label{sec:observed_systems:non-disc}

\begin{figure*}
    \centering
    \includegraphics[width=0.8\columnwidth]{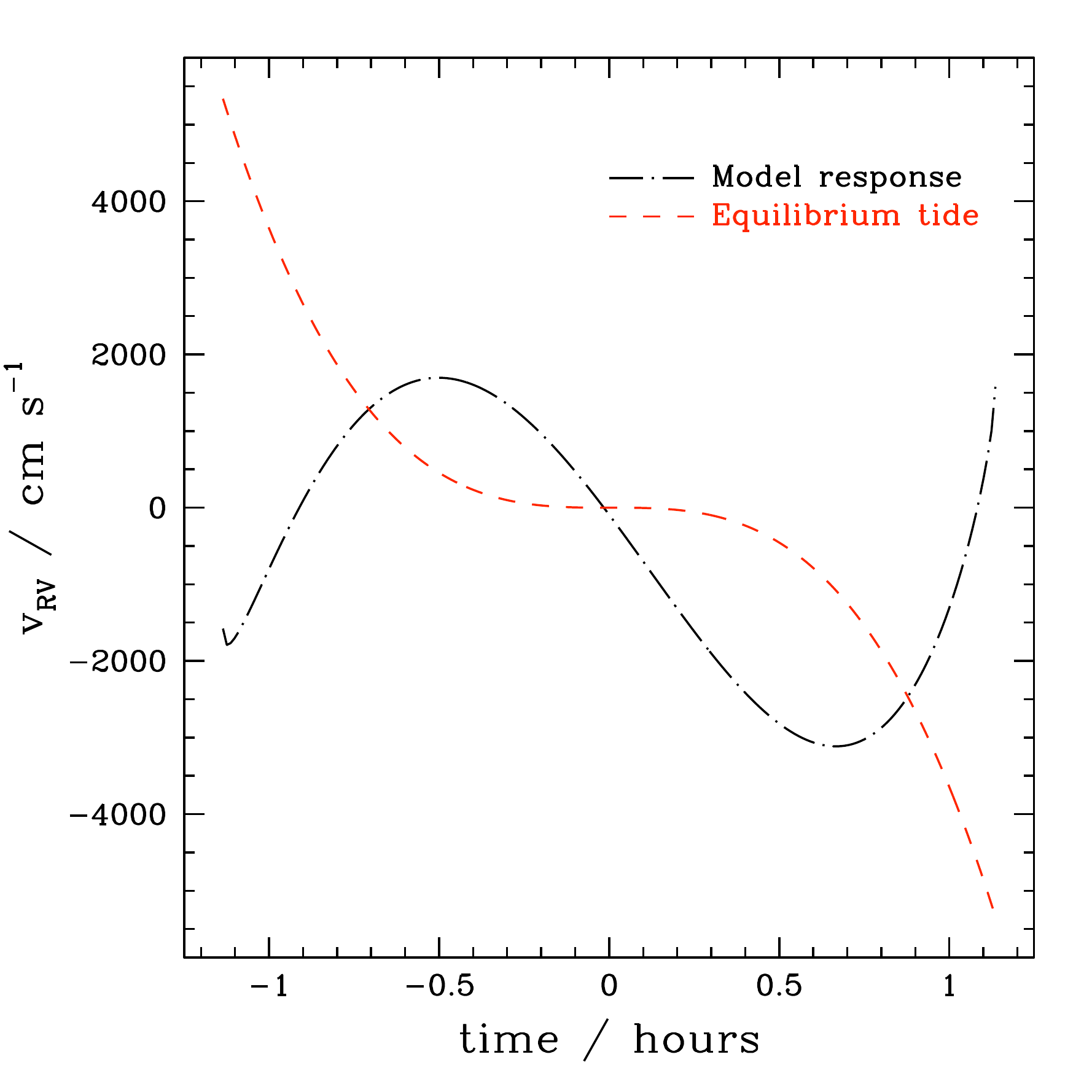}
    \includegraphics[width=0.8\columnwidth]{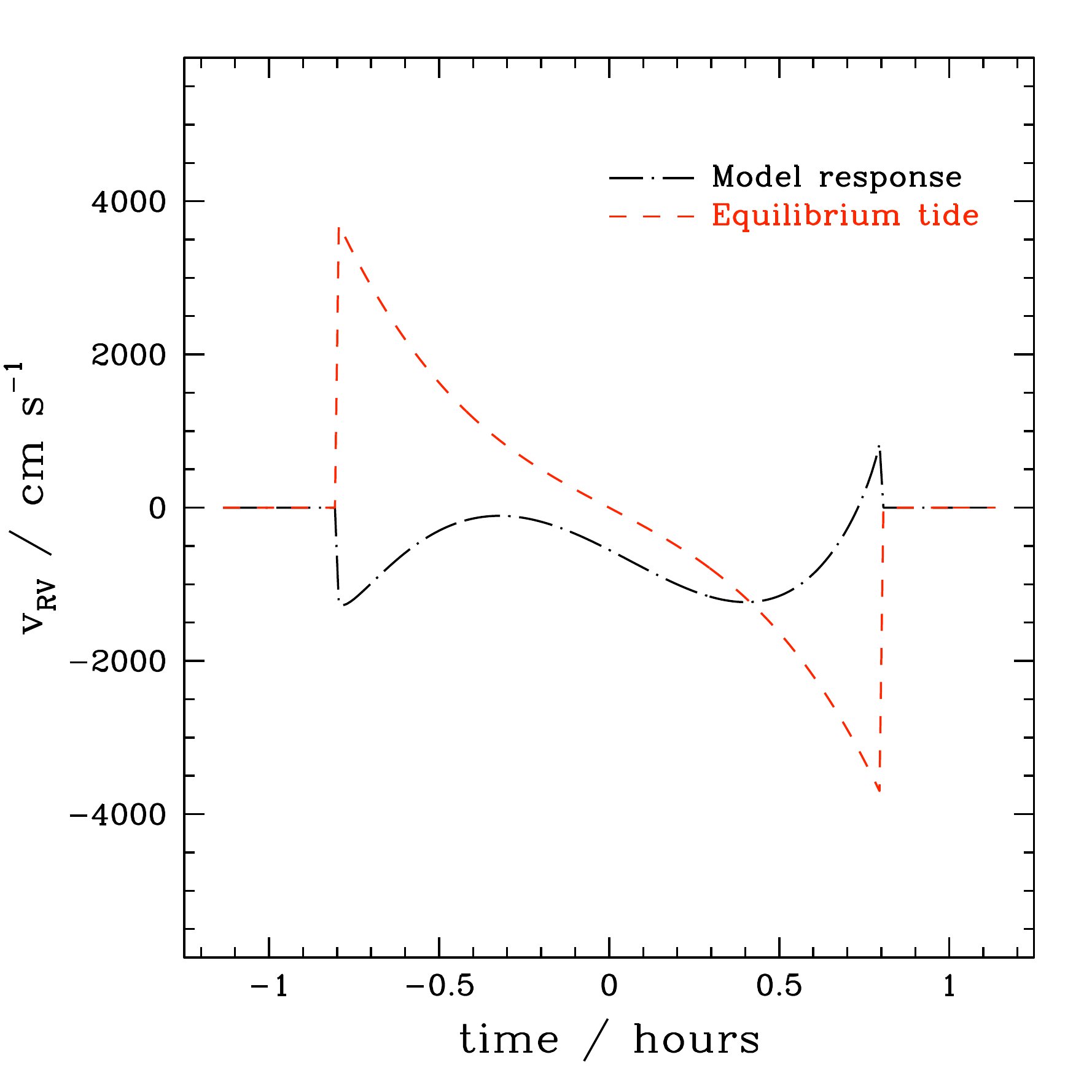} \\
    \includegraphics[width=0.8\columnwidth]{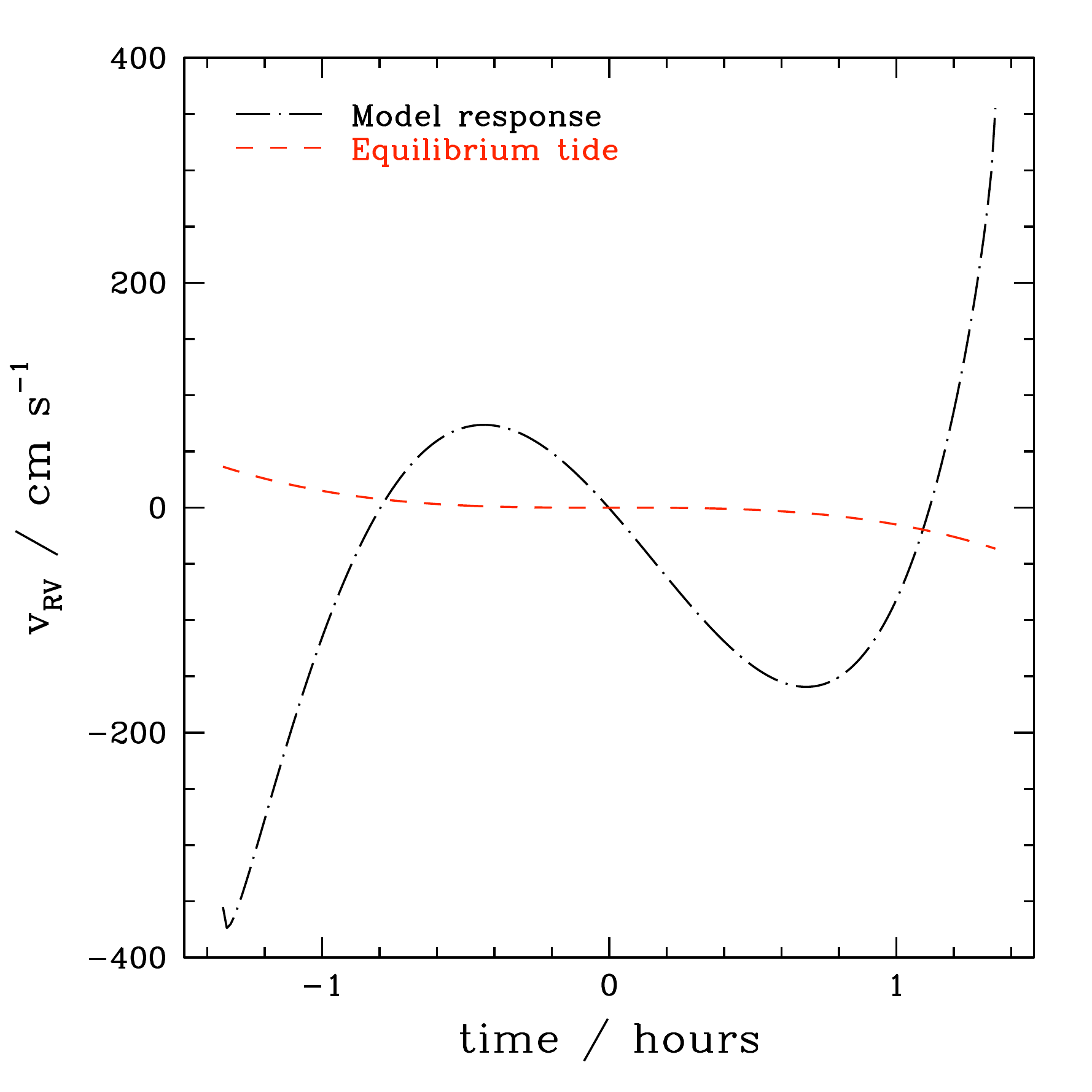}
    \includegraphics[width=0.8\columnwidth]{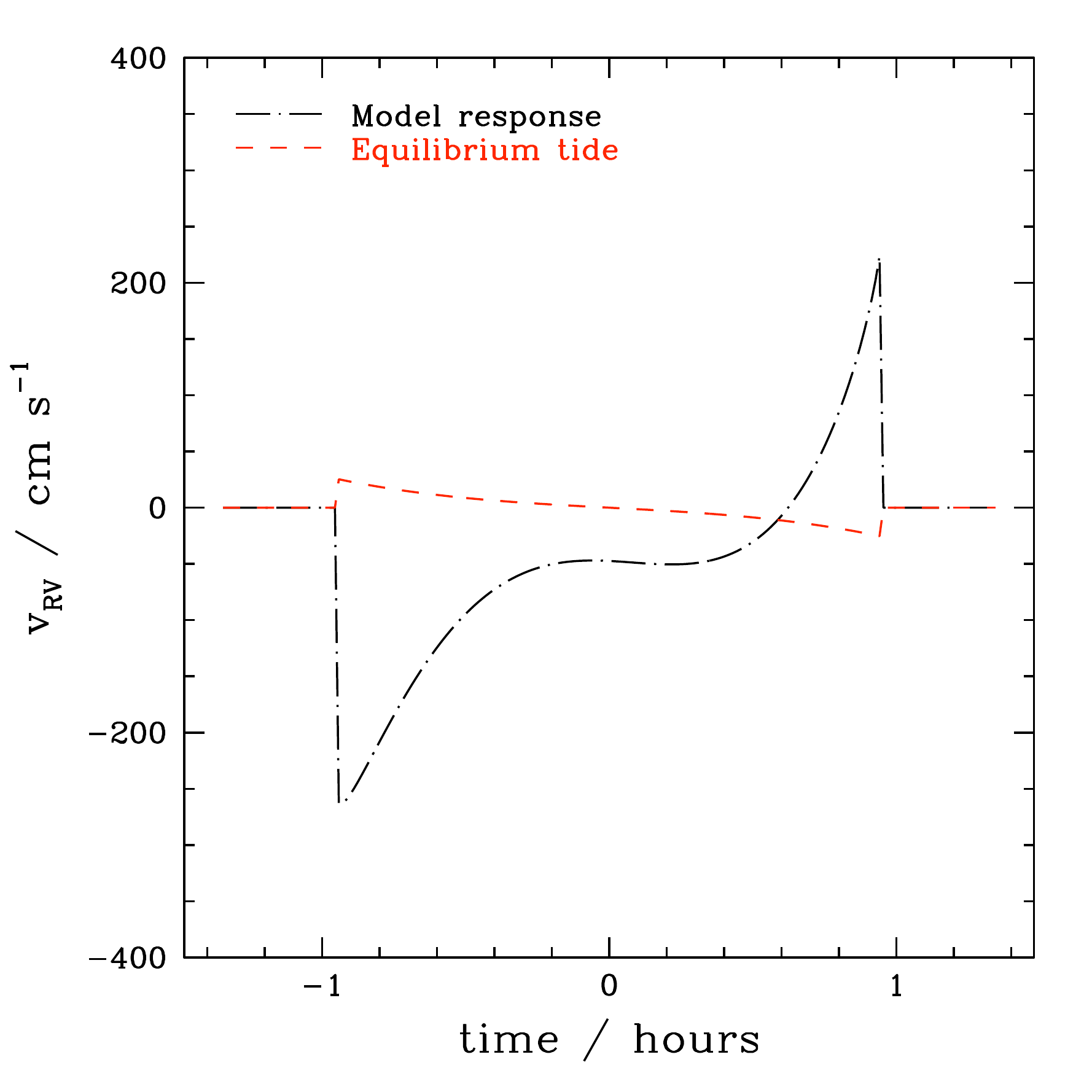} \\
    \includegraphics[width=0.8\columnwidth]{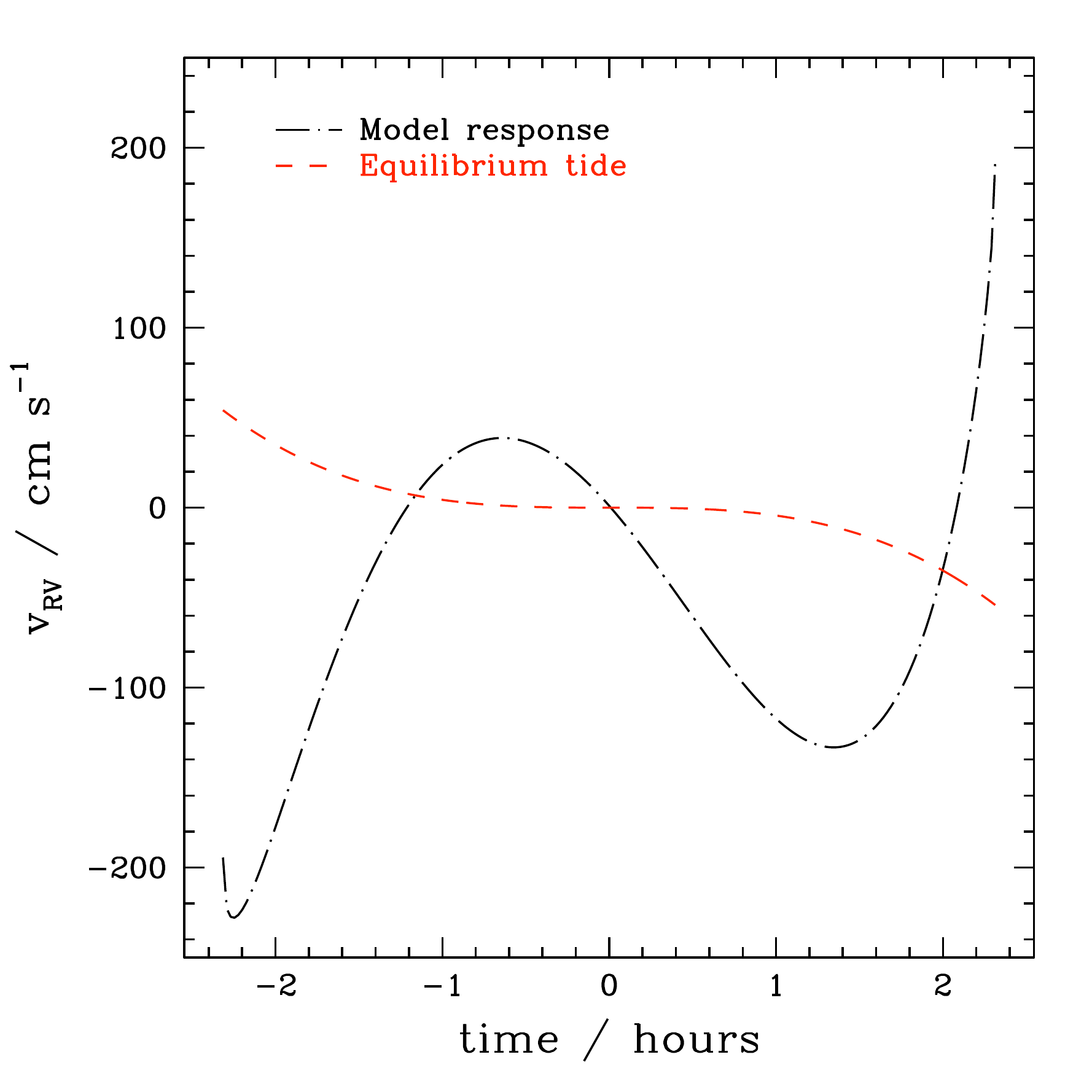}
    \includegraphics[width=0.8\columnwidth]{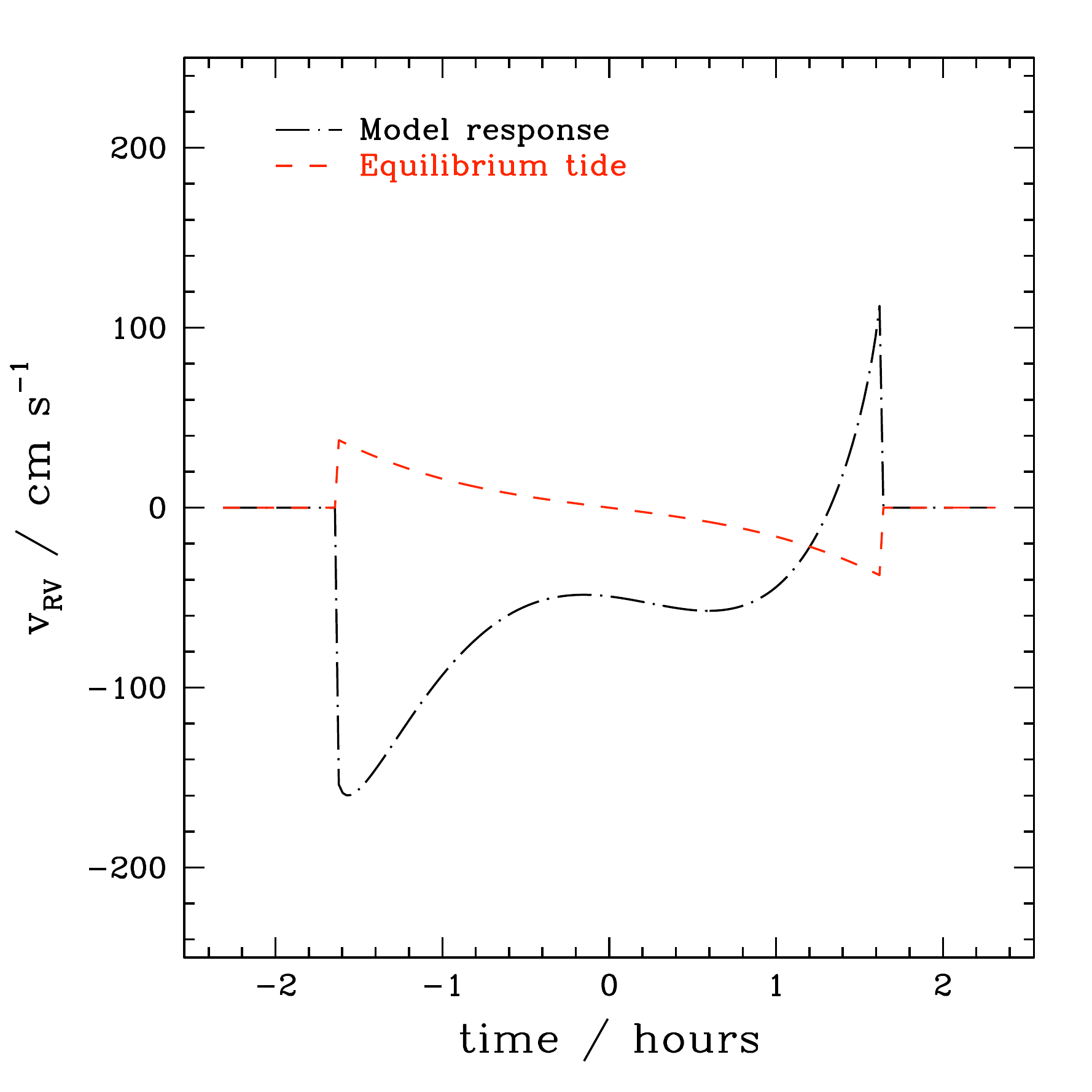}
    \caption{This figure shows the radial velocities blocked by the planets over the course of a transit, causing a dip in brightness at the corresponding location in the line--shape, for the perturbed convection model applied to WASP-18 (top), Qatar 5 (middle) and CoRoT-17 (bottom). The left column shows the case that the planet is transiting exactly edge-on, such that the blocked portion of the star is at $\theta_{*} = \upi/2$. The right column shows the case that the system is not exactly edge-on, such that the portion of the star that is blocked is at $\theta_{*} = \upi / 4$.  Note that this both affects the shape of the signal and shortens the duration of the transit. The contrast between the equilibrium tide and the modelled response (with perturbed convection) is significant in each shape, whilst the diversity of possible curves arising due to different surface behaviours and transit locations is apparent.}
    \label{fig:non-disc-integrated}
\end{figure*}

The profiles for the radial velocity signal blocked by the planet over the course of a transit are shown in Figure~\ref{fig:non-disc-integrated}. This shows only the value of the radial velocity blocked against time, and not the depth of the transit at that value of ${\rm v}_{RV}$.

 This signal could be observed by subtracting the signal during transit from the signal during the secondary eclipse. In an ideal case, this would produce a peak centred on the blocked radial velocity, and it is the variation of this central value that is plotted against time.

The curves in Figure\,\ref{fig:non-disc-integrated} show the predictions for the perturbed convection model and the equilibrium tide prediction applied to WASP-18, Qatar 5 and CoRoT-17. In each case the perturbed convection curves are significantly different to the equilibrium tide prediction, whilst similarities between the different systems are apparent. 

The deviation from the equilibrium tide comes from the difference in $V$ from the equilibrium tide prediction, as in each of these systems the imaginary component of $V$ is not negligible, and the magnitude of $V$ differs from $\xi_{r, \text{eq}}$ (for WASP-18 it is smaller than expected, for the other two systems it is significantly larger than expected). If time-resolved spectra are able to detect signals such as these, the radial velocity signal originating from a specific, known point on the stellar surface could be isolated. This would enable the profile of the radial velocity along the path of the transit to be broken down into its spatial components in order to separate the radial and tangential contributions and directly exposing the form of the oscillations.

\section{Discussion}
\label{sec:Discussion}

The response of a star being perturbed by a nearby companion may be significantly different to that predicted by a simple equilibrium approximation, due to the presence of non--adiabatic behaviour towards the stellar surface. This could lead to observable signals which greatly differ from those predicted from the equilibrium tide.

At the surface, the horizontal displacement is found to tend to a constant value, independent of the orbital period.  For long orbital periods, where the horizontal displacement dominates over the radial displacement, this results in the magnitude of the radial velocity  perturbation scaling as $P^{-1}$, instead of $P^{-3}$ for the equilibrium tide. Therefore, for long orbital periods, the  radial velocity  perturbation is much greater than expected from the equilibrium tide.

This constant value is attained for orbital periods where $\left| \xi_{r, \text{eq}} \right| \lesssim |V|$, and therefore breaks down for ultra--short period planets.  For those, the predicted RV signal may be smaller than expected from the equilibrium tide, and the scaling with period reverts to roughly match the $P^{-3}$ of the equilibrium tide. The phase of the RV signal is also generally found to differ from the equilibrium tide, as the RV signal is dominated by the real and negative component of $V$, leading to a signal which is roughly inverted compared to the equilibrium tide.

The photometric variation is found to have the same scaling as predicted by the equilibrium tide, being proportional to $P^{-2}$, which holds whether $F'_r$ is taken into account or not in the calculation of  the change in observed flux. If $F'_r$ is included, the magnitudes predicted can become very large, which are likely to be overestimates of the real signal. If only the contribution from the radial displacement is included, the  observed flux variation is smaller by three orders of magnitude. In this case, the perturbed convection model matches well with the equilibrium tide prediction, as $\xi_{r} \approx \xi_{r, \text{eq}}$, though the frozen convection prediction is several times smaller than this. For very short period orbits, the value of $F'_r$ is likely inaccurate, as it can become large and the assumption of small perturbations in the model would no longer be valid. The phase of the prediction using perturbed convection matches the equilibrium tide prediction (whether  $F'_r$ is included or not), though the frozen convection case does not -- this could potentially be used to distinguish between the two models using observations.

Near resonances the response of the stellar surface can depend strongly on the orbital frequency of the companion. In order to correctly capture this resonant behaviour, the resonant frequencies of the model must match those of the real star. As the resonances depend upon the structure of the star as a whole, matching the stellar model to the real resonant frequencies can be difficult, particularly if the stellar properties are not tightly constrained.
In general, we would still expect deviation from the equilibrium tide near resonances. Whilst this is unlikely to be observed for periods greater than a day, where resonances are very narrow,  it could be seen at  very short periods, where the resonances are wider.

Whilst the above behaviours are found to be common to both the frozen convection and perturbed convection models, the specific predictions made for a system depend strongly upon the choice of model used for the convective flux. Both models have different virtues, with the frozen convection approach providing a baseline comparison without complicating the model by perturbing a process as non--linear and non--local as convection. However, the artificial suppression of the stellar response within the convection zone does give rise to very large gradients in a thin region just below the surface, once the radiative flux begins to become significant. The perturbed flux approach was designed to describe the behaviour in the superadiabatic zone towards the top of the convective zone, and therefore is well suited to model the non--adiabatic effects in the region where convection becomes inefficient. Deep in the convection zone, this approach may not be applicable, as the entropy gradient is very shallow and errors may accumulate, giving rise to overly large flux perturbations at the surface. Overall, it is likely that both approaches fall short of reality, but they can be used to provide insight into the range of possible behaviours, and the dependence of the stellar response on the model of convection, whilst highlighting the deviation from the equilibrium tide.

The planets modelled in this work with periods under one day exhibited similar behaviour. Both WASP--19 and WASP--18 produced RV signals that were smaller than predicted by the equilibrium tide in the perturbed convection model, whilst the frozen convection model predicted a signal much larger than ${\rm v_{RV, \text{eq}}}$.  If observed,  this signal  would therefore enable the different models to be distinguished from each other. The predictions for the photometric variation were also very large, with the signal taking the perturbed flux into account on the order of $1-10$~per~cent. If the flux were overestimated and the change in  observed flux were mainly due to the radial displacement, the signal  would still be potentially  detectable, at $10-100$ ppm.  A photometric tidal signal has been observed for WASP--18 \citep{Shporer2019}, with a semi--amplitude of $\sim 190$~ppm, which is similar in both phase and amplitude to the $\xi_{r}$-only prediction in the perturbed convection case. This confirms that our calculation has greatly overestimated the perturbation to the flux in this case,  as expected from the discussion in section~\ref{sec:observed_systems:WASP-18}.

A radial velocity signal at twice the orbital frequency has been detected for WASP--18 b \citep{Triaud2010}. This was attributed to a non--zero eccentricity, although this explanation was disputed by \citet{Arras2012} who favoured the tidal oscillation as the source. Later,  work by \citet{Maciejewski2020a} has suggested that the tidal signal has an amplitude of $\sim 18$ m s$^{-1}$, which is smaller than predicted by the equilibrium tide by a factor of two, and approximately a factor of three larger than the perturbed convection prediction in this work. There is a phase difference between the two predictions,  and the equilibrium tide phase more closely matches the observations. However, the fact that the model is close to resonance can lead to a large change in the phase for a small change in forcing frequency, which may account for such a large discrepancy between the phase of the perturbed convection prediction and the observed tidal signal.

Similarly, \citet{Maciejewski2020} suggest the presence of a tidal radial velocity signal in WASP-12, with an amplitude of $\sim 7$~m~s$^{-1}$, which lies between the equilibrium tide amplitude of $\sim 4$~m~s$^{-1}$ used in this work, and the perturbed convection case, which gives an amplitude of $\sim12$~m~s$^{-1}$. In the perturbed convection result there is a phase lag compared to the equilibrium tide, although this is fairly sensitive to the response in the radiative zone, due to the relatively thin convection zone of the star. Stars with thin surface convection zones may therefore be more difficult to model, as the behaviour of the radiative zone is likely to have a greater impact upon the surface response, and at short periods this may bring in effects due to resonances. The observed flux variation of WASP-12 would be expected to be observable, with a fractional semi--amplitude of $\sim10^{-3}-10^{-4}$. If only the signal arising due to $\xi_{r}$ is present, it may be on the edge of being detectable, ranging from $\sim10^{-6}-10^{-5}$, depending on the model used for convection.

For the longer period planets, Qatar 5 and CoRoT--17, the perturbed convection model predicts signals $\sim 1 $ m s$^{-1}$, with the prediction from the frozen convection model being an order of magnitude larger.  Both values are larger than the RV signal expected from the equilibrium tide.   The photometric variation from the flux is around $0.1$~per~cent, and would therefore be expected to be observable, whilst the signal arising only from the radial displacement would be much more difficult to detect, at $\sim 1 $ ppm. Measurements in the detection of both Qatar 5--b \citep{Alsubai2017} and CoRoT--17 b \citep{Csizmadia2011} do not constrain either the photometric variation or the RV signal to the level where either of these signal would be clearly visible.

In applying this to more systems, this approach could be used to provide an independent estimate of the mass of a transiting planet, or could even be combined with the RV signal from the star's motion about the system's common centre of mass to break the degeneracy between the planetary mass and the orbital inclination. If other planets in the same system were present, it could be helpful to remove the tidal signal from the Hot Jupiter in order to reduce the periodic background noise, especially if the planets were in resonance with each other.

The non--disc--integrated methods for observing these tidal oscillations could provide useful insight into the nature of the tidal oscillations themselves, as the signal during transit and the time--dependent broadening both give signals which depend separately on $\xi_{r}$ and $V$. Detecting the signal during transit would require spectra to be taken quickly, and with a short cadence, which would be best suited for nearby, bright stars. In order to build up the SNR it would also be useful to capture many transits, and therefore a short period orbit would be preferable. The time--dependent broadening signal would be difficult to detect due to the many other sources of line--broadening, with a comparatively small variation. However, the long coherence time expected from the tidal signal could be taken advantage of by using observations taken over a long time in order to average out the other sources of broadening, and detect the coherent underlying variation, small though it is.

Whilst it may be simple to show that a system is deviating from the equilibrium tide, more work would be required in order to investigate the scaling of the behaviour with orbital period, amongst other parameters. Future work would be required in order to make testable predictions for systems which could be observed, and particularly for distinguishing between the non--adiabatic models and the equilibrium tide approximation.

Observations of the photometric signal arising from a tidal perturbation would give very useful constraints on the model used for the convective flux, and would help improving the model as a whole.

\section{Conclusions}
\label{sec:Conclusion}

The response of a star to a tidal perturbation is strongly affected by non--adiabatic effects towards the stellar surface, resulting in observable signals which can differ greatly from the equilibrium tide prediction. Generally, it is found that the horizontal displacement tends towards a constant value, independent of the orbital period of the system. This results in larger RV signals than predicted by the equilibrium tide, particularly for longer period orbits such as Qatar 5 b and CoRoT--17 b. For ultra--short period orbits, such as WASP--19 b and WASP--18 b, the RV signal is  found to be lower than expected from the equilibrium tide. The photometric variation is predicted to scale as in the equilibrium tide approximation, proportional to $P^{-2}$, but the magnitude of the flux at the surface found here is very large, and may be an overestimate. Observations would be helpful in constraining the model used for the perturbation to the convective flux.

Non--disc--integrated methods could be used to give clear evidence of deviation from the equilibrium tide,  and to separate the contributions from the radial and horizontal displacements, either using the RV signal during a transit or by observing the time--dependent line--broadening signal.

Observations of these signals would provide independent mass estimates for transiting exoplanets, and would enable to  break the mass--inclination degeneracy for non--transiting exoplanets.

\section*{Data availability}

No new data were generated or analysed in support of this research.

\section*{Acknowledgements}

AB is supported by a PhD studentship from the Science and Technology Facilities Council (STFC), grant ST/N504233/1.  We thank the referee for a very thorough review  that has significantly improved the paper.


\appendix

\section{Euler angles and rotations}
\label{App:angles}

In calculating the response of the star, it is greatly simplified by working in the star's natural frame of reference, $(\theta_{*}, \phi_{*})$, with the planet orbiting in the plane defined by $\theta_{*} = \pi / 2$. In this frame, the observer  is taken to be in the direction given by $(\theta_{0}, \phi_{0})$. In the observer's frame, described by $(\theta_{\text{ob}}, \phi_{\text{ob}})$, the observer is at $\theta_{\text{ob}} = 0$, with $\theta_{\text{ob}} < \pi / 2$ visible to the observer.

To convert between the two frames, we use the properties of spherical harmonics and Euler angles, guided by \citet{Morrison1987}. The two frames of reference are related to each other by a rotation given by
\begin{equation}
\mathbfit{e}_{z_{*}} = \mathbfss{R}^{\text{(a)}} (\alpha, \beta, \gamma) \mathbfit{e}_{z_{\text{ob}}},
\end{equation}
with $\mathbfit{e}_{z_{*}}$ being a unit vector in the star's frame, and $\mathbfit{e}_{z_{\text{ob}}}$ being the corresponding unit vector in the observer's frame. The operator $\mathbfss{R}^{\text{(a)}} (\alpha, \beta, \gamma)$ is equivalent to $\mathbfss{R}_{z_{\text{ob}}}(\alpha) \mathbfss{R}_{y_{\text{ob}}}(\beta) \mathbfss{R}_{z_{\text{ob}}}(\gamma)$ where the active convention for rotations is being used, and the $y_{\text{ob}}$ and $z_{\text{ob}}$ axes are those of the observer's frame. The Euler angles are found to be $(\alpha, \beta, \gamma) = (0, -\theta_{0}, -\phi_{0})$.

This leads to the relation between a given spherical harmonic, $Y_{l}^{m}$, between the two frames as
\begin{equation}
\label{eq:App:sph_coord_change}
Y_{l}^{m} (\theta_{*}, \phi_{*}) = \sum_{\mu = -l}^{l} Y_{l}^{\mu} (\theta_{\text{ob}}, \phi_{\text{ob}}) D^{l}_{\mu, m} (0, -\theta_{0}, -\phi_{0})
\end{equation}
where $D^{l}_{\mu, m} (\alpha, \beta, \gamma)$ is an element of the Wigner D--matrix.

The relevance of this on the system of interest is that the tidal potential $\phi_{\rm P}$  of the hot Jupiter (or any companion) can be expressed as a sum of spherical harmonics, and the lowest order (in terms of $R/D$, where $R$ is the stellar radius and $D$ is the semi--major axis of the orbiting body) non--constant term is:
\begin{equation}
\label{eq:App:tidal}
\phi_{\text{P}} = \Re \left( - \frac{G m_{\text{p}}}{4 D} \left( \frac{r}{D} \right)^{2} P_{2}^{2} (\cos \theta_{*}) \text{e}^{2 i (\omega t - \phi_{*})} \right),
\end{equation}
where $G$ is Newton's gravitational constant, $\omega$ is the angular frequency of the planet's orbit, $t$ is the time, measured from the start point defining the coordinate system, $m_{\text{p}}$ is the planet's mass, $r$ is the radial position in the star,  $P_{2}^{2} (\cos \theta_{*})=3 \sin^{2} \theta_{*}$ is the associated Legendre polynomial, and $\Re$ denotes the taking of the real part.  The Legendre polynomial  relates to spherical harmonics, as $Y_{2}^{-2} (\theta_{*}, \phi_{*}) = \frac{1}{4} \sqrt{\frac{15}{2 \pi}} \sin^{2} \theta_{*} \text{e}^{- 2 \text{i} \phi_{*}}$.
Since this is the only source of time and angular dependence in the system of linear equations describing the response of the star, any perturbed quantity $q'$ can be written as:
\begin{equation}
  q'(r, \theta_{*}, \phi_{*}, t) = \Re \left( q'(r)
\times 3 \sin^{2} \theta_{*} \text{e}^{2 i (\omega t - \phi_{*})} 
\right).
\label{eq:pertquant}
\end{equation}
By using equation~\ref{eq:App:sph_coord_change}, this can be expressed in the coordinates of the observer's frame. The great benefit of this comes when integrating over the visible disc, as its limits are $\theta_{\rm ob} \in \{ 0, \frac{\pi}{2} \}$, and $\phi_{\rm ob} \in \{ 0, 2 \pi \}$. If integrating over the visible disc without any weighting in $\phi$, the complex expression in equation~\ref{eq:App:sph_coord_change} can be simplified by the fact that $\int_{0}^{2 \pi} \text{e}^{\text{i} \mu \phi} \text{d}\phi $ is non--zero (and equal to $2 \pi$) only when $\mu = 0$.  Therefore:
$$ \int_0^{2 \pi} Y^{-2}_2 (\theta_{*}, \phi_{*}) \text{d} \phi_{\rm ob} =
\int_0^{2 \pi} D^{2}_{0, -2} (0, -\theta_{0}, -\phi_{0}) Y_{2}^{0} (\theta_{\rm ob}, \phi_{\rm ob}) \text{d} \phi_{\rm ob} .$$
Using: $$ D^{2}_{0, -2} (0, -\theta_{0}, -\phi_{0}) = \sqrt{\frac{3}{8}} \sin^2 \theta_0 \text{e}^{- 2 \text{i} \phi_{0}}, $$ and: $$ Y_{2}^{0} (\theta_{\rm ob}, \phi_{\rm ob}) = \frac{1}{2} \sqrt{\frac{5}{4 \pi}} 
\left( 3 \cos^2 \theta_{\rm ob} -1  \right),$$
we obtain the useful expression:
\begin{equation}
\label{eq:App:spherical_useful}
\int_{0}^{2 \pi} \text{d} \phi_{\rm ob} \sin^{2} \theta_{*} \text{e}^{- 2 \text{i} \phi_{*}} = \pi \sin^{2}  \theta_{0} \text{e}^{- 2 \text{i} \phi_{0}}  ( 3 \cos^{2} \theta_{\rm ob} - 1 ).
\end{equation}
Therefore,  for a perturbed quantity with angular dependence given by equation~(\ref{eq:pertquant}), the integral over the visible circle of constant $\theta_{\text{ob}}$ is:
\begin{align}
  \int_{0}^{2 \upi} {\rm d}\phi_{\rm ob} q'(R, \theta_{\text{ob}}, \phi_{\text{ob}}, t) & =  \nonumber \\
  3 \upi \Re \left[ q'(R)  ( 3 \right. & \left. \cos^{2} \theta_{\rm ob} - 1 ) \sin^{2} \theta_{0} \text{e}^{2 \text{i} ( \omega t - \phi_{0} )}  \right],
\label{eq:App:example_variable:new_coords}
\end{align}
where it is important to note the change from $\theta_{*}$ and $\phi_{*}$ to $\theta_{0}$ and $\phi_{0}$ in the sine function and exponential respectively, compared to equation~\ref{eq:pertquant}.

\section{Observed flux variation derivation}
\label{App:Lum}

Here, the details of the derivation of the  observed flux variation are given and justified.  The results derived below are used in section~\ref{sec:Methods:lum_var}. The method followed is similar to that of \citet{Robinson1982}, the primary difference being that this work explicitly keeps track of the non--radial perturbations to displacement, $\bm{\xi}_{h}$.

\subsection{Surface normal}
\label{App:Lum:normal}

The change in surface normal due to the oscillations is calculated by normalising: $$ \frac{\partial \bm{r}}{\partial \theta_{\rm ob}} \bm{\times} \frac{\partial \bm{r}}{\partial \phi_{\rm ob}} ,$$ where $\bm{r}$ is the location of the surface, given by $R \hat{\bm{r}} + \bm{\xi}$, with $\bm{\xi}$ being the vector displacement of the surface. The calculation is done in the observer's frame.  This leads to an expression for the perturbed normal to the surface as $\hat{\bm{n}} = \hat{\bm{r}} + \Delta \hat{\bm{n}}$, where the perturbation is:

\begin{equation}
  \Delta \hat{\bm{n}}  = \frac{1}{R} \left[ \hat{\bm{\theta}}_{\rm ob} \left( \xi_{\theta} -  \frac{\partial \xi_r }{\partial \theta_{\rm ob}} \right)  + \hat{\bm{\phi}}_{\rm ob} \left( \xi_{\phi} - \frac{1}{\sin \theta_{\rm ob}} \frac{\partial \xi_r }{\partial \phi_{\rm ob}} \right)  \right],
\label{eq:App:Delta_normal}
\end{equation}
where $\xi_r$ and $\xi_{\theta}$ are evaluated at $(r, \theta_{\rm ob}, \phi_{\rm ob})$ and $\xi_{\theta}=\partial V / \partial \theta_{\rm ob}$.

Note that no rescaling is necessary to first order, and that the perturbation to the normal acts only perpendicular to the radial direction.

This yields a change of  observed flux:
\begin{equation}
\label{eq:App:L_n}
\Delta L_{n} = \int_{0}^{\pi / 2} \int_{0}^{2 \pi} h_{0} \bar{F}_{0} \Delta \hat{\mathbfit{n}} \mathbf{\cdot} \hat{\mathbfit{n}}_{\text{ob}} \text{d}S_{0},
\end{equation}
with ${\rm d}S_0=R^2 \sin \theta_{\rm ob} {\rm d} \theta_{\rm ob} {\rm d} \phi_{\rm ob}$, and $\hat{\mathbfit{n}}_{\text{ob}}$ is in the direction of the $z$--axis in the observer's frame,  so that $h_0=c \left[ 1-a \left(1 - \cos \theta_{\rm ob} \right) -b \left(1 - \cos \theta_{\rm ob} \right)^2\right]$.
Using:
\begin{equation}
  \Delta \hat{\mathbfit{n}} \mathbf{\cdot} \hat{\mathbfit{n}}_{\text{ob}} =
  \frac{- \sin \theta_{\rm ob}}{R} \frac{\partial}{\partial \theta_{\rm ob}} \left( V -   \xi_r  \right) ,
  \label{eq:deltandotnob}
\end{equation}
equation~(\ref{eq:App:L_n}) becomes:
\begin{align}
  \Delta L_n & = -  R  F_0  c   \int_0^{\pi /2}   {\rm d} \theta_{\rm ob}
\bigg\{ \sin^2 \theta_{\rm ob}  \times \nonumber \\
  & \left[ 1-a \left(1 - \cos \theta_{\rm ob} \right) -b \left(1 - \cos \theta_{\rm ob} \right)^2\right] 
   \times
  \nonumber \\
  & \frac{\partial }{\partial \theta_{\rm ob}} \int_0^{2 \pi} {\rm d} \phi_{\rm ob}
    \big[ V(R, \theta_{\rm ob}, \phi_{\rm ob},t) - \xi_r (R, \theta_{\rm ob}, \phi_{\rm ob},t) \big]
    \bigg\} .
  \end{align}
The integral over $\phi_{\rm ob}$ is given by equation~(\ref{eq:App:example_variable:new_coords}),
and after integration over $\theta_{\rm ob}$ we finally obtain:
\begin{equation}
\label{eq:App:DeltaL_n:complex}
\Delta L_{n} =\Re \left[ \frac{9}{2} \pi R c \left( 1 - \frac{7a + 4b}{15} \right)  F_{0} \left[ V(R) - \xi_{r}(R) \right] \sin^{2} \theta_{0} \text{e}^{2 \text{i}(\omega t - \phi_{0})} \right].
\end{equation}

\subsection{Limb-darkening}
\label{App:Lum:Limb-Darkening}

Limb-darkening is given as:
\begin{equation}
\label{eq:App:Lum:Limb-Darkening}
h = c \left[ 1 - a \left( 1 - \hat{\mathbfit{n}} \mathbf{\cdot} \hat{\mathbfit{n}}_{\text{ob}} \right) - b \left( 1 - \hat{\mathbfit{n}} \mathbf{\cdot} \hat{\mathbfit{n}}_{\text{ob}} \right)^{2} \right],
\end{equation}
where the variables are as described in section~\ref{sec:Methods:lum_var}. This is perturbed through the normal to the surface, giving: 
\begin{multline}
\label{eq:App:Lum:Limb-Darkening:pert}
h = c \left[ 1 - a \left( 1 - \hat{\mathbfit{r}} \mathbf{\cdot} \hat{\mathbfit{n}}_{\text{ob}} \right) - b \left( 1 - \hat{\mathbfit{r}} \mathbf{\cdot} \hat{\mathbfit{n}}_{\text{ob}} \right)^{2} \right] \\
+ c \left[ a + 2b- 2 b \hat{\mathbfit{r}} \mathbf{\cdot} \hat{\mathbfit{n}}_{\text{ob}} \right] \Delta \hat{\mathbfit{n}} \mathbf{\cdot} \hat{\mathbfit{n}}_{\text{ob}} + \mathcal{O} \left( (\Delta \hat{\mathbfit{n}} \mathbf{\cdot} \hat{\mathbfit{n}}_{\text{ob}})^{2} \right),
\end{multline}
in which the first term is $h_{0}$, the second is defined as $\Delta h$ (with
$\Delta \hat{\mathbfit{n}} \mathbf{\cdot} \hat{\mathbfit{n}}_{\text{ob}}$ given by equation~\ref{eq:deltandotnob}), and the third term is second order in a small quantity, so is neglected.

This results in a change in observed flux of the form
\begin{equation}
\label{eq:App:Lum:Limb-Darkening:integral}
\Delta L_{h} = \int_{0}^{\pi / 2} \int_{0}^{2 \pi} \Delta h \bar{F}_{0} \hat{\mathbfit{r}} \mathbf{\cdot} \hat{\mathbfit{n}}_{\text{ob}} \text{d}S_{0}.
\end{equation}
This integral is calculated in the same way as the integral~(\ref{eq:App:L_n}) which gives $\Delta L_n$, and this leads to the final expression:
\begin{equation}
\label{eq:App:DeltaL_h:done}
\Delta L_{h} = \Re \left[\frac{12 \pi}{5} R c \left(a + \frac{3b}{4}\right) F_{0}  \left[ V(R) - \xi_{r}(R) \right] \sin^{2} \theta_{0} \text{e}^{2 \text{i}(\omega t - \phi_{0})} \right] .
\end{equation}

\subsection{Flux}
\label{App:Lum:F}

The emergent flux is defined as $\bar{F} = \mathbfit{F} \mathbf{\cdot} \hat{\mathbfit{n}}$, which assumes that all flux which reaches the surface is radiated isotropically (or, more precisely, that the anisotropy is included through limb-darkening). This separates into the equilibrium and perturbed quantities as
\begin{equation}
\label{eq:App:Lum:F:pert}
\bar{F} = \left( {F}_{0} \hat{\mathbfit{r}} + \Delta \mathbfit{F} \right) \mathbf{\cdot} \left( \hat{\mathbfit{r}} + \Delta \hat{\mathbfit{n}} \right) = F_{0} + \Delta \mathbfit{F} \mathbf{\cdot} \hat{\mathbfit{r}},
\end{equation}
where second order terms in small quantities have been neglected, and the fact that $\hat{\mathbfit{r}} \mathbf{\cdot} \Delta \hat{\mathbfit{n}} = 0$ has been used. $\Delta \mathbfit{F}$ is the Lagrangian perturbed flux, equal to $\mathbfit{F'} + ( \mathbf{\xi} \mathbf{\cdot} \mathbf{\nabla} ) \mathbfit{F}_{0}$, where the prime indicates an Eulerian perturbation.

Therefore, the perturbation to the emergent flux can be finally expressed as:
\begin{equation}
\label{eq:App:Lum:F:DeltaF}
\Delta \bar{F} = \Delta \mathbfit{F} \mathbf{\cdot} \hat{\mathbfit{r}} =F'_{r} + \xi_{r} \frac{{\rm d} F_{0}}{{\rm d} r},
\end{equation}
where the subscript italic $r$ indicates the radial component.

This results in a change in observed flux as:
\begin{equation}
\label{eq:App:Lum:F:int}
\Delta L_{F} = \int_{0}^{\pi / 2} \int_{0}^{2 \pi} h_{0} \Delta \bar{F} \hat{\mathbfit{r}} \mathbf{\cdot} \hat{\mathbfit{n}}_{\text{ob}} \text{d}S_{0}.
\end{equation}
With $F'_r$ and $\xi_r$ in $\Delta \bar{F}$ being evaluated at $(R, \theta_{\rm ob},\phi_{\rm ob})$,   the integral over $\phi_{\rm ob}$ is here again  given by equation~(\ref{eq:App:example_variable:new_coords}).  
After integration over $\theta_{\rm ob}$ we then obtain: 
\begin{align}
  \Delta L_{F} = \Re \bigg[ \frac{3 \pi}{4} R^{2} c \left( 1 + \frac{a + 2b}{15} \right) \bigg( F_{r}'(R) +  &  \xi_{r}(R) \frac{{\rm d} F_{0}}{{\rm d} r} \bigg)  \nonumber \\
  & \times \sin^{2} \theta_{0}  \text{e}^{2 \text{i}(\omega t - \phi_{0})} \bigg]
  \label{eq:App:Lum:F:explicit}
\end{align}
 where ${{\rm d} F_{0}}/{{\rm d} r}$ is evaluated at the surface.

\subsection{Surface area}
\label{sec:App:Lum:S}

In the observer's frame, the surface area element is defined as $\text{d}S_{(\rm ob,P)} = r^{2}_{(\rm ob,P)} \sin \theta_{(\rm ob,P)} \text{d} \theta_{(\rm ob, P)} \text{d} \phi_{(\rm ob,P)}$, where the subscript ${(\rm ob, P)}$ indicates that the coordinates are measured in the observer's frame and take into account the perturbation.

\noindent At equilibrium, the vector position of a point at the surface of the star is given by $\mathbfit{r} = R \hat{\mathbfit{r}} $ and its coordinates are $(R,\theta_{\rm ob}, \phi_{\rm ob})$.  When the star is perturbed, this point is moved and its vector position becomes: 
\begin{equation}
  \label{eq:App:Lum:S:r_{P}}
  \mathbfit{r}_{\text{P}} = \mathbfit{r} + \mathbf{\xi} = (R + \xi_{r}) \hat{\mathbfit{r}} + \xi_{\theta} \hat{\mathbf{\theta}}_{\rm ob} + \xi_{\phi} \hat{\mathbf{\phi}}_{\rm ob}. \end{equation}
Writing the displacement  $\mathbfit{r}_{\text{P}} -\mathbfit{r}$ as $ \Delta r \hat{\mathbfit{r}} 
+ R \Delta \theta_{ob} \hat{\mathbf{\theta}}_{\rm ob} + R \sin \theta_{\rm ob} \Delta \phi_{\rm ob} \hat{\mathbf{\phi}}_{\rm ob}$ and identifying with equation~(\ref{eq:App:Lum:S:r_{P}})  yields the perturbed coordinates of the point:
\begin{align}
  r_{(\rm ob,P)} & \equiv R + \Delta r =  R+ \xi_r , \nonumber \\
  \theta_{(\rm ob,P)} & \equiv \theta_{\rm ob} + \Delta \theta_{ob} = \theta_{\rm ob} + \frac{\xi_{\theta}}{R} , \nonumber \\
  \phi_{(\rm ob,P)} & \equiv \phi_{\rm ob} + \Delta \phi_{ob} = \phi_{\rm ob} + \frac{\xi_{\phi}}{R \sin \theta_{\rm ob}} . \nonumber
  \end{align}

\noindent  To first order in the perturbation we then obtain:
\begin{align}
  & r_{(\rm ob, P)}^{2}  = R^{2} + 2 R \xi_{r}, \nonumber \\
 & \sin \theta_{(\rm ob, P)}  = \sin \theta_{\rm ob} +
  \frac{\xi_{\theta}}{R} \cos \theta_{\rm ob} , \nonumber
  \end{align}
  and we use the Jacobian to change the variables over which we are integrating as:
\begin{equation}
\label{eq:App:Lum:S:Jacobian}
\text{d}\theta_{(\rm ob, P)} \text{d}\phi_{(\rm ob, P)} = \left[ 1 + \frac{1}{R} \left( \frac{\partial \xi_{\theta}}{\partial \theta_{\rm ob}} + \frac{1}{\sin \theta_{\rm ob}} \frac{\partial \xi_{\phi}}{\partial \phi_{\rm ob}} \right) \right] \text{d}\theta_{\rm ob} \text{d}\phi_{\rm ob},
\end{equation}
where second order terms have been neglected.

\noindent Using $\Delta \text{d}S = \text{d}S_{(\rm ob, P)} - \text{d}S_{0}$, where ${\rm d}S_0=
R^2 \sin \theta_{\rm ob} {\rm d} \theta_{\rm ob} {\rm d} \phi_{\rm ob}$, gives the expression for the change in surface area element as:
\begin{equation}
\label{eq:App:Lum:S:pert}
\Delta \text{d}S = R \left[ \sin \theta_{\rm ob} \left( 2 \xi_{r} +   \frac{\partial \xi_{\theta}}{\partial \theta_{\rm ob}}  \right) + \cos \theta_{\rm ob} \xi_{\theta} + \frac{\partial \xi_{\phi}}{\partial \phi_{\rm ob}}  \right] \text{d}\theta_{\rm ob} \text{d}\phi_{\rm ob} ,
\end{equation}
where second order terms in the perturbation have been neglected.

The change in observed flux due to this effect is given by
\begin{equation}
\label{eq:App:Lum:S:int}
\Delta L_{S} = \int_{0}^{\pi /2} \int_{0}^{2 \pi} h_{0} \bar{F}_{0} \hat{\mathbfit{r}} \mathbf{\cdot} \hat{\mathbfit{n}}_{\text{ob}} \Delta \text{d}S.
\end{equation}
To calculate this integral,  $\xi_{\phi}(\theta_* , \phi_*)$ has to be transformed into $ \xi_{\phi}(\theta_{\rm ob} , \phi_{\rm ob})$ in the expression~(\ref{eq:App:Lum:S:pert}) for
$\Delta \text{d}S $.  This involves summing up over the spherical harmonics $Y_l^{\mu}$, as indicated in equation~(\ref{eq:App:sph_coord_change}).  However, only the non--zero values of $\mu$ contribute to $\partial \xi_{\phi} / \partial \phi_{\rm ob}$, and the corresponding spherical harmonics  give zero when  integrated over $\phi_{\rm ob}$ in equation~(\ref{eq:App:Lum:S:int}).  Therefore, the term involving $\xi_{\phi}$ does not contribute.   As above,  using $\xi_{\theta} = \partial V / \partial \theta_{\rm ob}$ and  equation~(\ref{eq:App:example_variable:new_coords}) to calculate the integral over $\phi_{\rm ob}$, we obtain 
after integration over $\theta_{\rm ob}$: 
\begin{equation}
\label{eq:App:Lum:S:explicit}
\Delta L_{S} = \Re \left[ \frac{3}{2} \pi R c \left( 1 + \frac{a + 2b}{15} \right)  F_{0} \left[ \xi_{r}(R) - 3 V(R) \right] \sin^{2} \theta_{0}
  \text{e}^{2 \text{i}(\omega t - \phi_{0})} \right].
\end{equation}

\subsection{Limits}
\label{App:Lum:Limits}

The limits of the visible disc, and therefore of any integrals, are given by the points at which the vector towards the observer is tangential to the surface, such that $\hat{\bm{n}}_{\text{ob}} \bm{\cdot} \bm{r} = 0$, which can be re-expressed as $\hat{\bm{n}}_{\text{ob}} \bm{\cdot} ( \hat{\bm{r}} + \Delta \hat{\bm{n}} ) = 0$, where $\Delta \hat{\bm{n}}$ is the change in surface normal, given in section~\ref{App:Lum:normal}. For simplicity of notation, we can rewrite this as $\Delta \hat{\bm{n}} = A \hat{\bm{\theta}}_{\rm ob} + B \hat{\bm{\phi}}_{\rm ob}$,
where $A$ and $B$ are both first order in the perturbation and can be found by identifying this expression with equation~(\ref{eq:App:Delta_normal}).

Since $\hat{\bm{n}}_{\text{ob}}= \hat{\mathbfit{z}}$ in the observer's frame, this leads to the expression for the limits to be given by $\hat{\mathbfit{r}} \mathbf{\cdot} \hat{\mathbfit{z}} + A \hat{\mathbf{\theta}} \mathbf{\cdot} \hat{\mathbfit{z}} + B \hat{\mathbf{\phi}} \mathbf{\cdot} \hat{\mathbfit{z}} = 0$, which becomes $\cos \theta_{\rm up} - A \left( \theta_{\rm up} \right) \sin \theta_{\rm up} = 0$, where $\theta_{\rm up}$ is the coordinate of the point which delimits the visible disc and we have made it explicit that $A$ depends on $\theta_{\rm ob}$.  Since the limit is $\pi /2$ at equilibrium, we write $\theta_{\rm up}= \pi /2 + \delta \theta$.  The equation above then becomes, to first order in $\delta \theta$, $\delta \theta = -A_0 $, where $A_0$ is the value of $A$ at $\theta_{\rm ob} = \pi /2$.  We have used $A \left( \theta_{\rm up} \right) = A_{0} + \frac{\partial A}{\partial \theta_{\rm ob}} \delta \theta$ and neglected the second term on the right--hand side, which is second order in the perturbation.

This leads to the change in  observed flux:
\begin{equation}
\label{eq:App:DeltaL_L}
\Delta L_{L} = \int_{0}^{2 \pi} \int_{\frac{\pi}{2}}^{\frac{\pi}{2} - A_0} h_{0} \bar{F}_{0} \hat{\mathbfit{n}}_{\text{ob}} \mathbf{\cdot} \hat{\mathbfit{r}} \, \text{d} S_{0} .
\end{equation}
Since $A_0$ depends on $\phi_{\rm ob}$, we have to integrate over $\theta_{\rm ob}$ first.  This yields: 
\begin{multline}
\label{eq:App:DeltaL_L_int}
\Delta L_{L} = -\int_{0}^{2 \pi} c R^{2} \bar{F}_{0} \bigg( \frac{1 - a - b}{2} \sin^{2} A_0 + \frac{a + 2b}{3} \sin^{3} A_0 \\ 
- \frac{b}{4} \sin^{4} A_0 \bigg) \text{d} \phi_{\rm ob}.
\end{multline}
 As $A$ is a small quantity, $\sin A_{0}$ will also be a small quantity. The integral above will therefore be equal to $0$ to first order in small quantities, as the $\sin^{2} A_{0}$, $\sin^{3} A_{0}$ and $\sin^{4} A_{0}$ terms can be neglected.
Therefore, any change in observed flux due to a change in the limits of the integral can be neglected to first order.

\section{Radial velocity variation derivation}
\label{App:RV_var}

Here, the details of the derivation of the radial velocity variation are given and fully justified. The results derived below are used in section~\ref{sec:Methods:RV_var}.

The periodic change in shape of the star results in a periodic change in the velocity of any given surface element. Projecting this along the observer's line of sight gives the radial velocity (RV) which is proportional to the shift in wavelength caused by the motion (for the very non--relativistic motions considered here). Expressing this formally gives:
\begin{equation}
\label{eq:App:RV:RV_original}
\rm{v}_{\rm{RV}} = - \bm{\dot{r} \cdot \hat{n}}_{\text{ob}},
\end{equation}
where $\bm{r}$ is the vector from the centre of the star to the surface element in question.

To first order in perturbed quantities, this becomes:
\begin{equation}
\label{eq:App:RV:RV_first_order}
\rm{v}_{\rm{RV}} = - \bm{\dot{\xi} \cdot \hat{n}}_{\text{ob}} = \Re \left( - 2 \rm{i} \omega \bm{\xi \cdot \hat{n}}_{\text{ob}} \right)
\end{equation}

This can be encapsulated by a single curve by integrating over the disk, weighted by the observed flux, as done by \citet{Dziembowski1977}:
\begin{equation}
\label{eq:App:RV:disk_integrated:general}
\rm{v}_{\text{disc}} = \frac{\iint h \hat{\bm{r}} \bm{\cdot} \hat{\bm{n}}_{\text{ob}}\bar{F}_{0} \rm{v}_{RV} \rm{d}S_0}{\iint  h \hat{\bm{r}} \bm{\cdot} \hat{\bm{n}}_{\text{ob}} \bar{F}_{0}  \rm{d}S_0} = \frac{1}{2 \pi R^2} \int_0^{\pi /2} \int_0^{2 \pi} h \hat{\bm{r}} \bm{\cdot} \hat{\bm{n}}_{\text{ob}} \rm{v}_{RV} \rm{d}S_0,
\end{equation}
which can be analytically solved.

Using ${\rm d}S_0 = R^2 \sin \theta_{\rm ob} {\rm d} \theta_{\rm ob} {\rm d} \phi_{\rm ob}$, $ \hat{\bm{n}}_{\text{ob}} = \hat{\bm{z}}$, and ${\xi}_{\theta} = \partial V / \partial \theta_{\rm ob}$, we obtain:
\begin{align}
  \rm{v}_{\text{disc}} = \Re \left\{ \frac{-{\rm i} \omega }{\pi}  \right. & \int_0^{\pi /2} {\rm d} \theta_{\rm ob} h \cos \theta_{\rm ob} \sin \theta_{\rm ob}   \nonumber \\
 &  \left[   \cos \theta_{\rm ob} \int_0^{2 \pi} \xi_r \left(R, \theta_{\rm ob}, \phi_{\rm ob} \right) {\rm d} \phi_{\rm ob}  \right.  \nonumber \\  -  \sin \theta_{\rm ob} &  \left. \left.  \frac{\partial}{\partial \theta_{\rm ob}} \int_0^{2 \pi} V \left(R, \theta_{\rm ob}, \phi_{\rm ob} \right) {\rm d} \phi_{\rm ob} \right] \right\}
\end{align}

\noindent The integrals over $\phi_{\rm ob}$ are given by equation~(\ref{eq:App:example_variable:new_coords}), which yields: 
\begin{multline}
    \label{eq:App:RV:v_disc:explicit_integral_theta}
    \rm{v}_{\text{disc}} = \Re  \left\{- 3 \rm{i} \omega \sin^{2} \theta_{0} \rm{e}^{2 \rm{i} (\omega t - \phi_{0})}  \right. \times  \\
  \int_{0}^{\frac{\upi}{2}}  \rm{d}\theta_{\rm ob} \mathnormal{c} \left[ (1 - a - b) + (a + 2b) \cos \theta_{\rm ob} - b \cos^{2} \theta_{\rm ob} \right] \cos \theta_{\rm ob} \sin \theta_{\rm ob} \times \\
  \left.   \left[  \xi_{r}(R) \left( 3 \cos^{2} \theta_{\rm ob}  - 1 \right) \cos \theta_{\rm ob}  +   6 V(R) \cos \theta_{\rm ob} \sin^{2} \theta_{\rm ob}  \right]  \right\}  .
\end{multline} 

\noindent Evaluating this integral gives the final expression for the disc-integrated radial velocity as: 
\begin{align}
\label{eq:App:RV:disk_integrated:final}
  \rm{v}_{\text{disc}} = \Re & \left\{- \frac{4}{5} \rm{i} \omega \right.  c \sin^{2} \theta_{0} \rm{e}^{2 \rm{i} (\omega t - \phi_{0})}  \nonumber \\
 & \left.  \left[ \left( 1 - \frac{a}{16} + \frac{b}{56} \right) \xi_r(R) + 3  \left( 1 - \frac{3 a}{8} - \frac{5b}{28} \right) V(R) \right] \right\} .
\end{align}

\section{Non--disc--integrated approaches}
\label{App:non-disc}

Whilst we are unable to spatially resolve the surface of the star, there are still some possibilities for observing signals without simply taking the average effect.

\subsection{Inhomogeneous line broadening}
\label{App:non-disc:inhomogeneous}

Because of the different motions across the visible surface of the stellar disc, different parts of the stellar surface will emit light which has been Doppler shifted differently. Instead of taking the average of the total Doppler shift by looking at the variation in the central wavelength, we can examine the variation in the line--broadening by tracking how the brightness at each wavelength changes over the course of the oscillation.

By summing the  observed flux contributions at each wavelength we determine the overall shape of the line--broadening that results from the non--uniform surface motion. This can be expressed as:
\begin{equation}
\label{eq:App:non-disc:broadening}
n_{\rm ti} = \frac{1}{L_{0}} \left( \frac{\text{d}L(\rm{v})}{\text{d}\rm{v}}  \right)_{\rm ti} = \frac{1}{L_{0} \Delta \rm{v}} \sum_{\rm{v}-\frac{\Delta \rm{v}}{2}}^{\rm{v}+\frac{\Delta \rm{v}}{2}} h \cos \theta_{\rm ob} \bar{F}_{0} \text{d}S_{0},
\end{equation}
where $\Delta \rm{v}$ is the width of the bin, ${\rm v}$ is the radial velocity equivalent to the change in wavelength, and $n_{\rm ti}$ is the normalised lineshape, such that its total area is equal to 1.  This sums over only the  visible area elements that fall within the bin centred on $\rm{v}$.

In general, the resulting shape will not be simple, and can have significant  flux at radial velocities much greater than the disc--integrated value. All viewing angles will produce this broadening, with greatest time--dependence when the system is viewed edge--on, and a steady state when viewed at either pole.

This simplified case only accounts for broadening arising due to the tidal motion of the stellar surface. If other sources of line broadening are present the overall lineshape will be the result of all of the different sources combined. As a simple example, the tidal broadening is here briefly discussed in the context of thermal broadening to set out the general method for incorporating different broadening sources.

The lineshape arising from thermal broadening for a stationary surface element is
\begin{equation}
\label{eq:app:thermal}
n_{\rm th}(T, {\rm v}) = \sqrt{\frac{\mu m_{\rm H}}{2 \pi k_{B} T}} \exp \left( - \frac{\mu m_{\rm H} {\rm v}^{2}}{2 k_{B} T} \right).
\end{equation}
where $\mu$ is the molar mass of the source particle for the line being studied. 
For thermal broadening, the intensity emitted at a given radial velocity also depends upon the temperature of the surface element.

To combine the effect of both the tidal motion and the thermal broadening for a given surface element, the two lineshapes must be convolved. The thermal lineshape is given by equation~\ref{eq:app:thermal}, and the tidal lineshape for a given surface element is a delta function, offset by the radial velocity of that surface element, ${\rm v}_{\rm RV}$, giving $n_{\rm ti, elem} = \delta({\rm v - v_{RV}})$. The total lineshape for a particular surface element, $n_{\rm tot, elem}$, is then the convolution of these two lineshapes, given by
\begin{equation}
n_{\rm tot, elem} = \int_{- \infty}^{\infty} n_{\rm th}(T, x-v) n_{\rm ti, elem}(x-v_{\rm RV}) {\rm d}x
\label{eq:app:convolution}
\end{equation}
which can be written explicitly as
\begin{equation}
n_{\rm tot, elem} = \sqrt{\frac{\mu m_{H}}{2 \pi k_{B} T}} \exp \left( - \frac{\mu m_{H} ( {\rm v} - v_{\rm RV})^{2}}{2 k_{B} T} \right).
\end{equation}

In order to calculate the observable lineshape, $n_{\rm obs}$, we integrate over the visible disc, giving
\begin{equation}
n_{\rm obs} = \frac{1}{L_{0}} \int n_{\rm tot, elem} {\rm d}L.
\end{equation}

If equation~\ref{eq:app:convolution} is substituted in, we arrive at a triple integral giving the observable lineshape, as
\begin{multline}
n_{\rm obs} = \frac{1}{2 \pi} \int_{- \infty}^{\infty} {\rm d}x \int_{0}^{2 \pi} {\rm d}\phi_{\rm ob} \int_{0}^{\pi / 2} {\rm d}\theta_{\rm ob} \cos \theta_{\rm ob} \sin \theta_{\rm ob} h_{0} \delta(x - v_{\rm RV}) \\
\times \sqrt{\frac{\mu m_{\rm H}}{2 \pi k_{B} T}} \exp \left( - \frac{\mu m_{\rm H} (x - {\rm v})^{2}}{2 k_{B} T} \right)
\end{multline}
where both ${\rm v}_{\rm RV}$ and $T$ are dependent upon $\theta_{\rm ob}$ and $\phi_{\rm ob}$. Due to this dependence on the location on the disc, the integral over $x$ must be undertaken first. Therefore although the final lineshape is a result of the two different mechanisms, it is not equivalent to the convolution of the disc-integrated lineshapes arising from the individual mechanisms. This makes it difficult to compute.

This is because the thermal lineshape of each surface element is much wider than the disc-integrated lineshape arising from the tidal motion. Therefore the numerical integration must be sufficiently fine at all levels to resolve the line broadening from both mechanisms. This may still be possible, but does not lend itself to a nice analytical expression.

Two extreme cases may be considered -- the case of uniform surface temperature, and the case of negligible surface motion. In the former case, $T$ is independent of the location on the disc, and the observable lineshape reduces to the convolution of the disc-integrated thermal and tidal lineshapes. In the latter $\rm v_{RV} \approx 0$, eliminating the need for the convolution. However, some time-dependent broadening will still occur, as the temperature distribution over the visible disc will change over time.

To investigate the perturbation to the thermal broadening, we express the temperature as $T = T_{0} [1 + \Delta T(R, \theta_{*}, \phi_{*}, t) / T_{0}]$, where $\Delta T(R, \theta_{*}, \phi_{*}, t)$ is the Lagrangian perturbation to the surface temperature, including the full spatial dependence as given in equation~\ref{eq:pertquant}, and $|\Delta T(R, \theta_{*}, \phi_{*}, t)| \ll T_{0}$. The thermal lineshape can then be expanded around the effective temperature, $T_{0}$, giving
\begin{multline}
n_{\rm th}(T, {\rm v}) = n_{\rm th}(T_{0}, {\rm v}) + \Delta T(R, \theta_{*}, \phi_{*}, t) \frac{\partial n_{\rm th}}{\partial T} (T_{0}, {\rm v}) \\ + \mathcal{O} ( \Delta T^2(R, \theta_{*}, \phi_{*}, t) / T_{0}^{2} ).
\end{multline}
This gives
\begin{equation}
n_{\rm th}(T, {\rm v}) \approx n_{\rm th}(T_{0}, {\rm v}) \left[ 1 + \frac{\Delta T(R, \theta_{*}, \phi_{*}, t)}{T_{0}} \left( \frac{\mu m_{\rm H} {\rm v}^{2}}{2 k_{B} T_{0}} - \frac{1}{2} \right) \right]
\end{equation}
where the only dependence on $\theta_{*}$ and $\phi_{*}$ is in $\Delta T(R, \theta_{*}, \phi_{*}, t)$.

This can be integrated over the visible disc to give 
\begin{equation}
n_{\rm obs, th} = n_{\rm th}(T_{0}, {\rm v}) \left[ 1 + \alpha(t) \left( \frac{\mu m_{\rm H} {\rm v}^{2}}{2 k_{B} T_{0}} - \frac{1}{2} \right) \right]
\end{equation}
where $\alpha(t)$ is a small quantity which oscillates over time, given by
\begin{equation}
\alpha(t) = \frac{3}{8} \frac{\Delta T(R)}{T_0} c \left( 1 + \frac{a + 2b}{15} \right) \sin^{2} \theta_{0} {\rm e}^{2 {\rm i} (\omega t - \phi_{0})}.
\end{equation}
Thermal broadening, or other sources of broadening  (see, e.g., \citealt{Gray2005}), may therefore cause time-dependent broadening signals due to the time-varying perturbation to the stellar surface.   An estimate for $\Delta T(R)/T_0$ can be obtained from the boundary condition at the surface: $ 4 \Delta T(R)/ T_0 =  \Delta F_r(R) / F_{r0},$  where $\Delta F_r(R)$ is given by equation~(\ref{eq:lagFr}).  On  the right--hand side of this equation,  the term involving $\xi_r$   is very small compared to $F'_r$ for the periods of interest, as can be seen in Figure~\ref{fig:photometric_comparisons}.  Therefore, $\Delta T(R)/ T_0 \simeq  F'_r(R)/(4 F_{r0})$.  From Figure~\ref{fig:log_fractional}, we then see that $\Delta T(R)/ T_0$ is on the order of a few times $10^{-4}$ or a few times $10^{-5}$ for the models with perturbed or frozen convection, respectively, and periods of a few days.

\subsection{Observations during transit}
\label{App:non-disc:transit}

In the case of a transiting planet, a portion of the star is blocked from view by the planet, enabling the behaviour of the surface at that specific location to be studied. As the stellar oscillations occur at twice the orbital frequency, subtracting the signal whilst the planet is occluded (secondary eclipse) from the signal during the transit will isolate the signal from the blocked portion of the stellar surface.

\begin{figure*}
	\centering
	\includegraphics[width=1.5\columnwidth]{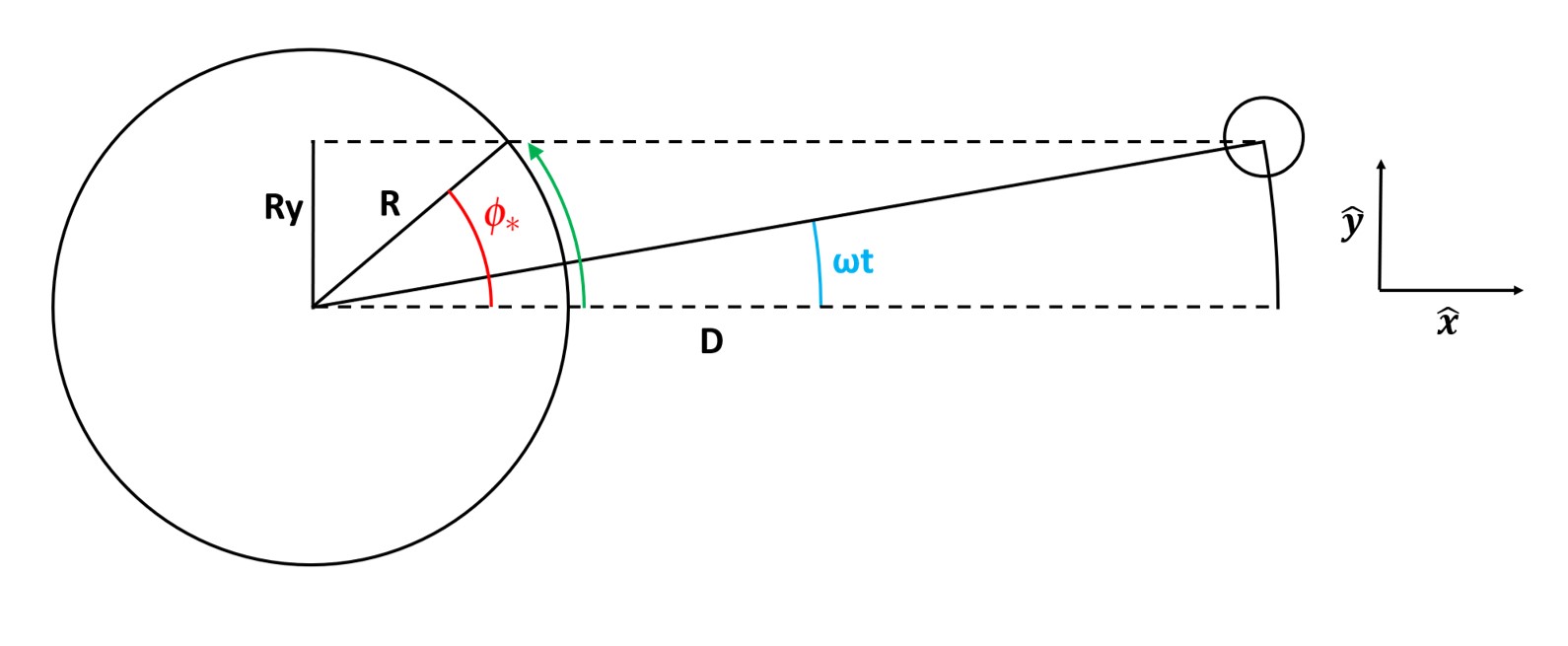}
	\caption{This figure depicts the coordinates used in section~\ref{App:non-disc:transit}, as viewed from above the plane of the planet's orbit, in the case that $\delta \theta = 0$. The observer is at infinity in the $\bm{\hat{x}}$ direction, and at $t=0$ the planet is at $(D, 0)$ in the $(x, y)$ plane. After time $t$ the planet has moved to be approximately at $(D, D \omega t)$ and the location which is blocked by the planet's silhouette has moved from $(R, 0)$ to $(R \cos \phi_{*}, R y)$, as shown by the green arrow, where $y$ is the the dimensionless coordinate introduced for eq.~\ref{eq:App:non-disc:v_RV_t}.}
	\label{fig:app:non-disc-integrated:transit}
\end{figure*}

We define $\delta \theta = \upi / 2 - i$, where $i$ is the orbital inclination. Since the planet is transiting,  $|\delta \theta| \le R / D \ll 1$ is a small quantity.  We therefore describe the stellar surface as if $\delta \theta = 0$ (that is, as if we view it exactly edge--on), which corresponds to $\theta_0=\pi /2$, but allow the planet's silhouette to deviate from being exactly edge--on.  As the transit occurs over an interval of time short compared to the oscillation period, we assume that each surface element moves with constant velocity for the duration of the transit, and that the planet's motion is linear. These introduce fractional errors of order $R/D$. This arrangement is shown diagrammatically in Figure~\ref{fig:app:non-disc-integrated:transit}.

The centre of the silhouette will cross the stellar surface according to:
\begin{equation}
    R \sin \theta_{*} \sin \phi_{*} = D \omega t ,
  \end{equation}
  with:
\begin{equation}
    R \cos \theta_{*} = D \delta \theta ,
  \end{equation}
   As we define the origin of the time coordinate as the epoch of inferior conjunction  $t=0$ corresponds to the centre of the transit. This gives:
  \begin{align}
  \cos \theta_{*} & = \frac{D}{R} \delta \theta, \nonumber \\
  \sin \phi_{*} & = \frac{D \omega t}{R \sqrt{1 - (\frac{D \delta \theta}{R})^{2}}} . \nonumber
  \end{align}

\subsubsection{Radial velocity}
\label{App:non-disc:transit:RV}

We have $\xi_{\theta}= \partial V / \partial \theta_* $, $\xi_{\phi} = (\partial V / \partial \phi_* ) / \sin \theta_*$ and ${\hat{n}}_{\text{ob}} = \bm{ \hat{x}}$.  Both $\xi_r$ and $V$ are given by equation~(\ref{eq:pertquant}), in which $\omega t$ is taken to be zero as the transit happens over a time interval short compared to the period of the oscillations, as already mentioned above.  Therefore, 
equation~\ref{eq:App:RV:RV_first_order} can be written as:
\begin{align}
    {\rm v_{RV}} (\theta_{*}, \phi_{*}) & = \Re \left\{ -6  \rm{i} \omega  \sin \theta_{*}
   \rm{e}^{ -2 \rm{i}   \phi_* }  \right. \nonumber \\
  & \left. \left[ \xi_{r}(r) \sin \theta_{*} \bm{\hat{r}} + 2 V(r) \left( \cos \theta_{*} \bm{\hat{\theta}_{*}} - \rm{i} \bm{\hat{\phi}_{*}} \right) \right] \bm{ \cdot \hat{x}} \right\} ,
  \label{eq:App:non-disc:v_RV_basic}
\end{align}
which we now  re--express in terms of $\delta \theta$ and $t$. 

Introducing the coordinates $y = \frac{D \omega t}{R}$ and $z = \frac{D \delta \theta}{R}$ we get: 
\begin{multline}
\label{eq:App:non-disc:v_RV_t}
{\rm v_{RV}} (y, z) = \Re \Bigg( \frac{- 6 \omega}{1 - z^{2}} \left[ 2 y \sqrt{1 - y^{2} - z^{2}} + {\rm i} \left( 1 - 2 y^{2} - z^{2} \right) \right] \\
\times \left\{ \sqrt{1 - y^{2} - z^{2}} \left[ \xi_{r} + z^{2} \left( 2 V - \xi_{r} \right) \right] + 2 {\rm i} V y \right\} \Bigg),
\end{multline} 
 which is valid over the range $y^{2} + z^{2} < 1$.

To describe the approximate width of the obscured ${\rm v_{RV}}$, we can use
\begin{equation}
\label{eq:App:non-disc:v_RV:width}
    \Delta {\rm v_{RV}} \approx max \left(   \frac{\partial {\rm v}_{RV}}{\partial y} \frac{R_{P}}{R},  \frac{\partial {\rm v}_{RV}}{\partial z} \frac{R_{P}}{R}  \right)
\end{equation}
where $R_{P}$ is the planetary radius.  The observed flux of the blocked region can also be found, as
\begin{equation}
\label{eq:App:non-disc:v_RV:depth}
    L_{\rm blocked} = \upi R_{P}^{2} h \bar{F}_{0}.
\end{equation}

Equations \ref{eq:App:non-disc:v_RV:width} and \ref{eq:App:non-disc:v_RV:depth} are valid once the entirety of the planet's silhouette is visible, or equivalently whilst $y^{2} + z^{2} < (1 - \frac{R_{P}}{R} )^{2}$.

The disc-integrated RV signal (given in equation~\ref{eq:Methods:RV:disk_integrated:final}) defines a line of solutions for $\xi_{r}$ and $V$, as it is one measurement being used to determine two complex values. By observing the variation in the radial velocity signal which is blocked during the transit the components can be separated, giving an independent measure of $\xi_{r}$ and $V$. Exactly how this works out in practise with real and imaginary components may not be straightforward, but it would be another way to look at the system and gain information.

\subsubsection{Luminosity}
\label{App:non-disc:transit:L}

The same technique could also be applied to variations in observed flux arising as a result of the transit, however this signal would be the result of the blocking of both the equilibrium flux and the perturbed flux. As such, the extra blocking due to the obscured perturbed flux would be very small, on the order of $\frac{R_{P}^{2} F'}{R^{2} F_{0}}$.

\bsp	
\label{lastpage}
\end{document}